\PassOptionsToPackage{dvipsnames}{xcolor}
\documentclass[preprint]{aastex631}
\pdfoutput=1
\usepackage{amsmath}
\usepackage{listings}
\lstdefinelanguage{pseudocode}{}
\usepackage{capt-of}   % don't use the package caption -- it causes some kind of conflict.  7/29/15
\usepackage{empheq}
\usepackage{enumerate}
\usepackage{hyperref}
\usepackage{hhline}
\usepackage{pythonhighlight}
\usepackage{longtable}
\usepackage{numprint}
\usepackage{pifont}
\usepackage{xspace}
\usepackage{mwe}
\usepackage{multirow}
\usepackage[mathlines]{lineno}
\usepackage[section]{placeins}
\lstnewenvironment{mypython}[1][]{\lstset{style=mypython,#1}}{}

\newcommand{\ed}{\textcolor{black}}
\newcommand{\edr}{\textcolor{black}}

\newcommand{\ho}{\ensuremath{H_0}\xspace}

\newcommand{\ser}{S\'{e}rsic\xspace}

\newcommand{\hst}{\emph{HST}\xspace}
\newcommand{\jwst}{\emph{JWST}\xspace}
\newcommand{\rst}{\emph{RST}\xspace}
\newcommand{\euc}{\emph{Euclid}\xspace}

\newcommand{\RST}{\emph{Roman Space Telescope}\xspace}

\newcommand{\tE}{\ensuremath{\theta_E}\xspace}

\newcommand{\numlomassresnet}{\ensuremath{17}\xspace}
\newcommand{\numlomassunet}{\ensuremath{1.1}\xspace}
\newcommand{\numlomassunetninenine}{\ensuremath{7.0}\xspace}

\newcommand{\numsmallthetae}{240\xspace}

\newcommand{\twopr}{\ensuremath{^{\prime \prime}}\xspace}

\submitjournal{ApJ}

\shorttitle{
ML-Driven Strong Lens Discoveries}
\begin{document}
% \title{ML-Driven Strong Lens Discoveries:\\ 
% Down to \tE $\sim 0.03''$ and $M_\mathrm{halo}^{lens}< 10^{11} M_\odot$}
\title{ML-Driven Strong Lens Discoveries:\\ 
Down to \tE $\sim 0.03 \twopr$ and $M_\mathrm{halo}< 10^{11} M_\odot$}
% \title{ML-Driven Strong Lens Discoveries:\\ 
% From ``Conventional'' to $< 10^{11} M_\odot$ Halo Mass}
% \title{Toward Finding Every Strong Lens}

% \correspondingauthor{Xiaosheng Huang}
% \email{xhuang22@usfca.edu}
% %, august.muench@aas.org}
\author[0000-0002-1804-3960]{Ethan~Silver}
\affiliation{Department of Physics, Harvard University, Cambridge, MA 02138}
\author[0009-0007-9369-2227]{R.~Wang}
\affil{Department of Computing, Data Science, and Society, University of California, Berkeley, Berkeley, CA 94720}
\author[0000-0001-8156-0330]{Xiaosheng~Huang}
\affiliation{Department of Physics \& Astronomy, University of San Francisco, San Francisco, CA 94117-1080}
\author[0000-0002-9836-603X]{A.~Bolton}
\affiliation{SLAC National Accelerator Laboratory, Menlo Park, CA 94025, USA}
\author[0000-0002-0385-0014]{C.~Storfer}
\affil{Institute for Astronomy, University of Hawaii, Honolulu, HI 96822}

\author{S.~Banka}
\affil{Department of Electrical Engineering \& Computer Sciences, University of California, Berkeley, Berkeley, CA 94720}

% \doublespace

\begin{abstract}
We present results on extending the strong lens discovery space down to much smaller Einstein radii ($\theta_E\lesssim0.03^{\prime\prime}$) and much lower halo mass ($M_\mathrm{halo}<10^{11}M_\odot$) through the combination of \textit{JWST} observations and machine learning (ML) techniques. First, we forecast detectable strong lenses with \textit{JWST} using CosmoDC2 as the lens catalog, and a source catalog down to 29th magnitude. By further incorporating the VELA hydrodynamical simulations of high-redshift galaxies, we simulate strong lenses. We train a ResNet on these images, achieving near-100\% completeness \textit{and} purity for ``conventional" strong lenses ($\theta_E\gtrsim 0.5^{\prime\prime}$), applicable to \textit{JWST}, \textit{HST}, the \textit{Roman Space Telescope} and \textit{Euclid} VIS. \textit{For the first time}, we also search for very low halo mass strong lenses ($M_{halo}<10^{11}M_\odot$) in simulations, with $\theta_E\ll 0.5^{\prime\prime}$, down to the best resolution ($0.03^{\prime\prime}$) and depth (10,000~sec) limits of \textit{JWST} using ResNet. A U-Net model is employed to pinpoint these small lenses in images, which are otherwise virtually impossible for human detection. Our results indicate that \textit{JWST} can find $\sim 17$/deg$^2$ such low-halo-mass lenses, with the locations of $\sim 1.1$/deg$^2$ of these detectable by the U-Net at $\sim100$\% precision (and $\sim 7.0$/deg$^2$ at a 99.0\% precision). To validate our model for finding ``conventional" strong lenses, we apply it to \textit{HST} images, discovering two new strong lens candidates previously missed by human classifiers in a crowdsourcing project \citep{Garvin_2022}. This study demonstrates the (potentially ``superhuman") advantages of ML combined with current and future space telescopes for detecting conventional, and especially, low-halo-mass strong lenses, which are critical for testing CDM models.

\end{abstract}

\section{Introduction}\label{sec:introduction}

Strong lensing is playing an increasingly significant role in astrophysics, cosmology, and fundamental physics.
It has been used to study how dark matter (DM) is distributed in galaxies and galaxy clusters 
\citep[e.g.,][]{bolton2006a, meneghetti2020a}
and constrain the inner density profile of the dark matter halo  \citep[e.g.,][]{auger2010a, shajib2021a}.
It is uniquely suited to probe ``dark" low-mass DM halos (i.e., without baryonic matter)
beyond the local universe \citep[e.g.,][]{vegetti2014a, cagansengul2020a}, and determine the nature of DM.
\edr{
Constraints on DM from strong lensing are therefore powerful and complementary to \citep[e.g.,][]{boddy2022a, vegetti2024a} the limits terrestrial experiments can place on DM particle candidates.\footnote{\edr{Although, at this time, to establish direct connection between astrophysical constraints on (typically phenomenological) DM models  with underlying particle physics constraints, often further specific assumptions need to be made \citep[][e.g., assumed a dark photon model]{kaplinghat2016a}.}}} 
% [\textcolor{blue}{add one citation?}].
% ---where terrestrial direct DM detection experiments so far have failed.

Given the tension in the \ho measurement at $\sim 5\sigma$ \citep[e.g.,][]{riess2019a, freedman2020a, planck2020a}, 
multiply-lensed quasars and supernovae (SNe) are 
ideal for measuring time delays and \ho 
and in the case of Type~Ia SNe, with the added benefit of standardizable luminosity \citep[e.g.,][]{refsdal1964a, kelly2023a, suyu2024a}.

\edr{For the properties of dark energy,} strong lensing can provide independent measurements of its equation of state
\citep[e.g.,][]{linder2011a, treu2016a, li2024a}.
In this respect, lensing systems with sources at more than one redshift are especially powerful, both at the galactic \citep[e.g.,][]{Collett_2012, collett2014a, Linder_2016, Sharma_2022} and cluster scales \citep[e.g.,][]{caminha2022a, sheu2024a}.
Future LSST and \euc surveys will likely find many such compound lenses \citep[e.g.,][]{Sharma_2023}.

\edr{Furthermore,} 
by combining strong lensing time delay of multiply-imaged time-varying sources 
and SNe~Ia distance measurements
one can measure the spatial curvature \citep[e.g.,][]{collett2019a}, 
and test the 
Friedmann-Lema\^{i}tre-Robertson-Walker (FLRW) metric \citep{rasanen2015a}.
For nearby strong lensing galaxies, 
extra-galactic tests of General Relativity can be performed 
by combining lens modeling with spatially resolved 
stellar kinematic observations \citep{collett2018a}. 

\edr{Finally,} gravitationally lensed gravitational waves can be used to determine cosmological parameters \citep{liao2017a, wei2017a} and test fundamental physics \citep{collett2017a, fan2017a}.

\edr{For lensed source science,} strong lensing as a cosmic telescope magnifies spectral \citep[e.g.,][]{cornachione2018a} 
and spatial features \citep[e.g.,][]{vanzella2020a} of the lensed distant galaxies,
providing the only way to study the morphology and internal structures of galaxies at sub-kpc scales at high redshifts.

This of course is not an exhaustive list, 
and therefore every strong lens is worth discovering, especially those with high-resolution imaging data from space telescopes.
% \hst and \jwst.}
\citet{marshall2009a} proposed an early, lens-modeling based, approach for automated lens search in \hst  observations. 
% \ed{In order to find strong lenses, we contrast the current
In recent years, the two main approaches for lens search are crowd sourcing and machine learning (ML).
% , which nowadays nearly always mean neural networks.  
The most notable recent results with the first approach are those of \citet{Garvin_2022}.
They found 198 new lenses in the \hst archive data from 2002 - 2018 using crowdsourcing. However, as we will show in this paper, there are two related limitations to this approach:  1) crowdsourcing is not scalable; and 2) crowdsourcing lens discoveries are incomplete. 
\citet{Metcalf_2019} presented the first strong lens search challenge using simulated data for LSST and \euc.
\citet[][]{lanusse2018a}, using the Residual Network (ResNet) architecture,
outperformed other algorithms, 
both supervised or unsupervised \citep[e.g.,][]{cheng2020a},
in terms of the area under the ROC curve (or AUC, a commonly used metric) for simulated \euc data.
 
% L18 was the first to use the Residual Network (ResNet) architecture for strong lens search. 
Among the many successful lens searches in ground-based observations \citep[e.g.,][]{jacobs2017a, jacobs2019a, canameras2021a, shu2022a},  
\citet[][H20]{Huang_2020} applied ResNet \citep[using the architecture in][]{lanusse2018a} to ground-based data for the first time. 
\citet[][H21]{Huang_2021} further improved it by adding ``shielding'' layers. 
Using ResNet, H20 and H21, together with \citet{Storfer_2024}, found $\sim 3500$ new lens candidates.
\edr{With ground-based data, human inspection is still necessary. For example, after evaluating the ResNet model on all the data, \cite{Storfer_2024} chose a probability threshold that gave $\sim47,000$ initial candidates. From these, human inspection identified $1895$ lens candidates. Human grading is based on criteria that 
%indicate lensing, typically 
include curvature, elongation, possible counter-images, and color difference between the lens and the lensed image, typically with the former being redder than the latter.}

In this paper, we apply this architecture to space-based observation, both simulated and real.
We will show below that using ResNet effectively addresses the two issues with crowdsourcing pointed out above.
This has significant implications for both current (\hst, \jwst, and \euc) and future datasets.
%For example, with a field of view that is $100 \times$ that of \hst, the Roman Space Telescope is expected to detect about 17,000 strong gravitational lenses \citep{Weiner_2020}.
To search for strong lenses in these datasets with the highest possible completeness and purity, ML is the only viable approach.

\ed{
In this work, we will first demonstrate the high efficacy of ResNet models in identifying ``conventional'' strong lenses, 
\edr{which we define as lenses with Einstein radii $\gtrsim0.5''$ and halo masses $\sim10^{13}M_\odot$}.
In addition, for the first time, we show the ResNet's ability to detect lenses with masses far lower than conventional strong lensing systems.
Such lenses will have much smaller Einstein radii.
% The lensed Type~Ia supernova SN 2016geu has a notably low Einstein radius of $\theta_E=0.29''$ \citep{Goobar_2017, More_2017, Mortsell_2020}.
So far, the smallest \tE for strong lensing systems is $\sim 0.17''$ \citep[See Figure~4 of][]{Goobar_2023}, namely, the lensed SN Zwicky \citep{Goobar_2023, Pierel_2023} and the lensed blazar B2 0218+35 \citep{Barnacka_2016}.}

As we will show below our ResNet models are expected to find many lenses with \tE an order of magnitude smaller, down to the resolution limit of \jwst, with halo mass reaching the lower bound of the lens catalog used in this work (CosmoDC2), $10^{10} M_\odot$.
The vast majority of these small lenses are virtually impossible to identify with crowdsourcing.
% \edr{There is no available real training data of these small lenses with which to train models, and they would likely be quite difficult to identify in lower-resolution ground-based observations (though that may not be impossible (Hsu et al. in prep)).
\edr{A viable path to being able to detect these small lenses involves simulated data of high-resolution space-based observations.
We then use the ResNet model trained on simulated \jwst data to demonstrate the possibility of confidently identifying them.}

\edr{We further employ the U-Net model to detect the \emph{locations} of these small \tE systems in image cutouts. U-Net was first designed by \cite{Ronnegerger_2015}
and utilized for the medical field to segment
human organs, such as the liver and kidney, on
medical scans. 
In the astrophysics field, \cite{Ostdiek_2022_2} 
used the U-Net to find dark matter subhalos in simulated conventional strong lenses and \cite{Caldeira_2019} used the U-Net for lensing reconstruction.
We therefore decide to apply the U-Net for detecting the location of small Einstein radius lenses.}

\edr{Some of these systems have lenses with halo mass down to $M_{\textrm{Halo}} \lesssim 10^{11} M_{\odot}$.
This potentially opens the door for a new way of testing the CDM model. CDM is virtually untested below the mass scale of typical galactic halos, at $M_{\textrm{Halo}} \sim 10^{13} M_{\odot}$. 
For example, Figure~11 in \citet{driver2022a} shows the halo mass function measured down to $M_{\textrm{Halo}} \sim 10^{13} M_{\odot}$ and that more than half of
the DM mass in the universe is stored in halos with $M_{\textrm{Halo}} \lesssim 10^{13} M_{\odot}$, as CDM predicts an abundance of them \citep[e.g.,][]{jenkins2001a, warren2006a, springel2008a}.
Finding low halo mass lenses thus represents the first step toward testing CDM below the typical galactic scale.
However, discovering these small systems in a systematic way is not humanly possible. An ML/AI approach is a must. 
As we will show in this work, many of these lenses are imminently discoverable and are at the frontier of testing CDM.}

By applying these ML techniques, 
we expect to not only find nearly every conventional lens from \hst, \jwst, the \RST (\rst),
and \euc (VIS) as well, but also a large number of low-mass galaxy lenses with halo mass down to $10^{10} M_\odot$ in \jwst data. 
\emph{This opens up a new discovery window.}
Prior application of ML in strong lens search has only been for time-saving purposes, i.e., it narrows down the image cutouts for human inspection.  
In this work, using ResNet and U-Net, we show that we can discover strong lens systems that most likely will escape human notice, specifically, ``conventional''-sized strong lenses with low arc-to-lens contrast and systems with Einstein radii down to the detection limits of \jwst ($\lesssim 0.03''$).
That is, for the first time, ML techniques are poised to make ``superhuman'' strong lens discoveries.

\ed{
We divide this paper into three main sections, covering the three major parts of our work:
1. A forecast for discoverable strong lenses in \emph{JWST} observations (Methods: Section \ref{sec:methods_forecast}, Results: Section \ref{sec:results_forecast}, Discussion: Section \ref{sec:disc_forecast}).
2. A pipeline for simulating strong lensing systems using realistic hydrodynamic simulations followed by applying ResNet models to identify them (Methods: Section \ref{sec:methods_trained_class_model}, Results: Section \ref{sec:results_trained_class_model}, Discussion: Section \ref{sec:discuss}).
3. Using the U-Net model
% architecture applied to our simulations, in order 
to detect and localize strong lenses with the lowest possible halo mass (Methods: Section \ref{sec:methods_unet}, Results: Section \ref{sec:results_unet}, Discussion: Section \ref{sec:pipeline_performance}).
}
\section{Strong Lensing Forecast with \jwst}\label{sec:forecast}

The first part of our work is performing a forecast for the numbers of strong lenses that could be observed by \jwst. In this section, we first present our methods of how we count the strong lenses, and then present our predicted results.

\subsection{Methods: Lensing Forecasts}\label{sec:methods_forecast}

In order to forecast the numbers of strong lenses, we combine the density of sources with the areas on the sky covered by strong lenses (essentially, cross section). 
Because a halo's Einstein radius increases with source redshift, we calculate the lensed area and integrate our forecast over many redshift bins.

\subsubsection{Integration over redshift bins}\label{sec:methods_integration}
% \begin{figure}
%     \centering
%     \includegraphics[width=.9\textwidth]{images/rectangle.png}
%     \caption{Rectangular cutout from the Williams et al. (2018) \citep{Williams_2018} catalog used for integration}
%     \label{fig:rectangle}
% \end{figure}

In order to forecast the frequency of lensing, we count the number of sources at each redshift and multiply it by the fraction of sources that will be lensed at that redshift. The source catalog used is the \textit{JWST} extragalactic mock catalog JAGUAR (JAdes extraGalactic Ultradeep Artificial Realizations), generated according to \citet[][W18]{Williams_2018} at the depth of the JADES (\emph{JWST} Advanced Deep Extragalactic Survey) survey. In the JAGUAR catalog, all galaxies are modeled with S\'{e}rsic profiles \citep{Sersic_1968}, and their morphological parameters follow the redshift evolution characterized by \emph{HST} surveys \citep{Williams_2018}.

In addition to a catalog of galaxies and their properties, the JAGUAR catalog also provides simulated images in different \textit{JWST} bands. For this work, we simply take the F115W band, and use this band throughout for consistency.

\begin{figure}
    \centering
    \includegraphics[width=\textwidth]{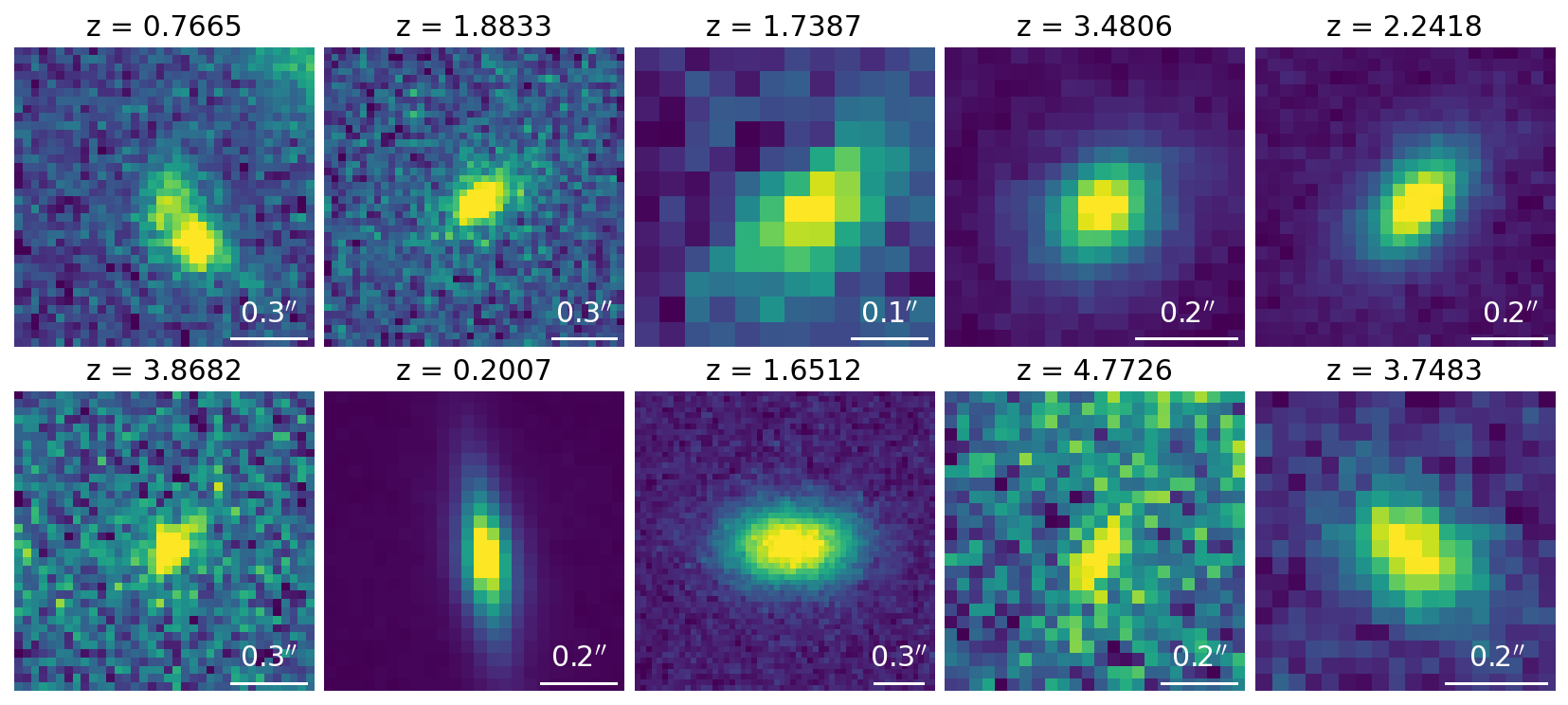}
    \caption{Random cutouts from the \cite{Williams_2018} catalog around sources with flux $>5\sigma$}
    \label{fig:cutouts}
\end{figure}

In order to perform an integration over the redshift range, the sources are divided into 20 redshift bins, separated by log spacing, from $z=14$ to $z=0.2$. In each redshift bin, the sources within a rectangular area on the sky are selected, and for each, we extract the $x$ and $y$ coordinates. We use the software package SEP\footnote{Python and C library for SEP can be found at \url{https://github.com/kbarbary/sep/tree/v1.1.x}.} to calculate the level of background noise $\sigma_{BKG}$ of the image, and to extract the pixels for each source.\footnote{The background can also be calculated from the levels provided with the catalog.} If the source has at least one pixel with $I>5\sigma_{BKG}$, then that source is counted as a potentially observable source. This counting gives the total density of sources on the sky in each redshift bin.

\subsubsection{Counting methods}\label{sec:counting_methods}

For the lenses, the total fraction of the area within the Einstein radius $\theta_E$ of a lens is calculated.\footnote{Here, we simply sum the total area of every lens. But this is not exactly equal to the total area on the sky covered by lenses, because it theoretically double counts any area on the sky that has two overlapping lenses. This of course is very rare and will not significantly affect the forecast.} 
% Even if this hypothetical scenario occurs, we can just decide to count a system with one source being lensed by two lenses as two ``lensing systems.''} 
At each redshift bin, the total ``lens-able area'' for sources is summed together due to all lenses at lower redshifts ($z_{lens}<z_{source}$). This is the area around each lens, within which sources have a small distance $D$ from the center of the lens. 
% If a source is close enough, we count it as lensing. 
We examine three criteria on the lens-source distance $D$:
\begin{enumerate}[A]
    \item \uline{Source in lens area}: $D<\theta_E$, where $\theta_E$ is the Einstein radius of the lens evaluated at the source redshift bin  \citep[this matches the method of][]{Collett_2015}. This method is mostly applicable to ``conventional''-sized strong lenses ($\theta_E > 0.5''$).
    \item \uline{Lens in source area}: $D<r_{source}$, where $r_{source}$ is the radius of the region of pixels that are $>5\sigma_{BKG}$. This method is mostly applicable to strong lenses with  $\theta_E < 0.5''$, i.e., systems with Einstein radii smaller than those of conventional lenses. We believe \emph{JWST} opens the door for finding these smaller systems and can lead to the discovery of low-halo-mass strong lenses.
    % (new small lenses for \textit{JWST})
    \item \uline{Lens in source area OR source in lens area}: $D<\textrm{max}(\theta_E, r_{source})$, \ed{i.e., the union of Methods A and B.}
    % (combining Methods A and B). 
    This is our final count of strong lensing systems.
\end{enumerate}

\begin{figure}
    \centering
    \includegraphics[width=.95\textwidth]{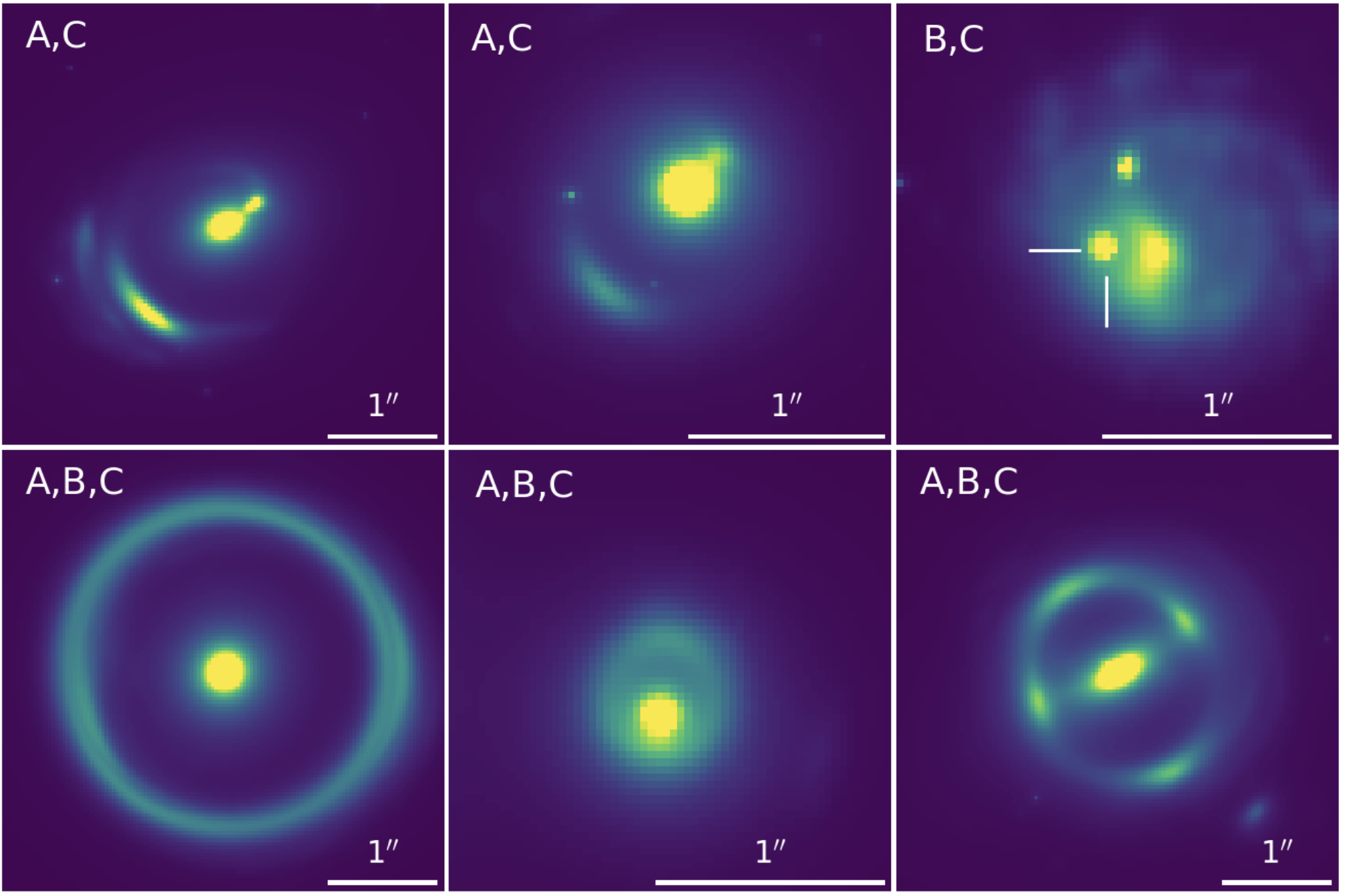}
    \caption{Examples of different types of systems for the purposes of counting (here we show simulated lensing systems from those generated in Section \ref{sec:methods_trained_class_model}). \edr{The letters in the upper left corners correspond to the criteria under which the system would be counted to contribute to the “lens-able area”, explained in \S~\ref{sec:counting_methods}}. 
    % We use Method A and B for counting, with Method C being the union of A and B and encompassing all scenarios in this figure (see text).
    % Hence the final counting is from Method C.
    % Images 1 and 2 are counted in Methods A and C, while image 3 is counted in Methods B and C.
    % Images 4, 5, and 6 are counted in both Methods A and B \ed{(i.e., the intersection of A and B}, corresponding to very small lens-source separation $D$), and of course,  C. 
    Image 3 shows cross-hairs pointing at the lens. Note that image 1 is not a double source: it has one source which has a complex source structure from the VELA simulations.}
    \label{fig:ABC_examples}
\end{figure}

Hereafter, we refer to these three lens counting criteria as Methods A, B, and C. Examples illustrating these three categories are shown in Figure \ref{fig:ABC_examples}. In this figure, images 1 and 2 show typical strong lensing systems where the source falls within the Einstein radius of the lens in projection, while image 3 shows an example of a smaller lens that falls onto the area of the source, but the center of the source does not fall within the Einstein radius. 
Finally, images 4, 5, and 6 show a lens and source that are closely aligned, so the center of the source is within the Einstein radius, and the center of the lens falls onto the area of the source (these images would be counted by all three counting methods). 
%If a lensing system like this is counted by all three methods, it is generally an Einstein ring for low ellipticity.

Then, the expected number of lensing events within some area of the sky in each redshift bin is given by multiplying the number of sources by the fraction of the sky covered by the total ``lens-able area''. 

\ed{For forecasting the numbers of lensing events with double sources, we only use Method~A.}
% These are calculated in a similar way, in which the number of sources is used to calculate the expected probability of a double source for each lens at each source redshift.
% In particular, now that We have 
\ed{We use} the \ed{same} density of sources on the sky in each redshift bin \ed{as before, and assume} 
% we assume 
they are randomly 
distributed on the sky. They then follow a Poisson random scatter, and for each lens, we calculate the probability of at least two sources falling in the same lens area, assuming a Poisson distribution.

In detail, we forecast the average number of sources that fall into the area of every lens in each redshift bin. We take this average, $\mu$, in the Poisson distribution to calculate the probability of $k$ sources falling within that lens area:
\begin{equation}
    P(k)=\frac{\mu^{k}e^{-\mu}}{k!}
\end{equation}
We then sum the expected number of single and double sources over all of the lenses in every redshift bin to obtain the total forecast numbers of lensing systems with single and double sources.

\subsubsection{Lens Sampling Procedure}\label{sec:methods_lens_sampling}
The lenses are initially sampled from the CosmoDC2 catalog \citep{Korytov_2019}, which uses a large cosmological N-body simulation to create an extensive catalog of galaxies out to $z=3$ with many galactic properties provided. 
In this paper, we use the example subset of the data, as the full dataset has far more halos than necessary for our purposes. Due to computing time constraints, $10^7$ lenses are chosen at random from the total catalog in the mass range of $10^{10} M_\odot <M_\mathrm{halo}<10^{14} M_\odot$. 

The CosmoDC2 catalog provides the lens halo masses (dark matter), and their redshifts. Testing demonstrated that this sample size is sufficient to produce consistent number forecasts at the accuracy we are seeking. However, the CosmoDC2 catalog does not include velocity dispersions of the galaxies.

We therefore use the tight relationship between the dark matter halo mass and the stellar velocity dispersion of quiescent galaxies provided by Equation~8 in \citet[][Z18]{Zahid_2018}, based on simulations (Illustris-1):

\begin{equation}\label{eq:zahid}
    \log\Big(\frac{M_\mathrm{halo}}{10^{12}M_\odot}\Big)=\alpha+\beta\log\Big(\frac{\sigma_{h,*}}{100\textrm{ km s}^{-1}}\Big)
\end{equation}

\noindent
where $\alpha=0.16$, $\beta=3.31$, and $\sigma_{h,*}$ is defined as the line-of-sight stellar velocity dispersion measured within the half-light radius, which is most analogous to observations.
The RMS scatter around this relationship is 0.17. Velocity dispersion can then be directly converted to $\theta_E$ assuming a singular isothermal ellipsoid (SIE) model, according to the following equation:

\begin{equation}
    \theta_E^{\textrm{SIE}}=4\pi\frac{\sigma_v^2}{c^2}\frac{D_{\textrm{ls}}}{D_{\textrm{s}}}
\end{equation}
\noindent
where $\sigma_v$ is the stellar velocity dispersion, $c$ is the speed of light, $D_{\textrm{ls}}$ is the distance from the lens to the source galaxy, and $D_{\textrm{s}}$ is the distance from the observer to the source galaxy.

\ed{CosmoDC2 does not explicitly classify elliptical galaxies, so we select only elliptical galaxies as those with a bulge-to-total luminosity ratio B/T$>0.9$. Here, $B$ is the luminosity of the galaxy bulge and $T$ is the galaxy's total luminosity, provided in the \emph{i}-band in CosmoDC2. This criterion ensures that the galaxy is bulge-dominated, and is a common way to select for elliptical galaxies \citep[e.g.,][]{Hoyle_2012,De_Lucia_2012,Mendez-Abreu_2017}. B/T$>0.9$ is similar to other criteria that have been used, such as B/T$>0.92$ used by \cite{Benson_2010} or bulge-to-disk ratio B/D$>8$ (equivalent to B/T$>0.889$) used by \cite{Sawala_2024}. However, this criterion to select ellipticals is arguably somewhat subjective and uncertain, as B/T$>0.8$ \citep{Saintonge_2005, Constantin_2021}, B/T$>0.75$ \citep{Pritchet_2024}, and B/T$>0.7$ \citep{Wilman_2013, Sesana_2014, Izquierdo-Villalba_2023} have also been used in previous work. There may very well be many observable lenses with B/T$<0.9$, but we choose a $0.9$ threshold in order to be conservative with our predictions, especially given that we are using an SIE lens model to match the Z18 relation.} 

\ed{The lens population generation in \citet{li2024b} took a similar approach as this work, but with some differences (due to different science goals). 
They also started from the dark matter halo mass in CosmoDC2 catalog and used Z18 for conversion from halo mass to velocity dispersion. 
But they only selected halos with a stellar mass $M_*> 10^{11} M_\odot$.
Also, instead of SIE, they used the sum of the dark matter \citep[modeled as a Navarro-Frenk-White, or NFW profile;][]{navarro1996a} and stellar (modeled as a \ser profile) components. 
To obtained the \ser index, they first converted the halo mass to \emph{DM velocity dispersion}, according to \emph{Equation (3)} in Z18 (instead of Equation~(8) in our case). 
For strong lensing systems, it is well-known that the combination of DM and stellar components can be approximated by an SIE model \citep[e.g.,][]{shajib2021a}.}

\begin{figure}
    \centering
    \includegraphics[width=.7\textwidth]{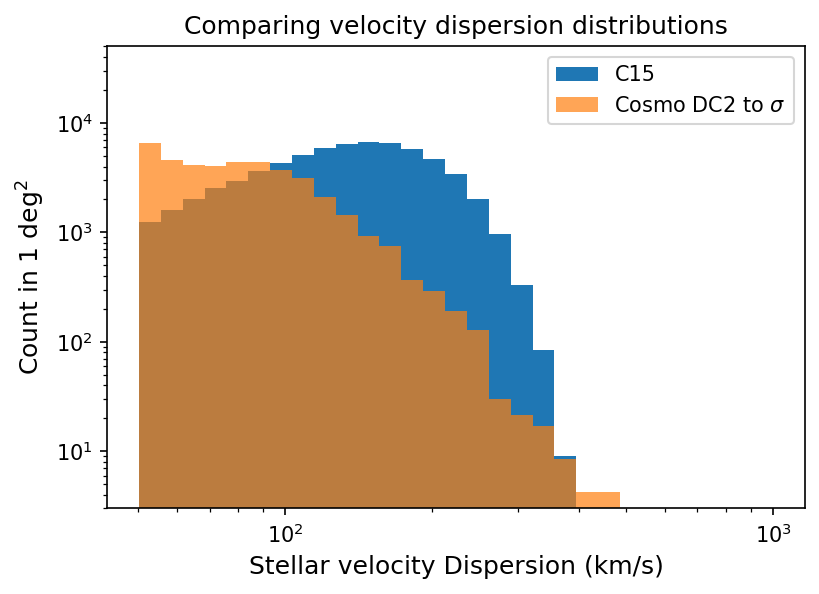}
    \caption{Distribution of $\sigma_v$ given by \cite{Collett_2015} and CosmoDC2 using \cite{Zahid_2018} on 1~deg$^2$ of sky (CosmoDC2 provides halo mass, which is converted to stellar velocity dispersion $\sigma_v$ using Equation \ref{eq:zahid}).}
    \label{fig:Collett}
\end{figure}

% \subsection{Magnification for small lenses}
% Here, we let ``source area'' refer to the area of the source with $I>5\sigma_{BKG}$. In methods B and C, we count lenses that fall within the source area, which is increased by the effect of lensing itself. We only count the lens if it falls within the pre-lensing source area, because we are trying to count small lenses that cause a distortion on top of a larger source. Thus, the relevant area over which the lens might fall is the area of the unlensed source. Furthermore, area magnification is a complicated problem that depends on many parameters beyond the scope of this paper.

\subsubsection{Collett (2015) lens distributions}\label{sec:collett_dist}
The LensPop code from \cite[][C15]{Collett_2015}  produces the distribution of lenses with a source within $\theta_E$ in projection (same as our Method A). The velocity dispersion and $\theta_E$ distributions of the whole set of original lenses (disregarding sources) can also be saved.

There is a notable discrepancy between the velocity dispersion distributions, shown in Figure \ref{fig:Collett}. C15 appears to produce numbers of lenses (when averaged over the range of Einstein radii) roughly 6 times greater than the distributions of halo masses provided by CosmoDC2 in combination with \cite{Zahid_2018}. This is likely due to the differences in sampling halo masses, based on N-body simulations, or sampling velocity dispersion, based on local observations from SDSS \citep[e.g.,][]{Abazajian_2003}, as done by C15.
% (see also the end of Section \ref{sec:results_forecast}).
\edr{This discrepancy can be seen in Table \ref{tab:forecasts} and causes the numbers we calculate to be smaller than C15. 
On the other hand, the greater (than C15) depth of \jwst we use increases the numbers from our calculations. These two effects end up mostly canceling out,
so the overall numbers of lenses using Method A are similar to C15.}

We decide to use the method of sampling halo mass 
% instead of velocity dispersion 
partly to take advantage of the large simulations of DM halos like CosmoDC2 that have been done since the time of C15, and partly
% We want to take advantage of these updated simulations, in part 
because later in this paper we specifically investigate 
% and focusing in on the 
the low halo mass, low Einstein radius regime of strong lensing.

\subsection{Results: Lensing Forecasts}\label{sec:results_forecast}

\begin{deluxetable}{cccccc}
\tablehead{\multicolumn{6}{c}{Method A: Source in lens area} \\
\colhead{$\theta_E$} & \colhead{N} & \colhead{percent} & \colhead{per deg$^2$} & \colhead{on sky} & \colhead{C15 equivalent}}
    \startdata
        \multirow{2}{*}{$>0.02''$} & 1 & 93 & 280 & 12 million & 11 million\\
        & 2 & 7 & 23 & 940,000\\ \tableline
        \multirow{2}{*}{$>0.05''$} & 1 & 92 & 270 & 11 million & 11 million\\
        & 2 & 8 & 23 & 940,000\\ \tableline
        \multirow{2}{*}{$>0.1''$} & 1 & 92 & 250 & 10 million & 10 million\\
        & 2 & 8 & 23 & 940,000 \\ \tableline
        \multirow{2}{*}{$>0.5''$} & 1 & 85 & 110 & 4.5 million & 6.4 million\\
        & 2 & 15 & 19 & 780,000\\ \tableline
        \multirow{2}{*}{$>1.0''$} & 1 & 80 & 45 & 1.9 million & 2.2 million\\
        & 2 & 20 & 11 & 470,000\\
    \enddata
    \caption{Lensing forecasts for the number of sources that fall within lens area for \textit{JWST} (Method A from \S~\ref{sec:counting_methods}). $N=1$ refers to single sources and $N=2$ refers to double sources. The ``C15 equivalent'' column is estimated from Figure 1 of \cite{Collett_2015}. Note that although the numbers are similar, they are not directly comparable due to different observation depths between \hst and \jwst (see text, \S~\ref{sec:collett_dist}). Results are given to 2 significant figures.}
    \label{tab:forecasts}
\end{deluxetable}

\begin{deluxetable}{ccccc}
\tablehead{\colhead{} & \multicolumn{2}{c}{Method B: Lens in source area} & \multicolumn{2}{c}{Method C: Source in lens area/lens in source area} \\ \colhead{$\theta_E$} & \colhead{per deg$^2$} & \colhead{on sky} & \colhead{per deg$^2$} & \colhead{on sky}}
    \startdata
        $>0.02''$ & 450 & 19 million & 690 & 29 million \\
        $>0.05''$ & 210 & 8.6 million & 450 & 19 million \\
        $>0.1''$ & 110 & 4.5 million & 340 & 14 million\\
        $>0.5''$ & 7.3 & 300,000 & 130 & 5.5 million\\
        $>1.0''$ & 1.2 & 50,000 & 59 & 2.5 million\\
    \enddata
    \caption{Lensing forecasts for \textit{JWST}, counting the number of systems where the lens falls within the source area (Method~B; see \S~\ref{sec:counting_methods}), and where either  the source falls within the lensing area or the lens falls within the source area  (Methods~C, which is the union of Method A and B). Results are given to 2 significant figures.}
    \label{tab:more_forecasts}
\end{deluxetable}

The results of the lensing forecasts can be divided up by the different counting methods, as detailed in Section \ref{sec:counting_methods}. The basic predictions for the number of possible lenses that lie in front of a source, that are observable with \textit{JWST}, are shown in Table \ref{tab:forecasts} for Method A and Table \ref{tab:more_forecasts} for Methods B and C. Table \ref{tab:forecasts} also shows the fraction of double sources, which is about $7\%$, for sources that fall within the lensing area, as well as equivalents for the counts from C15. 

We do not include the double sources for Methods B and C because for Method B, if there is a double source behind a lens with small $\theta_E$, that is just two aligned sources. It would likely be impossible to distinguish them, because they will not be lensed much into separated arcs. Therefore, to only count systems that are usually thought of as double sources, we only include those from Method A. 
Our final predictions for double sources are those shown in Table \ref{tab:forecasts}.

As previously mentioned, for our final predictions, we use Method C, shown on the right half of Table \ref{tab:more_forecasts}. This method provides the most comprehensive statistics of strong lensing systems that could be observed.

\subsection{Discussion: Lensing Forecasts}\label{sec:disc_forecast}

As discussed in Section \ref{sec:collett_dist}, we observe a discrepancy between the method of sampling halo masses from CosmoDC2 and C15's method of sampling velocity dispersions based on SDSS observations. To compare with the results from C15, we also make a prediction by sampling velocity dispersions using LensPop instead of halo masses using CosmoDC2. We use the same brightness cutoff as C15, at 27$^{\textrm{th}}$ magnitude in the same filter, the SDSS $i$-band. We obtain 10.4 million lenses with $\theta_E>0.05''$, compared with 11 million in C15.
Finally, as a consistency check, we have applied reasonable selection cuts appropriate for the SLACS lens and obtained the empirical, often-cited 1 in 1000 early-type galaxies being a strong lens \citep{bolton2006a} in Appendix~\ref{sec:lens_fractions}.

\ed{One notable result of these forecasts is the prediction of $940,000$ double source lenses (23 per square degree), with $\tE > 0.1''$ (Table~\ref{tab:forecasts}), 
due to the large source density with \jwst. These compound lenses can be used as a powerful probe of cosmology, as discussed in the Introduction (\S~\ref{sec:introduction}).}

\ed{The largest source of uncertainty in the forecast is from the significant uncertainty in the velocity dispersion function (VDF), for which there appears to be differences at about 0.5 dex at $z=0$ and at least 1.0~dex at $z>0.5$ 
when comparing different estimates \citep[e.g.,][]{Taylor_2022, Yue_2022}. These uncertainties are larger at lower velocity dispersions, especially at $\sigma_v<100$km/s, where there are few observations, and thus simulations are often relied upon. 
Below this limit, a few relevant factors are not precisely known observationally, including the dark matter to baryon ratio, the proportion of quiescent galaxies, and the redshift evolution of the VDF. This paper is a first attempt at quantifying how many of these lenses will have an observational effect. 
For this purpose, we have used some of the best tools currently available: the CosmoDC2 catalog and the simulation results of Z18.}
% . We believe this represents a best approach at this time.}
% we acknowledge the large uncertainties in our calculated forecasts.}

% This validates our procedure of sampling sources and calculating the number of lenses.

\section{\edr{Classification with the ResNet}}\label{sec:sim-resnet}
\edr{In the previous section (\S~\ref{sec:forecast}), we have forecast that \jwst is able to observe many strong lenses all the way down to its diffraction limit. }
\ed{In this section, we describe our strong lens simulation pipeline and ResNet training to identify the lenses in \S\ref{sec:methods_trained_class_model} and present our results in \S\ref{sec:results_trained_class_model}.
\edr{In this work, we define three categories of lenses: ``conventional-sized'' or``large'': $\tE > 0.5''$, ``intermediate'': $0.5''>\tE > 0.15''$, and ``small'': $0.15'' > \tE > 0.02''$.}
We not only simulate the ``conventional'' lenses, but also lenses that have far smaller Einstein radii, all the way down to the diffraction limit of \jwst. 
We systematically explore the capability of our ResNet architecture. 
We first show that it performs extremely well for the ``conventional'' lenses (\S\ref{sec:larger_lens_models}). 
In fact, when testing our trained model on a small set of real \hst images, we identified two new lens candidates that were missed by a human search effort (\S\ref{sec:new-lenses}).
We then present results for lenses with smaller Einstein radii (\S\ref{sec:smaller_lens_models}).
First, in the \edr{``intermediate''} regime of $0.15'' < \tE < 0.5''$, 
 where the lower bound roughly corresponds to the smallest \tE strong lenses found so far (e.g., see below about SN~Zwicky), we show once again that the model performs very well. We believe this is an interesting but hitherto scarcely explored area in strong lensing. For example, one of the two systems with \tE near the low end of this range is the lensed SN~Zwicky, with $\tE \approx 0.17''$. \citet{Pierel_2023} reported the projected mass within the Einstein radius to be $\sim 8 \times 10^{9} M_\odot$. This translates to a halo mass in the high $10^{11} M_\odot$ range. 
We then show that even in the ``small'' regime of $0.02'' < \tE < 0.15''$, our ResNet model continues to perform reasonably well \edr{when evaluated on simulated systems}. Here we particularly focus on the performance for identifying those lenses with the lowest mass in the smallest $\tE$ bin. 
Many of these are expected to be dwarf galaxies.
The ability to find these systems makes it possible to apply the strong lensing approach to understand the mass distribution of small galaxies, down to $M_\mathrm{halo} \lesssim 10^{11} M_\odot$, or two orders of magnitude smaller than the halo mass of conventional strong lenses.
This potentially opens the door for a new way of testing the CDM model. 
Finally, a series of discussions pertaining to the results presented above are given in \S~\ref{sec:discuss}.}

\subsection{Methods: ResNet Model}\label{sec:methods_trained_class_model}

\ed{ResNet remains one of the best algorithms for finding strong lenses in imaging data. In a recent lens search in ground-based observations, 
% by other members of our team, 
ResNet models continue to perform competitively relative to other more recent neural network models (J.~Inchausti Reyes et al. in prep).}

To train classification models for lenses, 20,000 total images are generated (10,000 lensed and 10,000 unlensed) and trained using a ``shielded" ResNet model, based on work in \citet[][H21]{Huang_2021}, with 4 A100 GPU's on Perlmutter at NERSC. To our knowledge, this is the first time neural networks have been trained on multiple GPU's in astrophysics applications. This ability, with its increased speed, allows us to test and iterate models quickly. We present the training of the following models using simulations, for three ranges of Einstein radius and different levels of noises for \emph{\hst} and \emph{\jwst} (the two telescopes with publicly available data), with a summary given in Table \ref{tab:models}:
\begin{itemize}
    \item Model 1: $0.50''<\theta_E<1.5''$
    \begin{itemize}
        \item Model 1a \edr{(HST-long)}: ``typical" level of noise for multi-exposure \hst images ($\sim2000-7200$s exposure time)
        \item Model 1b \edr{(HST-short)}: ``high" level of noise for single-exposure \hst images ($\sim420$s exposure time)
    \end{itemize}
    \item Model 2: $0.15''<\theta_E<0.50''$
    \begin{itemize}
        \item Model 2a \edr{(JWST-long)}: 10,000s exposure time with \jwst
        \item Model 2b \edr{(JWST-short)}: 1,000s exposure time with \jwst
    \end{itemize}
    \item Model 3: $0.02''<\theta_E<0.15''$
    \begin{itemize}
        \item Model 3a \edr{(JWST-small)}: 10,000s exposure time with \jwst
    \end{itemize}
\end{itemize}

\begin{figure}
    \centering
    \includegraphics[width=0.9\textwidth]{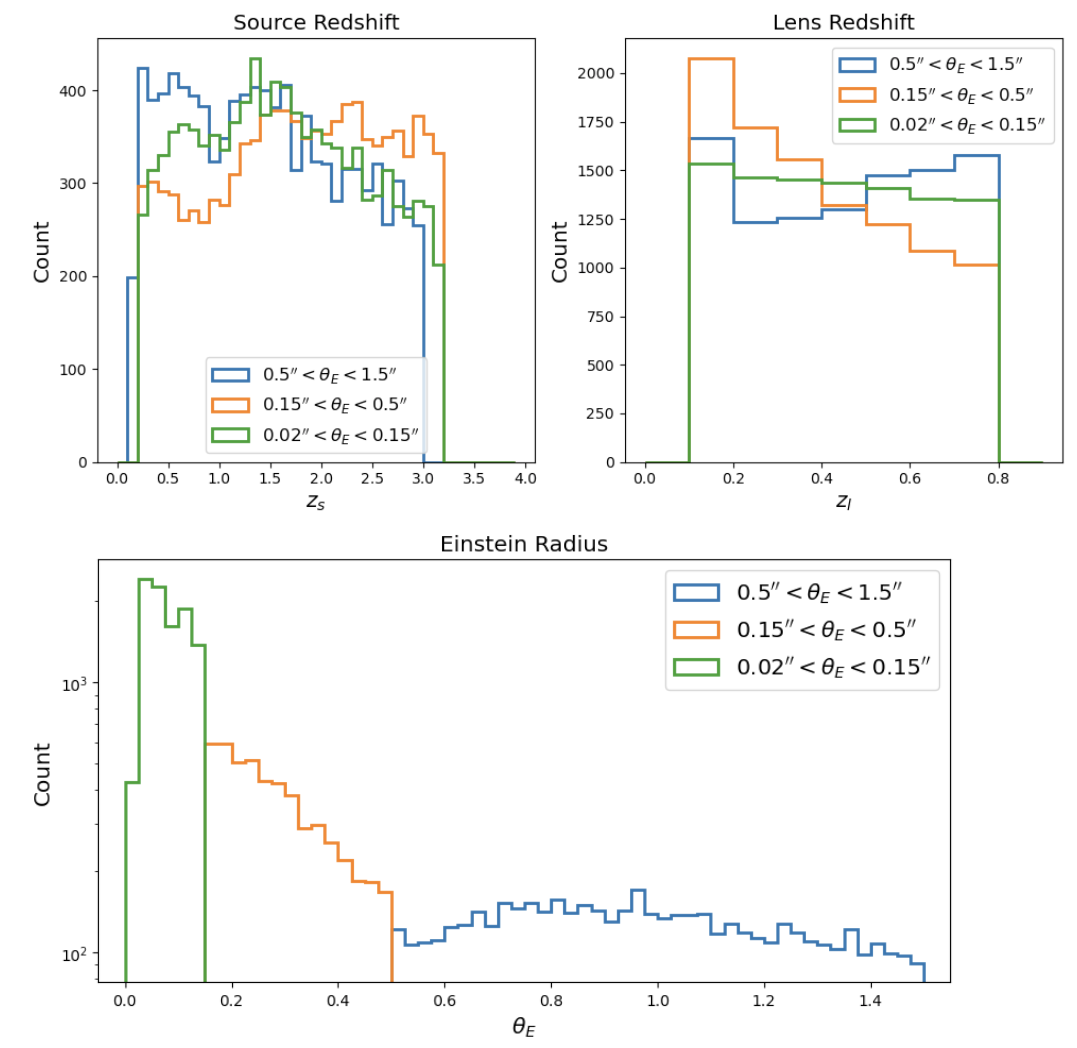}
    \caption{Distributions of 
    $z_l$, $z_s$, and $\theta_E$
    for the three Einstein radius ranges.
    %all of the models, dividing between the lenses in each of the three ranges of $\theta_E$.
    }
    \label{fig:sim_details}
\end{figure}

\begin{deluxetable}{ccccc}
\caption{The Three ResNet Models}    \label{tab:models}
\tablehead{\colhead{Model} & \colhead{$\theta_E$ range} 
& \colhead{Training Images}
% & \colhead{Images evaluated on} 
& \colhead{Model version} & \colhead{Exposure time}}
    \startdata
        \multirow{2}{*}{Model 1} & \multirow{2}{*}{(0.50'', 1.5'')} & \multirow{2}{*}{Simulated \hst} & Model 1a \edr{(HST-long)} & $\sim2000-7200$s \\ &&& Model 1b \edr{(HST-short)} & $\sim420$s\\ \hline
        \multirow{2}{*}{Model 2} & \multirow{2}{*}{(0.15'', 0.50'')} & \multirow{2}{*}{Simulated \jwst} & Model 2a \edr{(JWST-long)} & 10,000s \\ &&& Model 2b \edr{(JWST-short)} & 1,000s\\ \hline
        Model 3 & (0.02'', 0.15'') & Simulated \jwst & Model 3a \edr{(JWST-small)} & 10,000s\\
    \enddata
\end{deluxetable}
\noindent
Note that Model~1b \edr{(HST-short)} is only applied in Appendix \ref{sec:single_exp_test} (also see the end of \S\ref{sec:first_new_lens}). The distributions of the simulated lenses in these three ranges are shown in Figure~\ref{fig:sim_details}.

\ed{Below, in \S\ref{sec:sim}, we provide an overall description for our simulation pipeline, with an emphasis on lensed images.
The simulations of unlensed images are described in \S\ref{sec:methods_unlensed_images}.
Furthermore, for training models to identify smaller Einstein radius systems ($\tE < 0.5''$), modifications were introduced in the simulations for the unlensed images (\S\ref{sec:methods_unlensed_images} and 
\S\ref{sec:small_lens_changes}).
We describe the simulations of environmental galaxies in \S\ref{sec:env_gals}.
Finally, in \S\ref{sec:noise_preprocessing}, we show how we add noise and preprocess the images before training.}
% \begin{deluxetable}{ccccc}
% \tablehead{\colhead{Model} & \colhead{$\theta_E$ range} & \colhead{Images evaluated on} & \colhead{Exposure time}}
%     \startdata
%         Model 1 & $0.50''<\theta_E<1.5''$ & Real, multi-exposure \hst & $\sim2000-7200$s\\
%         Model 2 & $0.50''<\theta_E<1.5''$ & Real, single-exposure \hst & $\sim420$s\\
%         Model 3 & $0.15''<\theta_E<0.50''$ & Simulated \jwst & 1,000s and 10,000s\\
%         Model 4 & $0.02''<\theta_E<0.15''$ & Simulated \jwst & 10,000s\\
%     \enddata
%     \caption{Descriptions of the four different models we present.}
%     \label{tab:models}
% \end{deluxetable}

\subsubsection{Simulations}\label{sec:sim}

\begin{figure}
    \centering
    \includegraphics[width=\textwidth]{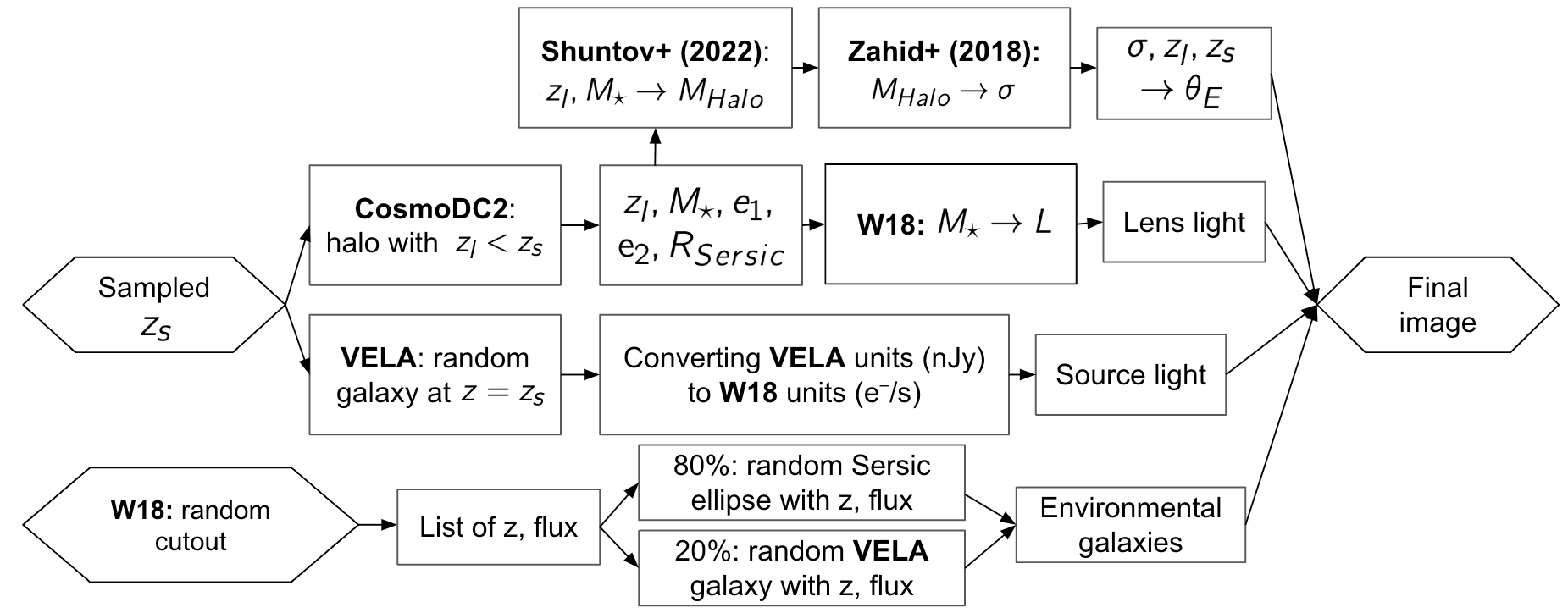}
    \caption{Flowchart describing the steps and sources used to simulate lensed and unlensed images in this training sample. To simulate unlensed images: for $0.5''<\theta_E<1.5''$, we turn off source light and set $\theta_E=0$; for $\theta_E<0.50''$, we still set $\theta_E=0$, but keep the source light on (see \S~\ref{sec:methods_unlensed_images}).}
    \label{fig:flowchart}
\end{figure}
% \newpage

Figure \ref{fig:flowchart} shows a flowchart of the steps used to generate the simulated images. We can now describe each step in detail. We simulate three types of galaxies in each image: the lens galaxy and source galaxy are described in this section, and any other foreground or background galaxies are called ``environmental galaxies'', described in Section \ref{sec:env_gals}.
\edr{First, source redshift $z_s$ is selected randomly from a normal distribution ($\mu=1,\sigma=3$) (truncated to be positive).
Then, a random galaxy from the CosmoDC2 catalog is drawn uniformly from those with $z_\ell<z_s$, selecting only central galaxies for simplicity. The drawn CosmoDC2 galaxy gives us the lens stellar mass ($M_\star$), lens redshift $z_\ell$, and ellipticity. To prevent overfitting, random scatter is applied to $z_{\ell}$ ($\epsilon\sim U(-0.5,0.5)$). We convert $M_\star$ to $M_{\textrm{halo}}$ using Equation 9 in \cite{Shuntov_2022}, and then calculate $\theta_E$ from $M_{\textrm{halo}}$ using Equation~\ref{eq:zahid}.\footnote{A simpler method for calculating $M_{\textrm{halo}}$ would be to use the halo mass provided in the CosmoDC2 catalog.} Then, we apply a cut on Einstein radius, based on which of the three $\theta_E$ ranges we are using, as well as the following cuts on redshifts:}
\begin{enumerate}
    \item $0.1<z_{l}<0.8$
    \item $z_{l}<z_{s}<\begin{cases}3.0&\textrm{Model 1 \edr{(HST-long/short)}}\\3.2&\textrm{Models 2 and 3 \edr{(JWST-long/short/small)}}\end{cases}$
\end{enumerate}
We make these redshift cuts (see Figure \ref{fig:sim_details}) to focus on the systems that can most reliably be observed. 
%because in this work, we are applying these simulations and models for the first time. 
\ed{For example, the SLACS lenses are in the range of $0.05<z_l<0.5$ \citep{Bolton_2008} and the BELLS lenses are in the range of $0.4<z_l<0.7$ \citep{Brownstein_2012}}. 
However, it is certainly possible for strong lensing systems to have lenses and sources that fall outside these ranges, especially with \emph{\jwst}. A simple and useful extension of this work would be to apply the same techniques to broader ranges of lens and source redshifts. Our preliminary investigation did show that for source redshift above $\sim 6$, even with long exposure ($\sim$10,000s), and with magnification from galaxy-scale strong lensing, the lensed arcs can often be fainter than the level of \jwst background noise. 
% However, extending both the lens and source redshifts beyond those currently used in this paper is possible.

The images are simulated using the Lenstronomy gravitational lensing package \citep{Birrer_2018,Birrer_2021}\footnote{Available at \url{https://github.com/lenstronomy/lenstronomy}.}, at a pixel scale of 0.031$''$. \ed{The PSF used is Gaussian, truncated at $3\sigma$.} The sources used are from the VELA hydrodynamic simulations \citep{Snyder_2018, Simons_2019}, a high-quality series of cosmological galaxy simulations. The source image is a randomly selected VELA simulation at $z_{s}$. There are 34 VELA galaxies, each at a wide range of redshifts, and at each redshift, they are provided at many random 3D angles.
\edr{Even though this is a relatively small number, the VELA simulations span the
typical stellar mass range of seeing-limited kinematics samples
in the literature \citep{Simons_2019}.
They have often been used to study different aspects of high-redshift galaxies including morphology \citep{Snyder_2015, Rubin_2021}, which is particularly relevant to this study.
This represents a step forward from merely using S\'{e}rsic or \hst images of low-redshift galaxies as sources in lensing simulations \cite[e.g.,][]{Wagner-Carena_2022}.
Nevertheless, in future work, our simulations would benefit from a larger simulation sample that captures a greater range of galaxy dynamics, star formation, and merger histories.}

Thus we select a random galaxy at a random angle, at the desired redshift $z_s$. The lens is placed at $z_l$ with an SIE mass profile. For Model 1 \edr{(HST-long/short)} with $0.5''<\theta_E<1.5''$, the lens is placed at the center of the cutout image, while the source is given a random position drawn from a normal distribution centered at $0$, with $X,Y\sim \mathcal{N}(\mu=0'',\sigma=0.25'')$. 
We use this set-up because when applied to real images, the cutout of the image can be centered on the lens. 

For Models 2 and 3 \edr{(JWST-long/short/small)} with $\theta_E<0.50''$, now it no longer makes sense to center the lens and randomly displace the source, because the source is now generally the dominant light source, not the lens. Thus, if applied to real images, where the source and lens positions are unknown, the image cutouts could not be centered on the lens. Thus, we now center the cutout image on the source and draw the lens position with $X,Y\sim \mathcal{N}(\mu=0'',\sigma=0.25'')$. 

Lens brightness is scaled to match W18. For Model 1 \edr{(HST-long/short)}, arc brightness is scaled up by a factor chosen as \large$\sim10^{U(0.5,2.0)}$\normalsize. This is done to replicate the selection bias inherent in real lenses, where lenses with bright sources/arcs are more likely to be  observationally identified, such as in the SLACS and BELLS samples. In fact, we choose a large range of arc brightnesses to simulate many different kinds of lensing systems. This is only done for Model 1 \edr{(HST-long/short)}, because it is designed to be later tested on real \hst images of lensing systems, chosen from the SLACS set.
% not just simulations.

For the lens light, we need to calculate the lens light amplitude given the generated halo mass and redshift. We fit a power law to the W18 catalog's values for the $M_{*}/F$ ratio ($M_\star$/total flux) as a function of $z$ for quiescent galaxies. The VELA catalog gives total flux already, so the amplitude can be found for the sources as well. We convert lens and source light amplitudes to the same units so we can combine the lens and source light.

For the sources, the S\'{e}rsic index $n_{\textrm{S\'{e}rsic}}$ is sampled randomly: $n_{\textrm{S\'{e}rsic}}\in U(2,6)$. This range is chosen to approximately match the distribution of indices as shown in \cite{Simard_2011}. 
% The lens light’s value of $n_{\textrm{S\'{e}rsic}}$ can significantly affect how easy it is to visually see lensing, because it affects how concentrated the light is. 
The ellipticity and angle of the galaxy are simply sampled from the CosmoDC2 catalog along with $M_\star$.

\subsubsection{Unlensed images}\label{sec:methods_unlensed_images}
In order to ensure that lensed and unlensed images are as similar as possible, unlensed images are generated in much the same way as lensed images. For Model 1 \edr{(HST-long/short)} with $0.5''<\theta_E<1.5''$, a central bright galaxy is chosen the same as the lens light for a lensed image, but there is no lensed source. The same environmental galaxies (\S\ref{sec:env_gals}) are also added.

However, for Models 2 and 3 \edr{(JWST-long/short/small)} with $\theta_E<0.50''$, because the source is now generally the dominant light source, a different definition of a ``non-lensed'' image should be used. With large $\theta_E$ ($> 0.5''$) systems , the corresponding non-lenses have a bright central elliptical galaxy, but no source. 
Whereas for a small $\theta_E$ ($< 0.5''$) system, the corresponding non-lens is a source with no lens galaxy. 

% In addition, 
For large $\theta_E$ systems, 
in the corresponding non-lensed image, both the lens light and lensing effects are removed.
% it is unrealistic for a large, bright lens to appear without a corresponding lensing effect, 
However, for small $\theta_E$ systems (due to the low mass of the foreground galaxy and/or unfavorable geometry), it is possible to have a foreground galaxy with lensing effect so small as to be non-detectable, which would make that cutout image a non-lens. 
Thus, in the non-lensed image, we retain the foreground galaxy light, but remove the lensing effect.
Thus this foreground galaxy essentially becomes an environmental galaxy that overlaps with the background galaxy (which was the source in the lensed image). 
% This small galaxy that produces light but no lensing effect could be further away than the source, or very close to the observer or the source, thus producing little to no lensing effect. 
We remind the reader that for the sources, 
both elliptical galaxies and VELA simulated galaxies are sampled from $z$ between 0 and 3,
\ed{whereas for the simulations of lenses, the elliptical galaxies have redshifts between 0.1 and 0.8 (see \S\ref{sec:sim})}.
Thus, in this situation, near the center of the cutout,
there are two scenarios in our simulations resulting from random sampling:
1) an elliptical galaxy is in the foreground and a simulated VELA galaxy in the background;
2) a VELA galaxy in the foreground, and an elliptical galaxy in the background.
We call the non-lenses of this type, consisting of overlapping galaxies, Type 1.
Of course, most real cutout images that do not have lensing typically will not have overlapping galaxies like these. 
We train and validate on this type of simulated non-lenses  because they are the most likely to be classified as \emph{false} positives by a neural network model (i.e., the most challenging non-lenses).
Later we will also test our trained model on the most common non-lens cutouts (those without overlapping galaxies; \S~\ref{sec:additional_tests})---we call these Type~2, and we construct these non-lenses by simply removing the overlapping elliptical galaxy from our existing non-lens validation images, so they have the same VELA galaxies as in our validation set.
\edr{A small percentage of such non-lens images may contain \emph{coincidentally} overlapping galaxies. These non-lenses are what is typically used in the training samples in the literature \citep[e.g.][]{Metcalf_2019,lanusse2018a}.}
\edr{Note that a simple relationship between these types of non-lenses is:}
\begin{equation*}
    \textrm{Type~1 non-lenses = Type~2 non-lenses + deliberate overlapping galaxies}
\end{equation*}
We choose to call the type of non-lenses used in the training set Type~1.
We detail these two types of non-lenses in Table~\ref{tab:non_lenses}.

\begin{deluxetable}{ccccc}
\tablehead{\colhead{Non-lens type} & \colhead{Description} & \colhead{Model~3 Training+Validation Sets} & \colhead{Model 3 Testing Set} & \colhead{FPR}}
    \startdata
        Type 1 & Overlapping galaxies & Yes & No & 4.45\% \\ \hline
        Type 2 & No overlapping galaxies & No & Yes & 2.2\%\\
    \enddata
    \caption{Descriptions of the two main types of non-lenses, whether they are used to train Model~3 \edr{(JWST-small)}, as well as the false positive rate (FPR) using Model~3 \edr{(JWST-small)} when evaluated on each group (with a 0.5 ResNet probability threshold). Unless otherwise specified, the reported metrics for Model~3 \edr{(JWST-small)} are only from the validation set using Type~1 non-lenses.}
    \label{tab:non_lenses}
\end{deluxetable}
% \ed{Therefore, we do not apply the redshift cuts to this overlapping galaxy that we did for lensed systems}. 

% \begin{deluxetable}{ccccc}
% % \scriptsize
% \tablehead{\colhead{Non-lens type} & \colhead{Description} & \colhead{Model~3 Training+Validation Sets} & \colhead{Model 3 Testing Set} & \colhead{FPR}}
%     \startdata
%         \multirow{2}{*}{Type 2} & Env. galaxies \& coincidental & \multirow{2}{*}{No} & \multirow{2}{*}{Yes} & \multirow{2}{*}{2.2\%} & &  overlapping galaxies\\ \hline
%         \multirow{2}{*}{Type 1} & \multirow{2}{*}{Type 2 + intentional} & \multirow{2}{*}{Yes} & \multirow{2}{*}{No} & \multirow{2}{*}{4.45\%} & & overlapping galaxies\\
%     \enddata
%     \caption{Descriptions of the two main types of non-lenses, whether they are used to train Model~3 \edr{(JWST-small)}, as well as the false positive rate (FPR) using Model~3 \edr{(JWST-small)} when evaluated on each group (with a 0.5 ResNet probability threshold). Unless otherwise specified, the reported metrics for Model~3 \edr{(JWST-small)} are only from the validation set using Type~1 non-lenses.}
%     \label{tab:non_lenses}
% \end{deluxetable}

\subsubsection[Extra changes for the smallest theta\_E range]{Extra changes for the smallest $\theta_E$ range}\label{sec:small_lens_changes}

For Model~3 \edr{(JWST-small)}, with the smallest range of Einstein radii ($0.02''<\theta_E<0.15''$), we are interested in low-mass lenses, particularly for $\theta_E<0.10''$. However, many of the lenses in that Einstein radius range are still high-mass lenses, with an unfavorable redshift setup (e.g., $z_l$ and $z_s$ being very close) that causes them to have a small Einstein radius. The primary effect is that these lenses have a large amount of lens light despite their small Einstein radii.
% , which decreases ResNet performance. 
These are not the kind of systems we are interested in finding.
% To better 
To focus on low-mass lenses, in the training sample, we remove any lenses with $\theta_E<0.10''$ with a brighter lens than the source itself. This is also done for the overlapping galaxies \ed{(see the above section \S\,\ref{sec:methods_unlensed_images})} for the non-lenses.

\begin{deluxetable}{cc}
\caption{Galaxy Models Used in Simulations}    \label{tab:galaxy_models}
\tablehead{\colhead{Component} & \colhead{Galaxy Type}}
    \startdata
        Lens galaxies & 100\% S\'{e}rsic ellipses\\
        Source galaxies & 100\% VELA galaxies \\
        Environmental galaxies & 80\% S\'{e}rsic ellipses; 20\% VELA galaxies\\
    \enddata
\end{deluxetable}

\subsubsection{Environmental galaxies}\label{sec:env_gals}
To add environmental galaxies (other foreground or background galaxies unaffected by lensing and producing no lensing effects themselves that fall within the image cutout) to lensed images, a random cutout from the W18 catalog is selected, and each object in the cutout is replaced by either a VELA image of the same redshift (with an 20\% chance) or a random S\'{e}rsic ellipse (with a 80\% chance), scaled to the same total flux at the same position, with brightness scatter added. Brightness is scaled by a factor chosen as \large$\sim10^{U(0,0.7)}$\normalsize. This makes bright environmental galaxies more common in training data, helping the neural network to learn to distinguish them. These are all added to the
% background of 
lensed images used in the training sample.
Table \ref{tab:galaxy_models} \ed{summarizes} the types of galaxy light models used in these simulations. 
% The lens galaxies are all S\'{e}rsic ellipses, the source galaxies are all VELA galaxies, and the environmental galaxies include both types.

\subsubsection{Noise treatment and pre-processing}\label{sec:noise_preprocessing}
No noise is added to the simulations at first. \edr{We normalize each image by performing the following, represented as pseudocode:
\begin{itemize}
    \item Subtract the mean of the image
    \item Divide the image by its standard deviation
    \item If Model~1 \edr{(HST-long/short)}:
    \begin{itemize}
        \item Calculate the 99$^{\textrm{th}}$ percentile of the image 
        \item Set all pixels greater than this value to the 99$^{\textrm{th}}$ percentile value
    \end{itemize}
\end{itemize}
% The image preprocessing steps done right before training are as follows:
We then split the training sample into an 80\% training set and a 20\% validation set.}

% The normalization steps we perform on each image are represented in Python code below:
% \begin{python}[caption={Normalization for ResNet Training}]
% def normalize(image):
%     image = (image - image.mean()) / image.std()
%     if Model_1:
%         cap = np.percentile(image,99)
%         image[image>cap]=cap
%     return image
% \end{python}

\begin{figure}
    \centering
    \includegraphics[width=.85\textwidth]{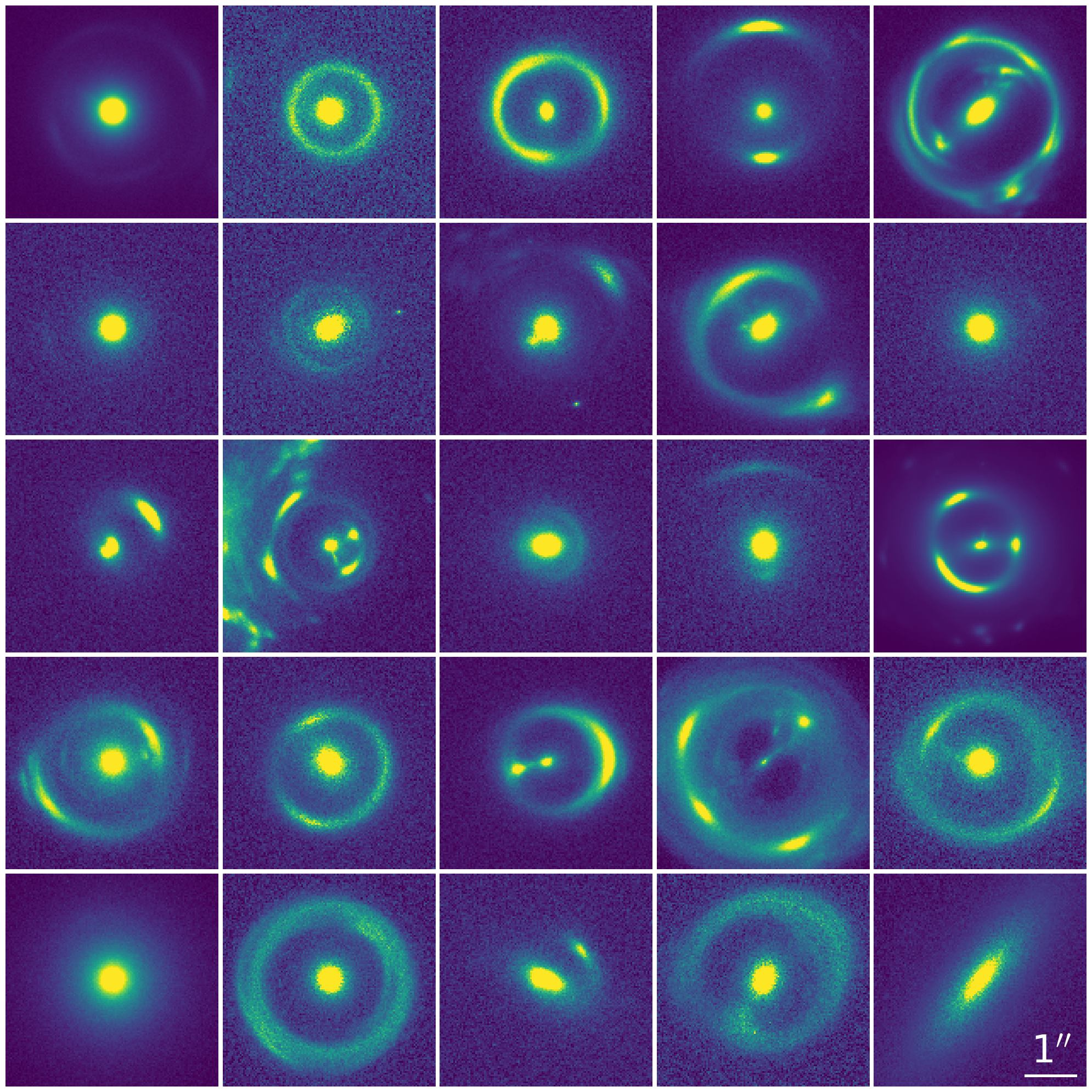}
    \caption{Examples of simulated lensed systems with $0.5''<\theta_E<1.5''$ (Model 1 \edr{(HST-long/short)}). Noise is added as described in \S~\ref{sec:noise_preprocessing} at the level of Model~1a \edr{(HST-long)} (Table~\ref{tab:models}).}
    \label{fig:sims}
\end{figure}

Noise is added to the images as an image augmentation layer in the ResNet model. Gaussian background noise and Poisson noise are added in a wide range with a random background $\sigma$ and exposure time for each training iteration.
The noise ranges we use replicate the levels of noise in the real \hst images that we will later test on, as explained in more detail below. In Model~1a \edr{(HST-long)}, for multi-exposure \hst images, $$\sigma_{BKG}\sim U(0,0.2)\textrm{ and }t_{exp}\sim 10^{U(2,6)}s$$ 

For Model 1 \edr{(HST-long/short)}, we perform the normalizing step of preprocessing before training, so the noise that is added is similar for bright and faint images. The reason 
% we do this 
is that when we evaluate on real \hst images, we can choose the noise levels to be similar to those real images. 
(If we use the same range of absolute noise levels, then because our simulations include some faint and some bright systems, a particular noise level might be very noticeable on faint systems and not at all affect bright systems. 
It is ineffectual to train a neural network with images where the noise makes the arcs unobservable or images where the noise is far too low.)
% In order to improve our performance on real \hst images, it is important to match the level of noise \textit{relative} to the brightness of the image, i.e. both faint and bright systems will ``appear noisy'' at similar levels as the real images. Thus, the noise range used replicates the range of relative noise levels in the real \hst images we will use to test on.
This scheme appears to work well. We show a sample of 25 simulated lensed systems with $0.5''<\theta_E<1.5''$ in Figure \ref{fig:sims}. 
Due to the complex sources we used and the large ranges for source position and redshift relative to the lens, 
our simulations are able to produce a diverse set of lensing configurations that visually match well observed systems.
Furthermore, as we will show below, trained on images simulated using this scheme, our ResNet model is able to perform well on out-of-sample, single-exposure \hst images with a very short exposure time, and even discover new strong lens candidates with faint arcs. 
% We recognize that it is possible that this scheme, when applied to a larger variety of real systems, may not be optimal.
% For example, hypothetically it does not provide enough examples of very faint arcs that are still above the noise level. 
% However, these systems, even if identified by a neural network.
% , may be hard to confirm, so it may not be necessary to find all of them.
% This normalization method 
% % might also be worse 
% may not be optimal when applied to real \jwst images, but a full investigation would require testing different images on real data, so we leave this to a future study.

For Models 2 and 3 \edr{(JWST-long/short/small)}, using noise appropriate to \jwst, 
% to accurately add this noise, 
we perform the normalizing step of preprocessing during training instead, right after the noise is added. That is, during training, the model adds the noise to and then normalizes each image in the batch.
Therefore, we use absolute noise in this case.
In this Einstein radius regime ($< 0.5''$), we have not tested our models on real data. 
Using absolute noise for now seems a sensible choice.
To apply our models on real \emph{\jwst} images may benefit from using relative noise as well. We leave this to a future investigation.
% This is because we do not have real images to test on, and thus want validation images that are as accurate as possible, using correct absolute levels of noise for \jwst, so we add the noise before normalizing. 

We use the background level from W18, and add Poisson noise using $t_{exp}=1000s$ for Model~2a \edr{(JWST-long)}, $t_{exp}=10,000s$ for Model~2b \edr{(JWST-short)}, and $t_{exp}=10,000s$ for Model~3 \edr{(JWST-small)}. While we train Model 2 \edr{(JWST-long/short)} on both exposure times, systems with the smallest Einstein radii of Model~3 \edr{(JWST-small)} are much more difficult to detect, and require the longer exposure time of $t_{exp}=10,000s$ for good performance.

\subsubsection{Training details}
Our training scheme is guided by H21, but we optimized the hyperparameters for each of the models presented in this work, including learning rate, learning rate decay schedule, batch size, number of filters in each shielding layer, and where we cap each image. We also tested both re-scaling images from 0 to 1, and normalizing with the standard deviation. In addition, we tried adding dropout layers to the models with a range of different values for the rate (the fraction of input units dropped), but this never improved validation performance, so we do not use any dropout.
For Model 1 \edr{(HST-long/short)}, the learning rate starts at $lr_0=1.25\cdot10^{-3}$ and decays by $5\times$ every $80$ epochs. Total training is done for $360$ epochs, with a batch size of 64. In each shielding layer (see H21) of the ResNet model, we use 32 filters.
For Models 2 and 3 \edr{(JWST-long/short/small)}, the learning rate also starts at $lr_0=1.25\cdot10^{-3}$, but does not decay. Total training is carried out for $500$ epochs, with a batch size of 16. In each shielding layer of the ResNet model, we use 16 filters.

\ed{Model 3 \edr{(JWST-small)}($0.02''<\theta_E<0.15''$) initially exhibited  overfitting.
By varying model hyperparameters, this has been significantly reduced (see Figure \ref{fig:auc_loss_small}).\footnote{Other strategies we tried include 
%adding dropout layers, 
changing the initial learning rate and removing the shielding layers entirely, but these changes did not lead to further improvements in validation performance.} 
% Overfitting was not completely eliminated, although it was greatly reduced.
This is of course is the most challenging classification problem in this work.
The ResNet from H21 was not designed to address this problem. 
But because this architecture continues to achieve high performance for smaller and smaller Einstein radius systems, we decide to push all the way to the diffraction limit of \emph{\jwst}.
\emph{Our current level of performance even at this limit remains excellent: achieving a validation AUC of 0.9644}.
% In order to achieve even better performance for this smallest Einstein radius range, it may be necessary to move to 
It may be possible to achieve even better performance for this problem by using a different machine learning architecture altogether, though this is outside the scope of this work.}

\subsection{Results: ResNet Model}\label{sec:results_trained_class_model}
In this section, we present the results of each of our trained classification models. We start with analyzing the performance on conventional strong lenses, followed by  testing the model on real \emph{\hst} images (\S\,\ref{sec:larger_lens_models}).
We then analyze the models on lenses with smaller $\theta_E$ and predict their real-world performance (\S\,\ref{sec:smaller_lens_models}).

\subsubsection{Conventional strong lenses: Toward finding every lens}\label{sec:larger_lens_models}
Every trained model is first evaluated on simulated validation data.
For Model 1 \edr{(HST-long/short)}, we also test it on real \hst images. For the real \hst images, the model is evaluated on selections of lenses and non-lenses. These tests demonstrate that our trained model can achieve near-100\% completeness \emph{and} near-100\% purity for conventional strong lenses ($\theta_E>0.5''$).

\paragraph{Model 1a \edr{(HST-long)}: validation on simulated training images}

\begin{figure}
    \centering
    \includegraphics[width=.49\textwidth]{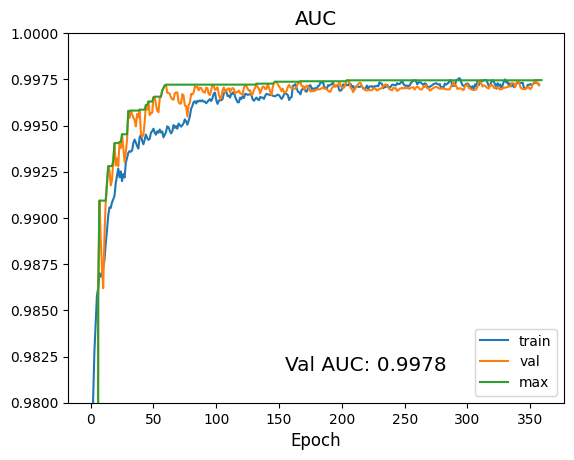}
    \includegraphics[width=.49\textwidth]{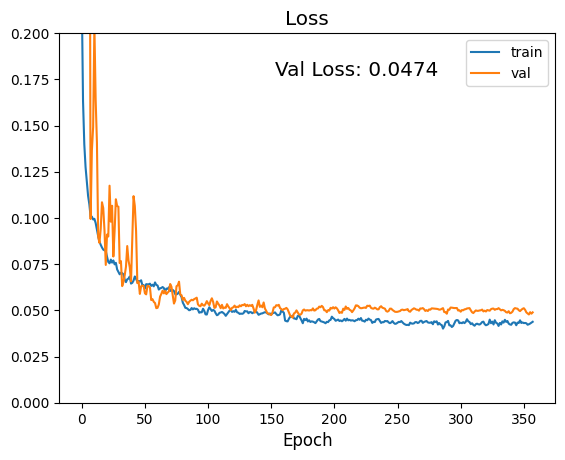}
    \caption{Validation and training loss and area under the ROC curve (AUC) over 360 epochs for Model 1a \edr{(HST-long)}. The validation AUC for the validation set  reaches a maximum of 0.9978. We also indicate the best validation AUC and loss. The curves are boxcar smoothed with a window size of 3.}
    \label{fig:auc_loss}
\end{figure}

Over the course of 360 epochs, both the training and validation AUCs steadily increase (Figure \ref{fig:auc_loss})
% , reaching $0.9973$ by the end, while the 
with the latter reaching a maximum of $0.9978$ by around epoch 150. 
The loss in Figure \ref{fig:auc_loss} presents a similar overall picture, where the validation loss reaches a minimum plateau of approximately $0.0442$ by epoch 150, and as expected, the training loss continues to decrease slowly, approaching $0.04$ by the end of the training.
In these two figures, the training AUC steps upwards and the training loss steps downwards at the first few epochs after the 80th and 160th epochs, due to the learning rate decay.

\begin{figure}
    \centering
    \includegraphics[width=.7\textwidth]{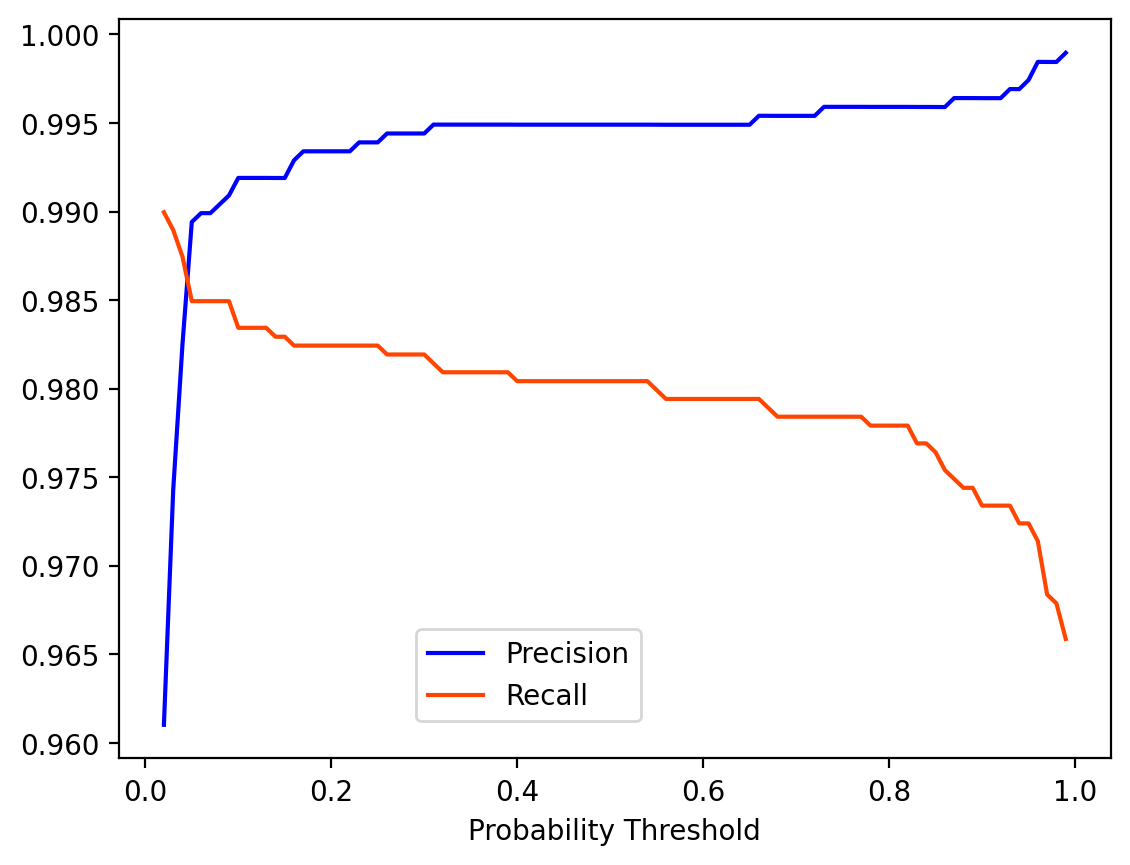}

    \caption{\edr{Precision and recall for Model 1a \edr{(HST-long)} as a function of probability threshold, with selected values shown in Table \ref{tab:model_1a_pr} below.}}
    \label{fig:prc}
\end{figure}
\begin{deluxetable}{ccc}
\caption{Model 1a \edr{(HST-long)} Precision and Recall}    \label{tab:model_1a_pr}
\tablehead{\colhead{Probability} & \colhead{Precision} 
& \colhead{Recall}}
    \startdata
    0.1 & 0.9919 & 0.9834 \\
    0.2 & 0.9934 & 0.9824 \\
    0.3 & 0.9944 & 0.9819 \\
    0.4 & 0.9949 & 0.9804 \\
    0.5 & 0.9949 & 0.9034 \\
    0.6 & 0.9949 & 0.9794 \\
    0.7 & 0.9954 & 0.9784 \\
    0.8 & 0.9959 & 0.9779 \\
    0.9 & 0.9964 & 0.9734 \\
    \enddata
\end{deluxetable}

To further evaluate the validation performance of the trained model, we can consider the precision and recall curves, shown in Figure \ref{fig:prc}. 
% At the shoulder of the curve, the model has a recall from $\sim0.975$ to $\sim0.985$ and a precision from $\sim0.99$ to $\sim0.995$. 
\ed{At a $50\%$ probability threshold, which represents a reasonable compromise between precision and recall, the model attains a 0.995 precision and 0.980 recall. 
That is, our model achieves 98\% completeness with a 99.5\% purity, a very high level of performance.
This result shows that for \edr{our simulated} conventional strong lenses ($\theta_E > 0.5''$), our ResNet model can nearly find all of them!
Therefore, in the next section, we apply our model to a small test set of real \emph{\hst} data.}

\paragraph{Model 1a \edr{(HST-long)}: testing on \hst observations}\label{sec:first_new_lens}

\begin{figure}
    \centering
    \includegraphics[width=.8\textwidth]{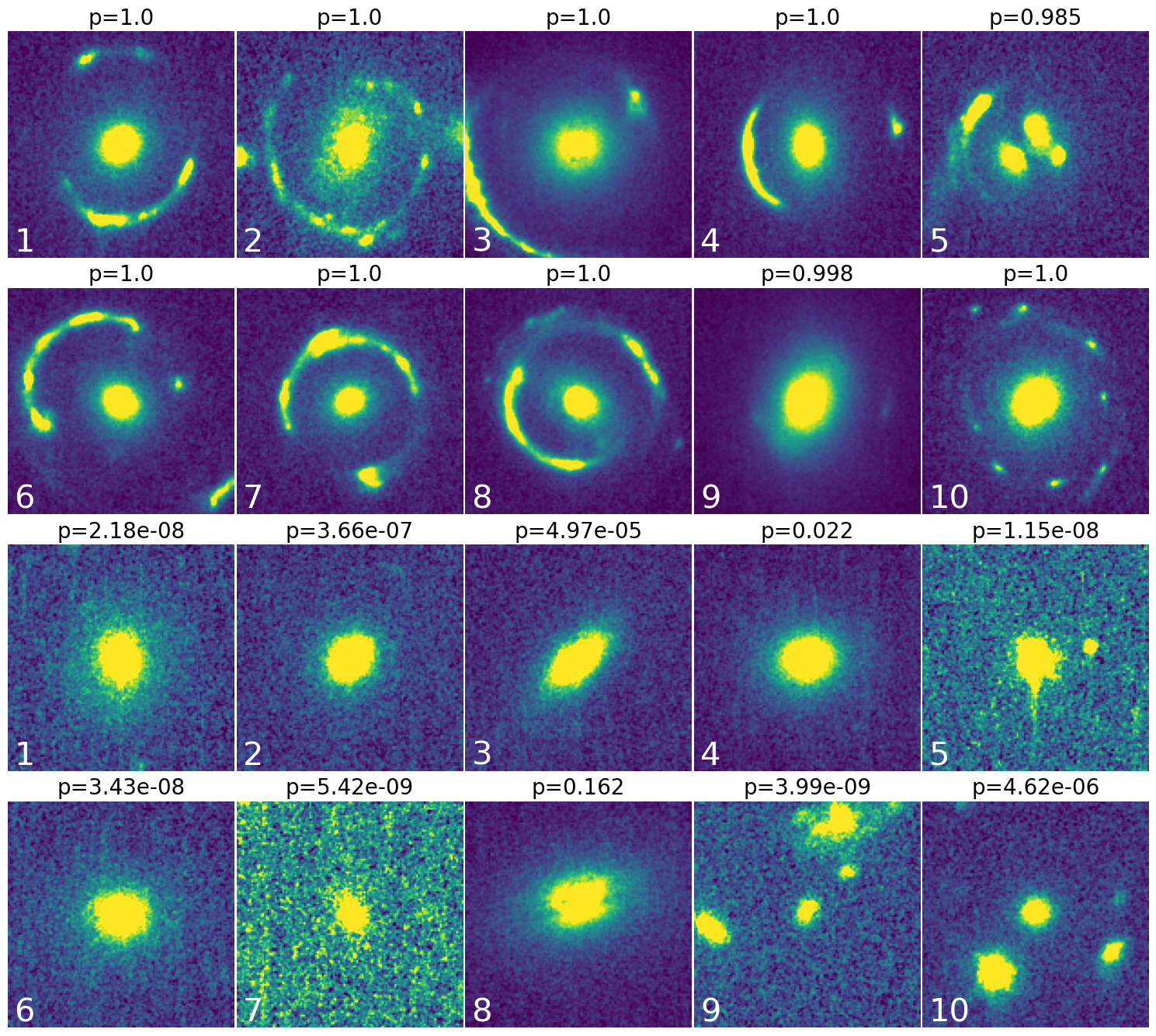}
    \caption{Trained Model 1a \edr{(HST-long)} tested on lenses with brighter arcs (top two rows) and non-lenses (bottom two rows). From \hst GO-15923, with exposure times between 4800 and 7800s. The model predicted probability is shown for each image.}
    \label{fig:vegetti}
\end{figure}
All the \hst observations we use for testing in this paper can be obtained from the MAST archive at doi:\dataset[10.17909/19gx-bz03]{http://dx.doi.org/10.17909/19gx-bz03}. 
The first test set of real images we use is a set of 10 lenses from the \hst program
%``Gravitational lensing by field halos: a clean test of dark matter models'' 
GO-15923 (PI: S.~Vegetti), which were observed in F606W on the WFC3 instrument. In this dataset, each image is observed with an exposure time of 4800-7600 seconds, and the arcs of each system are bright. The two or three exposures per system are drizzled to a pixel scale of $0.03''$ using Drizzlepac \citep{Hoffman_2021}.\footnote{\url{https://www.stsci.edu/scientific-community/software/drizzlepac.html}} 
This is consistent with the pixel scale of the simulations (which is $0.031''$ to match \emph{\jwst}). Finally, we follow the same image preprocessing as the simulated images (see Section \ref{sec:noise_preprocessing}).
In this test, we also include 10 non-lens galaxy cutouts from the same program, chosen randomly from the non-lens galaxies with a roughly similar brightness level to the lenses. For these cutouts, the galaxy at the center is not restricted to be an elliptical galaxy.
While it is possible that there might be some non-elliptical galaxies that this model performs less well on, 
on this set of test images, Model 1a \edr{(HST-long)} performs well.
Furthermore, determining which galaxies are ellipticals is not difficult for bright galaxies (which conventional strong lenses tend to be) with space-based observations \citep[see, e.g., ][]{Lang_2016}. 

The test images and the predicted probability values for each being a lens are shown in Figure~\ref{fig:vegetti}. The lowest-probability lens is lens \#5, at 98.51\%. This system, having three galaxies as the lens, is clearly an \emph{out-of-sample} positive example, since the simulated images used for training only have one galaxy as the lens. 
Thus, this is an encouraging result.
In future work, it is worth testing the potential of this model to find other out-of-sample lenses. 
The highest-probability non-lens is non-lens \#8, at 16.19\%, well below the threshold of 50\%, and this non-lens is not an elliptical galaxy. 
As all the cutouts in the training sample (lenses and non-lenses) have an elliptical galaxy at the center, but \ed{\emph{not}} all the galaxies at center of the test \emph{\hst} non-lens cutouts are ellipticals, 
this shows the model also performs well on out-of-sample \emph{negative} examples.
If we restrict to non-lenses with elliptical galaxies at the center, the highest probability is non-lens \#4, at 2.24\%, leading to an even wide separation between the probabilities given to lenses and non-lenses.
As mentioned earlier, in a real search, we can start by only using cutouts with an elliptical galaxy at the center.

The observed lenses include one lens (lens \#9, Figure \ref{fig:vegetti}), which was discovered to be a lens candidate accidentally. This image was initially included as a randomly-selected non-lens. The neural network assigned it a very high lensing probability of 99.84\%. We then noticed a possible faint arc and counter-arc in the image after varying the pixel scaling. Thus we reclassified it as a lens candidates. An \hst citizen science project appears to have missed this system \citep{Garvin_2022}. Thus, to our knowledge, this is the first lens candidate found by a neural network with ``superhuman performance'' (because it was originally missed by the human eye).

\begin{figure}
    \centering
    \includegraphics[width=.9\textwidth]{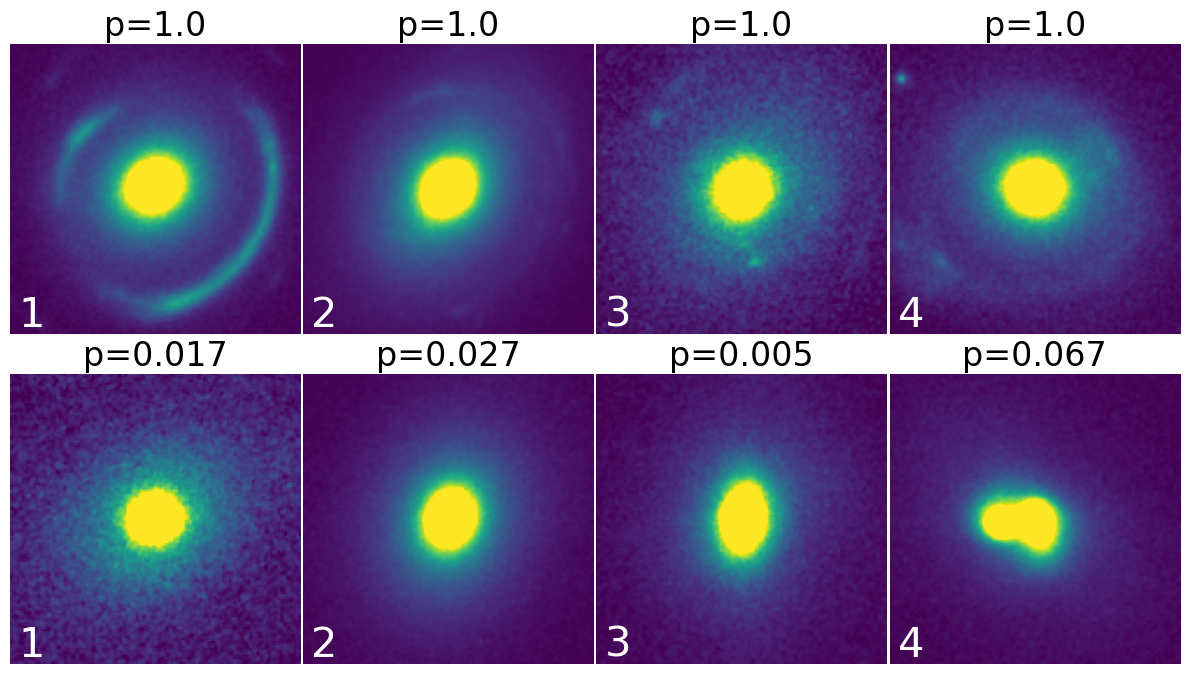}
    \caption{Trained Model 1a \edr{(HST-long)} tested on lenses with fainter arcs (top row) and non-lenses (bottom row). From \hst-GO 10886, with exposure time between 1500 and 2000s. The probability the model predicted is shown for each image.}
    \label{fig:bolton}
\end{figure}

Given the impressive performance of Model 1a \edr{(HST-long)}, we further examine its performance on a set of four lenses and four non-lenses from the an \hst program with significantly shorter exposure times,  
%``The Sloan Lens ACS Survey: Towards 100 New Strong Lenses'' (
GO-10886 (PI: A. Bolton), with 1500-2000~sec in the F814W band of WFC1. 
These systems also generally feature fainter arcs than the previous set from GO-15923. The non-lenses here are all elliptical, although they were also chosen randomly from the non-lenses with a similar brightness to the lenses. The images are again drizzled to a pixel scale of 0.03$''$. 

Model 1a \edr{(HST-long)} performs just as well on these lensed systems despite the fainter arcs in comparison to the lens light, with a wide separation between the probabilities given to lenses and non-lenses, as demonstrated in Figure \ref{fig:bolton}. The probabilities assigned to all 4 lenses round to 100\%. The highest-probability non-lens is non-lens \#4, assigned 6.7\%. 

\ed{\emph{We conclude that our model has performed excellently in these first two tests on real \hst observations}.
The vast majority of \emph{\hst} observations are at the depths of these two sets or deeper. 
In addition, to achieve the same depth, it would take \emph{\jwst} significantly less amount of exposure time. The results of these two tests therefore indicate that we can find many more lenses, 
\ed{\emph{probably all of them with comparable or deeper observations}}, 
in existing \jwst data and likely in \hst data as well, given that the human search did not find all the lenses (we will discuss this aspect more in \S~\ref{sec:new-lenses}).}

\ed{In order to push the limits of what is possible with this set of model and simulations, we analyze the model's ability to perform on single-exposure \hst images.
Given that the vast majority of existing and future observations from \hst, \jwst and \rst all have/will have much higher quality images, and we only perform this test to see how far we can push our ResNet model, we show the results in Appendix \ref{sec:single_exp_test}: even on this set, our model performs well.}

\subsubsection{Models for small 
lens searches}\label{sec:smaller_lens_models}

% Although we cannot test on real data, the
For models trained to classify strong lensing with small Einstein radii ($\theta_E<0.50''$), 
% because such observations are not available, we can 
though we have not tested them on real data in this work,
we examine their performance on the validation set. In this section, we analyze the validation performance of Models 2 and 3 \edr{(JWST-long/short/small)}, and then estimate the purity of Model 3 \edr{(JWST-small)} in real-world deployment.

\paragraph{Model 2 \edr{(JWST-long/short)}: validation performance}

\begin{figure}
    \centering
    \includegraphics[width=\textwidth]{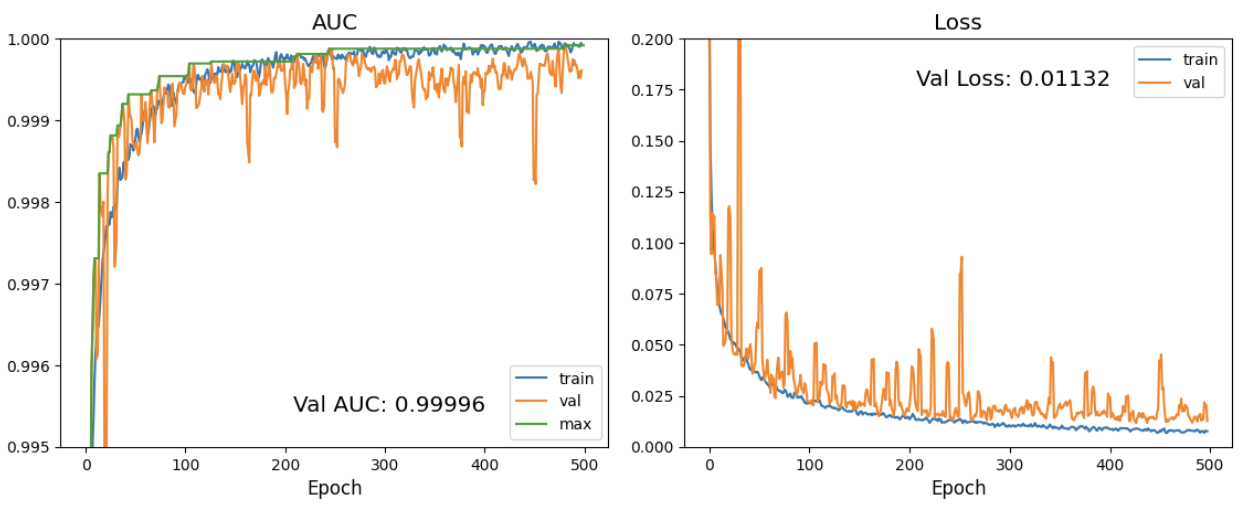}
    \caption{AUC and loss curves for the model trained on small Einstein radius systems ($0.15''<\theta_E<0.50''$) with $t_{exp}=10,000$~sec (Model 2a \edr{(JWST-long)}).}
    \label{fig:auc_loss_.15-.5_10000s}
\end{figure}

\begin{figure}
    \centering
    \includegraphics[width=\textwidth]{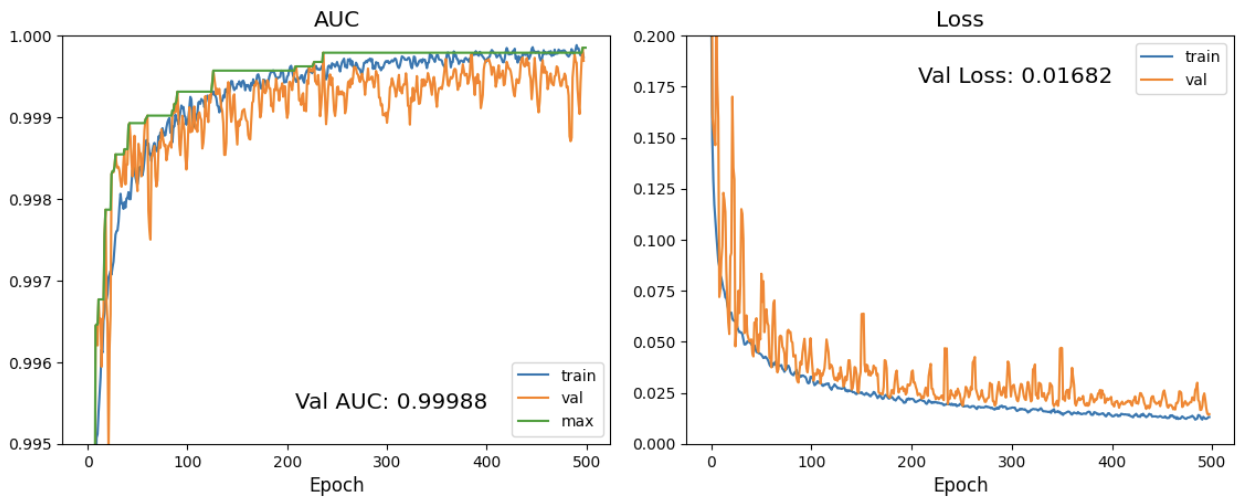}
    \caption{AUC and loss curves for the model trained on small Einstein radius systems ($0.15''<\theta_E<0.50''$) with $t_{exp}=1000$~sec (Model 2b \edr{(JWST-short)}).}
    \label{fig:auc_loss_.15-.5_1000s}
\end{figure}

For the intermediate range of Einstein radii ($0.15''<\theta_E<0.50''$) with $t_{exp}=10,000$~sec on \jwst, 
the trained model (Model 2a \edr{(JWST-long)}) performs well, with the AUC and loss curves shown in Figure \ref{fig:auc_loss_.15-.5_10000s}. The validation AUC reaches a maximum value of $=0.99996$ (Figure \ref{fig:auc_loss_.15-.5_10000s}).
% with a loss of $\sim0.01132$.

For the same $\theta_E$ range, the version Model 2b \edr{(JWST-short)}, trained and validated on simulated observations with a lower exposure time of $t_{exp}=1000$~sec (more common for \jwst) still performs well, with the AUC and loss curves shown in Figure \ref{fig:auc_loss_.15-.5_1000s}. The validation AUC reaches a maximum value of $=0.99988$, only slightly lower than the previous case.
% with a loss of $\sim0.01682$.

\paragraph{Model 3 \edr{(JWST-small)}: validation performance}\label{sec:model3_val_performanace}

Given that the performances for simulated systems with much smaller Einstein radii ($0.15'' < \tE < 0.5''$) than conventional lenses remain excellent,
we decide to push and see how the ResNet can perform all the way to the diffraction limit of \jwst.

\begin{figure}
    \centering
    \includegraphics[width=\textwidth]{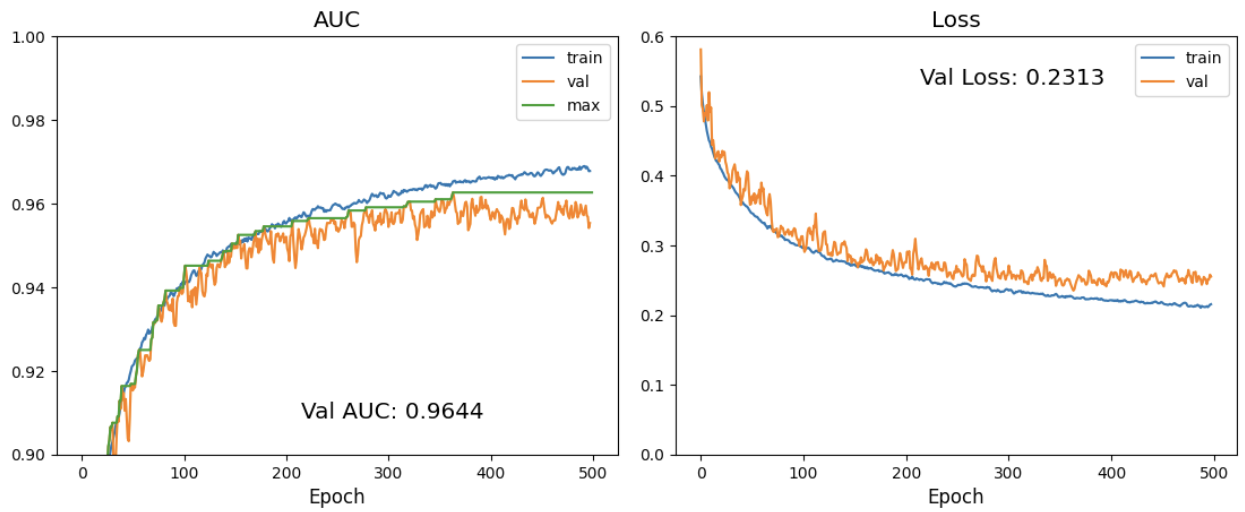}
    \caption{AUC and loss curves for Model 3 \edr{(JWST-small)} with $0.02''<\theta_E<0.15''$. For the non-lenses, light for the foreground galaxy is included in the simulation. This makes the training more challenging (see \S~\ref{sec:methods_unlensed_images}).}
    \label{fig:auc_loss_small}
\end{figure}

% For the smallest range of lenses ($0.02''<\theta_E<0.15''$), 
In this range, we are primarily interested in the very low mass galaxies and not the high-mass ones with unfavorable redshift combinations. 
Even for these systems, the trained model continues to perform well---reaching a maximum validation AUC of over $0.96$ (Figure \ref{fig:auc_loss_small})---even though the convergence is somewhat slower than the models for larger Einstein radii. 
% The loss curve presents a similar overall picture, where the validation loss reaches a minimum plateau by the end of training. 
% Both the AUC and the loss show a small degree of over-fitting, with the loss for validation slightly above the training, and the AUC for validation slightly below the training.

\begin{figure}
    \centering
    \includegraphics[width=\textwidth]{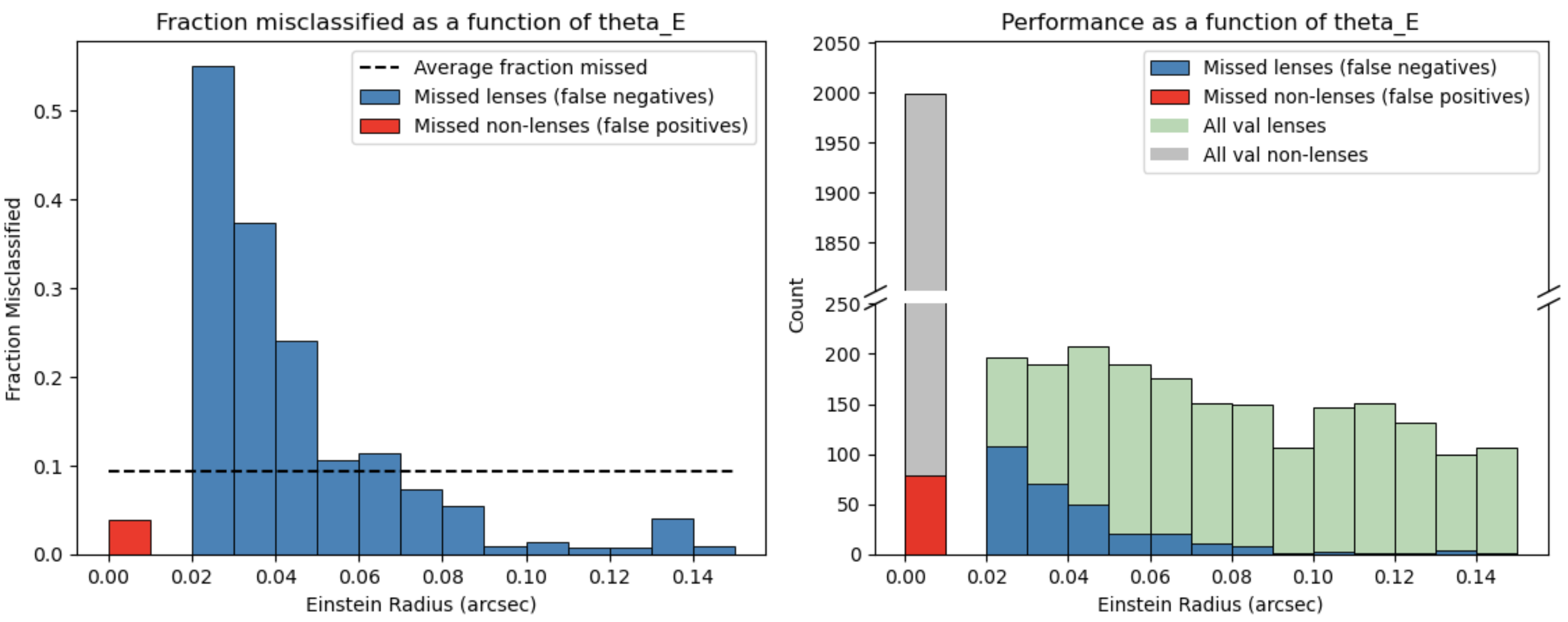}
    \caption{Performance of Model 3 \edr{(JWST-small)} with $0.02''<\theta_E<0.15''$ as a function of Einstein radius, at a threshold of 0.5. The columns at $\theta_E=0.0''$ correspond to non-lenses. Therefore, on the left plot, the column at $\theta_E=0.0''$ shows the fraction of non-lenses misclassified as lenses, and the other columns show the fraction of lenses misclassified as non-lenses.}
    \label{fig:small_performance}
\end{figure}

One of the most important ways to evaluate this model is how it performs on validation images as a function of Einstein radius. This allows us to determine the smallest $\theta_E$ systems that this method can correctly classify. 
Figure \ref{fig:small_performance} shows a strong dependence of the model's performance on Einstein radius. 
For $\theta_E \gtrsim0.10''$, the model performs very well on validation images. 
But the fraction of lenses that are misclassified drastically increases as Einstein radius decreases further. 
When $\theta_E$ reaches $0.02''$, only about 40\% of the lenses are correctly identified. 
It is important to note that this performance is highly dependent on the threshold chosen for classification. 
This could be adjusted to decide the trade-off between precision and recall, and with this threshold (0.5), 4.45\% of non-lenses are misclassified as lenses.

\begin{figure}
    \centering
    \includegraphics[width=\textwidth]{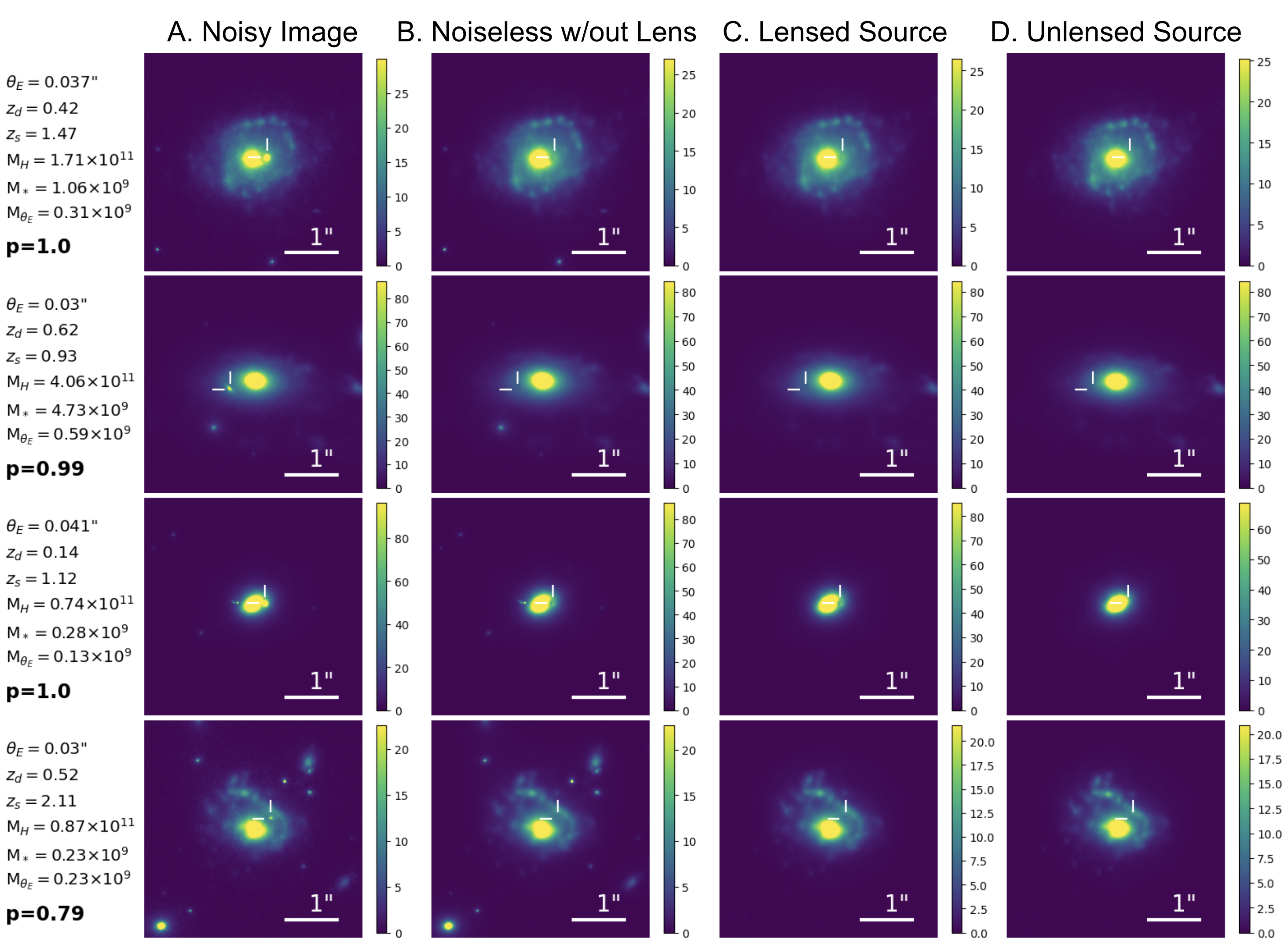}
    \caption{Correctly classified lenses for Model 3 \edr{(JWST-small)} ($0.02''<\theta_E<0.15''$), or true positives. 
    Four examples are shown, with one system in each row. 
    The four columns show A. the full image seen by the ResNet, B. the image with the lens light removed, C. the image with the lens light and environmental galaxies removed (i.e., just the lensed source), and D. the original unlensed source. 
    The white crosshairs show the location of the lens.
    With the lens light removed, one sometimes (e.g., for the systems in the second and third rows) can see a small ``dimple'' (essentially, a small Einstein ring) in the second and third columns. 
    On the left, for each image, we show the Einstein radius, the lens and source redshifts,
    the halo and stellar masses for the lensing galaxy as well as the projected mass within \tE ($M_{\theta_E}$),
    and the probability assigned by the ResNet.
    Note that for the system in the third row, the projected mass within \tE is nearly two orders of magnitude smaller than that for the lensed SN~Zwicky.}
    \label{fig:very_small_tp}
\end{figure}

\begin{figure}
    \centering
    \includegraphics[width=\textwidth]{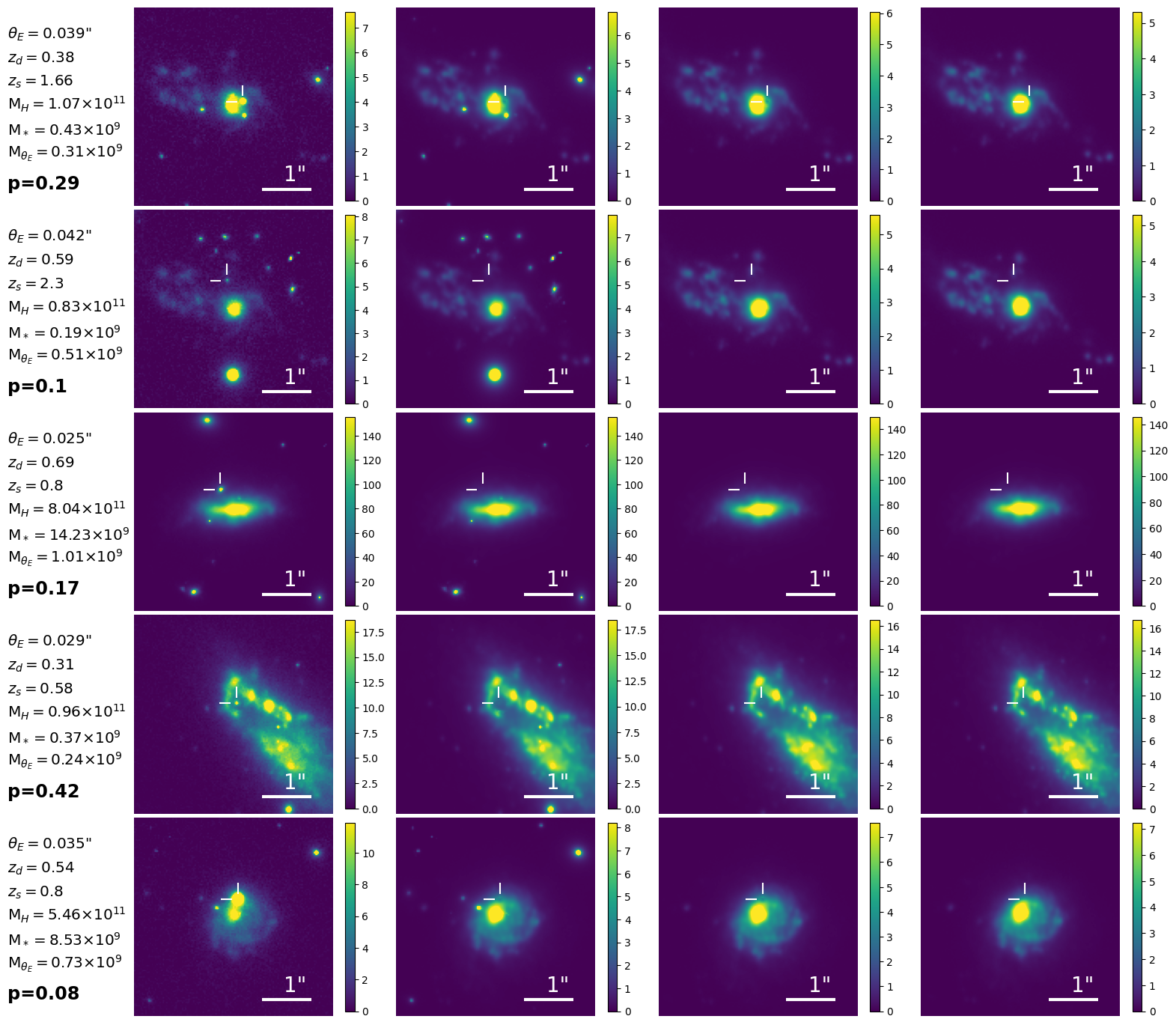}
    \caption{Incorrectly classified lenses for Model 3 \edr{(JWST-small)} ($0.02''<\theta_E<0.15''$), or lenses misclassified as non-lenses (false negatives).
    The arrangement of the columns and the information shown on the left are the same as for Figure~\ref{fig:very_small_tp}.
    In these cases the ResNet probabilities are all below the threshold (0.5).
    %Four examples are shown, with one system in each row. 
    %The four columns show A. the full image seen by the ResNet, B. the image with the lens light subtracted, C. the image with the lens light and environmental galaxies subtracted (i.e. just the lensed source), and D. the original unlensed source. The white crosshairs show the location of the lens. The titles of each image show the Einstein radius, the lens and source redshifts, and the probability assigned by the neural network.
    }
    \label{fig:very_small_fn}
\end{figure}

To better understand how the neural network model is performing on the smallest lens images, we have displayed four examples of correctly classified simulated lenses with $\theta_E<0.05''$ in Figure~\ref{fig:very_small_tp}.
% These systems have a range of lens-to-source brightness ratios, and
There are two factors that make these systems extremely challenging to identify by human eye.
First, they have a wide range of lens-to-source brightness ratios 
and the area around the lens is often only visible with careful image pixel value scaling. 
Second, these are very small lensing systems with barely perceptible lensing effects.
% , \emph{even with the knowledge of the location of the lens}.
For example, the systems shown in the third row has an Einstein radius of $\tE = 0.041''$, with a projected mass within \tE nearly two orders of magnitude smaller than that for the lensed SN~Zwicky.
The ResNet, however, readily identifies this system by assigning a high probability of nearly 1.
% First, these are very small lensing systems. are, 
% most of them would be quite hard to definitively identify with the human eye.
Even so, 
% it is not always clear that they are occurrences of lensing. 
it would be helpful to find the precise location of these lenses for further examination and possible follow-up observations. 
We will apply a U-Net model for this purpose in 
\S\,\ref{sec:unet}.

Finally, it is notable that in that system, with the lens light removed, one can see a small ``dimple'' (essentially, a small Einstein ring) in the second and third columns. The removal of the lens light, of course, is not possible in real observations.\footnote{With multi-band imaging, this may be possible.} However, based on this observation, we think this technique can one day be applied to identifying lenses with even smaller masses, so small that the lens itself is dark (see more discussion in \S\,\ref{sec:low_mass_halos}).

In Figure \ref{fig:very_small_fn}, 
we display five examples of incorrectly classified simulated lenses with $\theta_E<0.05''$ 
(i.e., they are misclassified as non-lenses, or false negatives). 
\ed{These represent typical causes for the model to misclassify a lens, as we lay out below.}
The first image is likely misclassified due to the relatively faint source (see the colorbar in Figure \ref{fig:very_small_fn}) as well as the relatively bright lens light compared to the source.
The second image has a lens that is far away from the brighter parts of the source, as well as a faint source. The third image has a far away lens and a very small Einstein radius, at $\theta_E=0.025''$. 
The fourth image has the lens in a source with very complex structure, 
and therefore it is difficult to distinguish the lensing effects.
Finally, the fifth image involves a lens that has a high mass for its Einstein radius, so its lensing effect is obscured by lens light.
%This is a very difficult problem, as the Einstein radii are so small, so nearly all of the false negatives are attributable to a high noise level, a complicated source, very low lensing effect, or a lensing effect obscured by lens light.
This is obviously a very difficult problem, as the Einstein radii are so small.
After looking at a number of examples, 
we do not see any cases where the lenses are discernible to the human eye (keep in mind that we know where the lenses are in the simulations) but are missed by the model. 
That is, in finding these very small Einstein radii systems, 
the trained model is performing at least as well as a human can. 
In fact, for a large fraction of the cases, the model is able to identify systems that will be missed by a human inspector.
That is, for this classification problem,
the model is performing at a ``superhuman'' level.

Even with high classification probabilities, it may still be hard to confirm these lensing systems with small Einstein radii because often it is hard to visually see where the lens is. 
For this reason, as stated earlier, in the next section (\S\ref{sec:unet}), we will train a U-Net model that can pinpoint the locations of these small lenses.
% as shown in the following Section \ref{sec:methods_unet}.

\begin{figure}
    \centering
    \includegraphics[width=\textwidth]{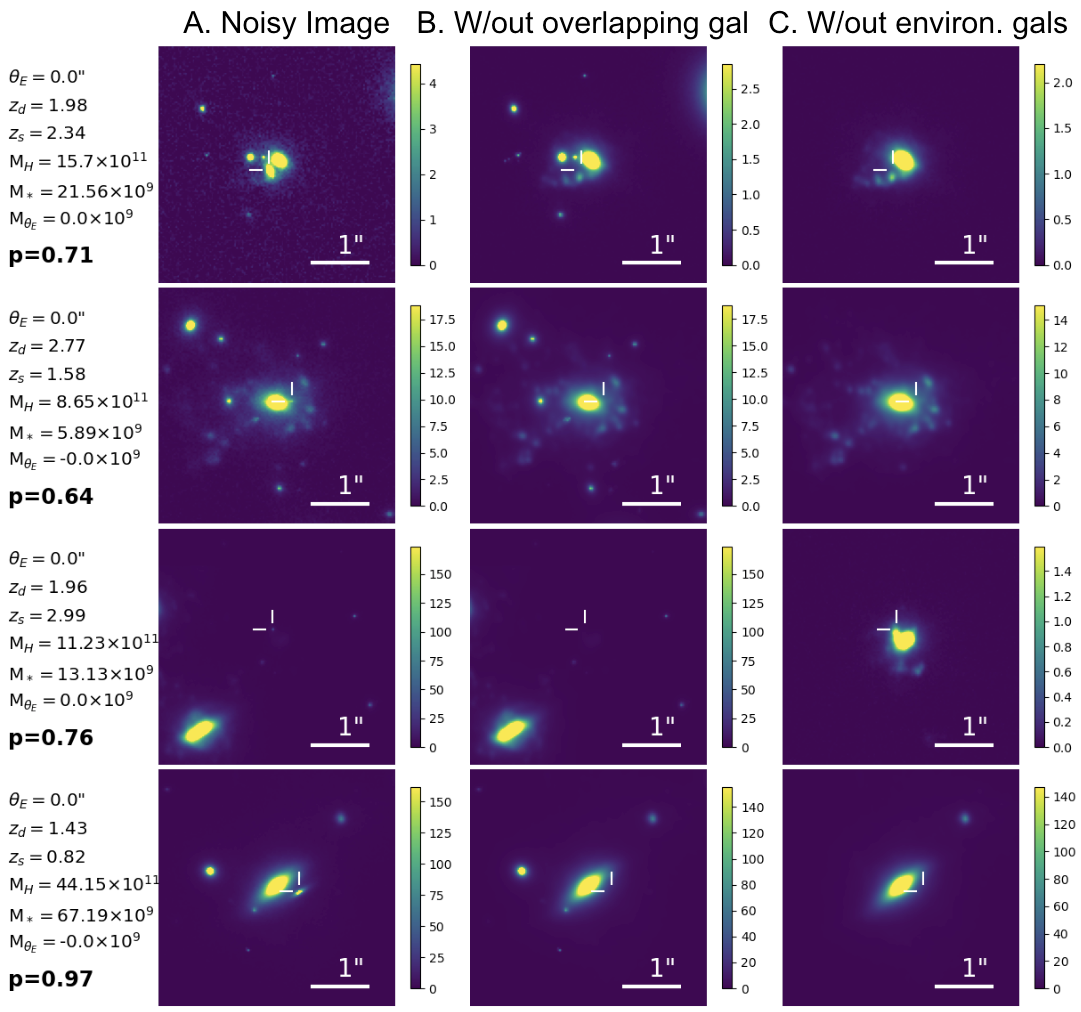}
    \caption{Incorrectly classified non-lenses for Model 3 \edr{(JWST-small)} ($0.02''<\theta_E<0.15''$), or non-lenses misclassified as lenses (false positives).
    The arrangement of the columns and the information shown on the left are nearly the same as for Figure~\ref{fig:very_small_tp}: here there are three columns: A. the full image seen by the ResNet, B. the image with the light of the overlapping galaxy removed, and C. the image with the overlapping galaxy light and environmental galaxies removed (i.e., just the source). \edr{Here, the white crosshairs show the location of the overlapping galaxy.}
    These are non-lenses, so $\theta_E=0$.}
    \label{fig:very_small_fp}
\end{figure}

\edr{In Figure \ref{fig:very_small_fp}, 
we display four examples of incorrectly classified simulated non-lenses
(i.e., they are misclassified as lenses, or false positives). 
These represent typical scenarios for the model to misclassify a non-lens, as we lay out below.
The first and second images are likely misclassified due to their source structure which could be interpreted as having lensed arcs. The first image also has multiple nearby environmental galaxies which could be interfering.
The third image has a VELA environmental galaxy that is much brighter than the source galaxy (clearly revealed in column C), which is the likely cause.
Finally, the fourth image does not have an obvious cause for misclassification.}

\edr{Overall, many false positives we have observed have an overlapping galaxy or environmental galaxies close to the source. VELA environmental galaxies also do not seem to cause significantly more false positives than S\'{e}rsic environmental galaxies. While it is not always obvious to the human eye why a non-lens was misclassified as a false positive, interpretability is difficult for neural networks, and we can only speculate.}

\paragraph{Additional test of Model 3 \edr{(JWST-small)}}\label{sec:additional_tests}

\ed{We perform an additional test on Model 3 \edr{(JWST-small)}, by examining its performance 
% is to evaluate its predictions 
on a modified version of the validation data.}
\edr{We define the variations of non-lenses in Table~\ref{tab:non_lenses}: while we only ever train the model on non-lenses of Type 1 (i.e. images with an overlapping elliptical galaxy), we also evaluate our model on the test set with the overlapping galaxy removed from all test images. Therefore, this test set has non-lenses that contain only the VELA source galaxy and environmental galaxies (We remind the reader that we call these Type~2 non-lenses.)}
% We define the variations of non-lenses in Table~\ref{tab:non_lenses}, which include 
% Type 1---they are, as stated before, the type of non-lenses used in our training.
% For this test, we evaluate on non-lens images without overlapping galaxies (Type 2 in Table~\ref{tab:non_lenses}; also see Section \ref{sec:methods_unlensed_images}).
Note that they were not used in the training and therefore represent out-of-sample images. 

We want to know the performance of our trained model on these images because  real image cutouts that do not have overlapping galaxies will be a strong majority of the negatives when this model is applied to real observations.
We find that without retraining the Model~3 \edr{(JWST-small)}, its performance on this out-of-sample set of non-lenses (Type~2) is in fact much better than the non-lenses in the original validation set (using only Type 1 non-lenses). 
This is perhaps not a surprise, as Type~2 non-lenses are actually more different from the lenses, but it nevertheless is important to check.
As seen in Figure \ref{fig:small_performance}, 4.45\% of Type 1 non-lenses are misclassified as false positive lenses (at a ResNet probability threshold of 0.5).
Meanwhile, only 2.2\% of the Type 2 non-lenses are misclassified as lensed. \edr{We believe this performance is more representative of the model performance on non-lenses in real images.}
% when the overlapping galaxies are removed.

\ed{
% because we are being general in simulating our non-lens overlapping galaxies. 
\edr{In summary: We emphasize that we have gone beyond 
% the standard ML performance metrics 
the typical practice of \edr{only using Type~2 non-lenses.}
% On balance, w
We have constructed a more challenging non-lens sample by using Type~1 non-lenses. To remind the reader, all of these non-lens images have an overlapping galaxy so that the ResNet can be \textit{specifically trained to recognize the lensing effect.}}
Then, \emph{without retraining}, 
we test our model 
% by testing against 
on out-of-sample Type~2 non-lenses. 
% Type~2 non-lenses are cutouts that do not contain overlapping galaxies. 
On these, the model actually performed better than on the non-lenses in the training sample (Type~1).
This is an important test because the vast majority of the non-lenses from actual observations will be of Type~2.
% 2) Type~1$\alpha$: to construct these, we remove the lensing effects in the lenses.
% We performed this test to see how well Model~3 can ``detect'' the lensing effect, everything else being equal between the lenses and non-lenses in this case.
% We recognize this is a somewhat contrived scenario.
% Given the near alignment, lensing effect would likely occur, but we allow for the cases where they merely overlap. Without a careful study, it is not clear how likely such a scenario would occur, though it is not unreasonable to estimate that the probability is on par with strong lensing (which is probably the case for Type~1 non-lenses as well).
% Therefore, these are much rarer than Type~2 non-lenses.
% %(here we also need to remove a bias; see the paragraph above). 
We are satisfied with Model~3 \edr{(JWST-small)} as it has passed this important test.
% , both of which use out-of-sample non-lenses.
% In summary, our challenging training sample does not cause the ResNet to sacrifice performance on the most abundant non-lensed cases (Type B), but helps us distinguish from these rarer non-lenses with an overlapping galaxy (Type A1). Such cases are rare, likely on a comparable order of abundance as true lenses. 
% Finally, we verify that the ResNet is indeed able to detect the lensing effect.
% Thus, our pipeline has not.
}

\ed{We perform one additional test on Model~3 \edr{(JWST-small)} by evaluating it on simulations using truncated SIS mass profile instead of SIE. The details are described in Appendix \ref{sec:trunc_lens_models}: we find that the model's performance is not degraded.}

\begin{figure}
    \centering
    \includegraphics[width=\textwidth]{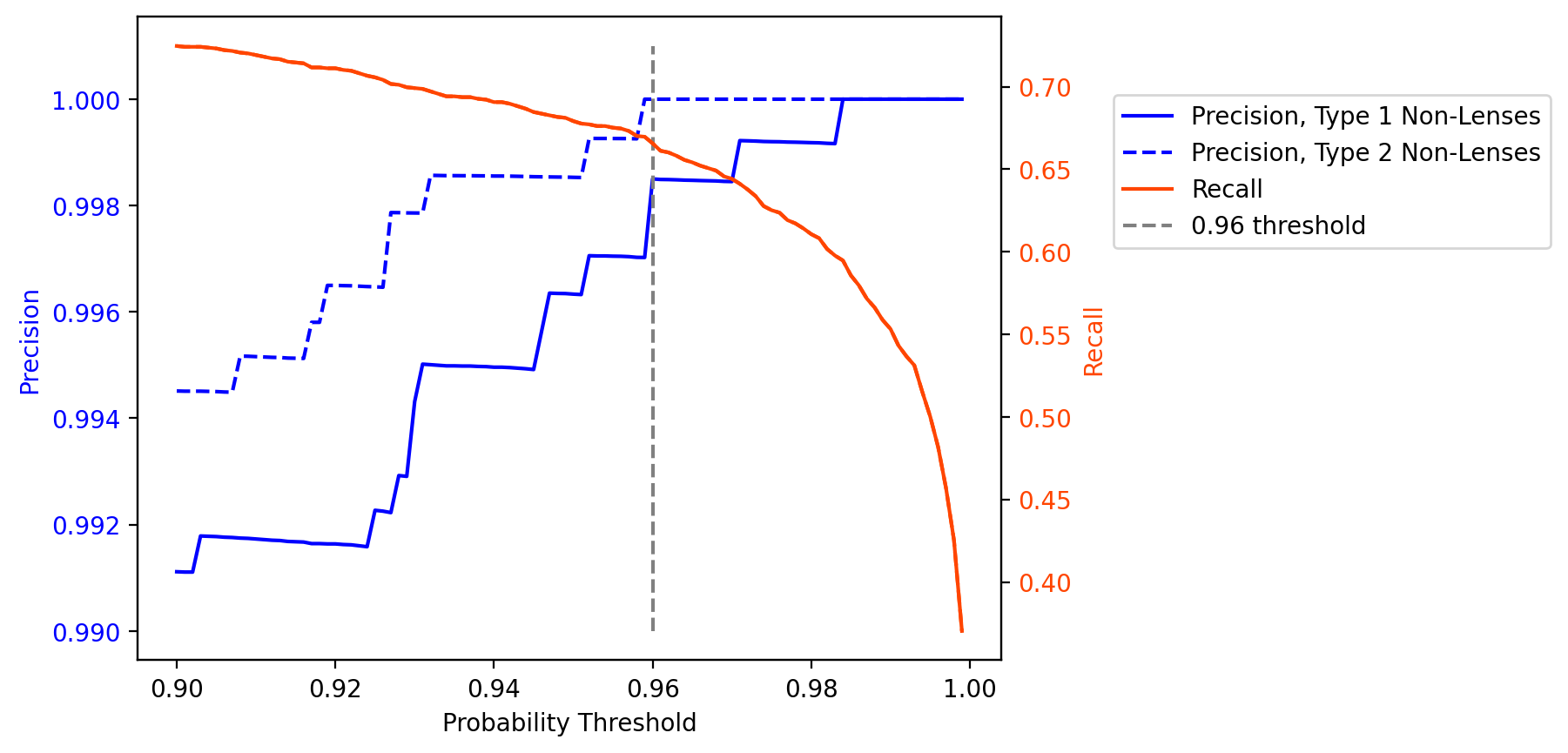}
    \caption{\edr{Precision and recall for Model 3 \edr{(JWST-small)} as a function of probability threshold, with selected values shown in Table \ref{tab:model_3_pr} below. The solid lines show the performance on the original validation set, using Type~1 non-lenses. For actual searches for these low-mass lenses, the more relevant metrics are the dotted lines, showing the performance on the original validation set with Type~2 non-lenses (see Section \ref{sec:methods_unlensed_images}). Because only the non-lenses differ, the recall is identical for the Type~1 and Type~2 non-lenses. Note that at these very high precisions, there are a very small number of false positives, so the jumps in precision are caused by individual false positives being removed.}}
    \label{fig:new_precision_recall}
\end{figure}
\begin{deluxetable}{cccc}
\caption{Model 3 \edr{(JWST-small)} Precision and Recall}    \label{tab:model_3_pr}
\tablehead{\colhead{Probability} & \colhead{Precision (Type 1)} & \colhead{Precision (Type 2)} 
& \colhead{Recall}}
    \startdata
    0.91 & 0.9917 & 0.9952 & 0.7191 \\
    0.92 & 0.9916 & 0.9965 & 0.7111 \\
    0.93 & 0.9943 & 0.9979 & 0.6992 \\
    0.94 & 0.9950 & 0.9986 & 0.6907 \\
    0.95 & 0.9963 & 0.9985 & 0.6792 \\
    \textbf{0.96}
 & \textbf{0.9985} & \textbf{1.0000} & \textbf{0.6657} \\
    0.97 & 0.9985 & 1.0000 & 0.6442 \\
    0.98 & 0.9992 & 1.0000 & 0.6107 \\
    0.99 & 1.0000 & 1.0000 & 0.5532 \\
    \enddata
    \tablecomments{The bold line denotes the threshold of 0.96 that gives 100\% precision on the validation set with Type~2 non-lenses, which we use going forward. The recall is identical for Type~1 and Type~2 as they only differ in the non-lenses.}
\end{deluxetable}

\paragraph{Estimates of completeness and purity for deployment}\label{sec:purity}

\ed{One important measure for the success of these models is the precision we expect them to achieve when deployed in real observations. For Model~3 \edr{(JWST-small)}, our validation set has 50\% lenses and 50\% non-lenses, but in real observations, we will be evaluating on all background sources. From our forecasts in Section \ref{sec:forecast}, we predict that $0.0002615\approx1/3800$ observable sources will have a lens with $\theta_E>0.02''$ in front of it.}

\ed{Our original validation set has overlapping galaxies in each non-lens cutout (Type 1 non-lenses). The precision and recall curves for the validation set are shown in Figure \ref{fig:new_precision_recall}.
However, in real observations, this type of non-lenses would probably be roughly as rare as lenses. 
Given how low the FPR is for this type of non-lenses (see the dotted lines in Figure \ref{fig:new_precision_recall}), they will not be a significant source of contamination.} 

\ed{The vast majority of non-lenses, however, do not have overlapping galaxies (Type 2 non-lenses). 
We thus want to carefully evaluate purity using Type 2 non-lenses.
The precision and recall for Model 3 \edr{(JWST-small)} on Type~2 non-lenses are shown as the dotted lines in Figure \ref{fig:new_precision_recall}. 
Based on the precision and recall alone, the model appears to perform better for Type~2 non-lenses than for Type~1 non-lenses, but we need to take into consideration that Type~2 non-lenses are far more prevalent than Type~1 non-lenses in real observations.}
% Even though it appears to be better than the PRC for Type 1 non-lenses, we need to take into consideration that this type of non-lenses is far more prevalent than Type~2 non-lenses.}
As an example (following \S\,4 of H21), at a 0.95 probability threshold, in the validation sample of 4000, 1358/2000 lenses (67.9\%) are found as true positives (TPs), and 2/2000 non-lenses (0.1\%) are false positives (FPs).
Consider a real-world sample of 10,000 sources. As we mentioned above, approximately $1/3800$ real sources has a lens in front of it, so we would expect roughly $10,000\times1/3800\approx2.6$ lenses in the sample. At a 0.95 threshold, we can find 67.9\% of these, or $2.6\times.679\approx1.8$ lenses (TPs). Also, $0.1\%$ of the 10,000 non-lenses are FPs, or $10,000\times 0.001\approx10$ FPs. Overall, our model would then give us $\sim11.8$ positive lens candidates, of which 10 are false positives and 1.8 are false positives. This gives a FP-TP ratio of 5.56-to-1 at the stage of human visual inspection or follow-up observations.

The table shown on Figure~\ref{fig:new_precision_recall} shows that we reach 1.0 precision at a 0.96 probability threshold and above (no false positives), with a completeness of 66.6\%.
% Due to our limited validation set size of 4000, we cannot properly estimate the false positive rate at these higher thresholds, but we can almost certainly improve our FP-TP ratio by increasing our threshold, decreasing the number of FPs even further while only slightly decreasing our completeness.
\emph{The potential discovery of such small Einstein radius systems would be unprecedented.}
% , and this is an acceptable purity.}

\ed{In the interests of being thorough, we also perform this purity estimate for Model 2 \edr{(JWST-long/short)}, with the range of $0.15''<\theta_E<0.50''$, using Type~2 non-lenses. For Model~2a \edr{(JWST-long)}, with a 10,000s exposure time, we reach 1.0 precision at a 0.999 probability threshold. At this precision, we reach a completeness of 94\%. For Model~2b \edr{(JWST-short)}, with a 1000s exposure time, we reach 1.0 precision at a 0.95 probability threshold. At this precision, we reach a completeness of 91\%. 
These results indicate that for lenses in this Einstein radius range, we reach $>$ 90\% completeness with no contamination (FPs).
Clearly, these represent a very high level of performance.}

\subsection{Discussion: ResNet Model}\label{sec:discuss}

\ed{Below we discuss the results presented so far, regarding:
identifying low-mass halos ($M_\mathrm{halo} <10^{11}M_\odot$) in \S~\ref{sec:low_mass_halos};
halo mass distribution of systems with low Einstein radii in \S~\ref{sec:small_thetaE_mass};
forecasts of detectable numbers of low-mass lenses in \S~\ref{sec:forecasts_detectable};
pixel value scaling and the related advantage of lens search with neural networks in \S~\ref{sec:pixel_val_scaling}, and in this context, we present two new lens discoveries in \S~\ref{sec:new-lenses}.}

\subsubsection[Low-mass halos below 10\^11 solar masses]{Low-mass halos below $10^{11}M_\odot$}\label{sec:low_mass_halos}

In the present form of this work, we have not explored searching for observational evidence for halos down to $10^{9} M_\odot$.
% but our methodology can be easily generalized to include these low-mass halos. 
However, our approach does allow us to discover lenses with halo mass down to $10^{10}M_\odot$; to fully test CDM, the abundance of halos in the mass range of $10^{10}M_\odot$ to $10^{11}M_\odot$ needs to be determined as well. In this work, we show that $\sim40$\% of systems with Einstein radius $<0.05''$ are in the $10^{10}M_\odot$ to $10^{11}M_\odot$ mass range, which translates to thousands in our simulation dataset. This opens up a whole new regime to study galaxy mass and dark matter models.

There are two other reasons we did not explore low mass halos down to $10^9 M_\odot$: 1) the CosmoDC2 catalog only goes down to $10^{10} M_\odot$; and 2) preliminary studies showed that the lensing effects would be so small that the Einstein radius would be $< 0.02''$, below the resolution limits of \jwst.
In fact, for halos down to $10^9M_\odot$, Einstein radii are expected to be up to $\sim0.0035''$, depending on how favorable the lensing configuration is. We believe that these might in fact be on the edge of being discoverable by the ResNet with location detection made possible using U-Net. 
We leave this investigation to a future study.
But we point out that the most likely of such candidates can then be checked with follow-up observations using, for example, ALMA interferometry for now, which recently achieved 5 mas resolution \citep{Asaki_2023}, and TMT or ELT with AO in the future. Even higher resolutions have been achieved with optical interferometry. The ESO Very Large Telescope (VLT) Interferometer (VLTI) can achieve resolutions of 2 mas \citep{Olofsson_2011, GRAVITY_2017}.  In the radio regime, \cite{Gomez_2022} has achieved the current limit for resolution in astronomy at 12 $\mu$as.
The challenge when attempting to find these very small halos with $<10^{10}M_\odot$ is of course their very small lensing effects. However, an advantage is that these are expected to eventually be fully ``dark'' halos with no lens light\footnote{Or emitting so little light as to be effectively ``dark''.} that might obscure the lensing effect, which would make the detection by ML techniques such as ResNet \emph{easier}.

\subsubsection{Masses of systems with small Einstein radii}\label{sec:small_thetaE_mass}

\begin{figure}
    \centering
    \includegraphics[width=.7\textwidth]{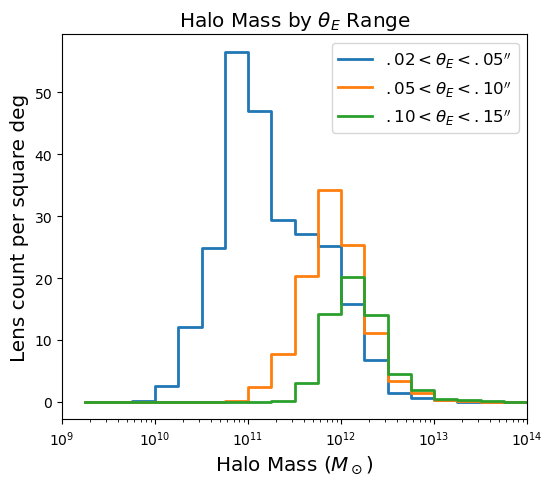}
    \caption{The forecast mass distribution of halos with $0.02''<\theta_E<0.15''$.
    Note that these histograms make it clear that the condition of $M_\mathrm{halo} < 10^{11} M_\odot$ essentially implies $0.02'' < \tE < 0.05''$.
    Though the inverse is not true: there are many systems with \tE in that smallest bin, but the lens halo mass is $ > 10^{11} M_\odot$.}
    \label{fig:forecast_masses}
\end{figure}

In our forecasts and simulations, we include lensing systems with small Einstein radii. 
% For example, Tables \ref{tab:forecasts} and \ref{tab:more_forecasts} include systems with $0.02''<\theta_E<0.10''$, and Model 3 has $0.02''<\theta_E<0.15''$. 
Figure \ref{fig:forecast_masses} shows the forecast mass distributions of halos with $0.02''<\theta_E<0.15''$, broken down further into three smaller bins, ($0.02'', 0.15''$), ($0.05'', 0.10''$), and ($0.10'', 0.15''$).

\begin{figure}
    \centering
    \includegraphics[width=.7\textwidth]{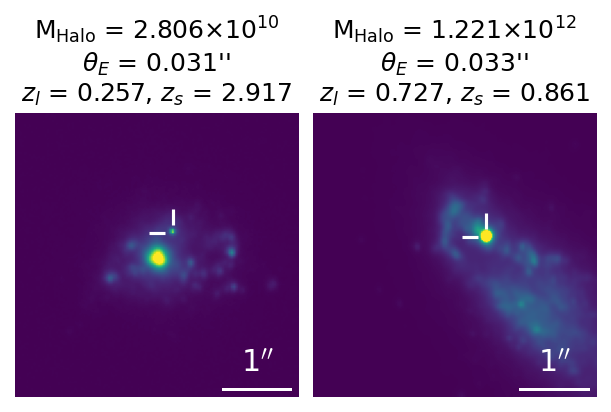}
    \caption{Two examples of simulated systems (in the context of Model 3 \edr{(JWST-small)}) with a similar Einstein radius around $0.03''$. The left system has a low mass, and the right system has a mass almost two orders of magnitude larger. This is because the left system has a redshift configuration that is much more favorable to lensing than the right system.}
    \label{fig:small_te_examples}
\end{figure}

We can roughly divide these systems into those with a low mass ($M_{\textrm{Halo}}\lesssim 10^{11}M_\odot$), and those with a high mass ($M_{\textrm{Halo}}\gtrsim 10^{11}M_\odot$) but an unfavorable source and lens redshift combination that results in a small Einstein radius.
Figure~\ref{fig:forecast_masses} shows two important understandings.
\ed{1) \emph{For the purpose of 
% digesting our results further on 
studying systems with $M_{\mathrm{Halo}}\lesssim 10^{11}M_\odot$,
the focus should be on the Einstein radius range of $0.02''<\theta_E<0.05''$.}}
That is, the condition of $M_\mathrm{halo} < 10^{11} M_\odot$ essentially implies $0.02'' < \tE < 0.05''$.
2) But the inverse is not true: in the Einstein radius range of $0.02'' < \tE < 0.05''$, $\sim39.6$\% of the simulated lenses have a halo mass below $10^{11}M_\odot$, and therefore there are many systems with lens halo mass $ > 10^{11} M_\odot$.
% It is important to recognize that many lensing systems with $\theta_E<0.05''$ actually have relatively high-mass halos (such as $M_{\textrm{Halo}}\sim 10^{12}M_\odot$). Two examples showcasing this difference are shown in 
Figure \ref{fig:small_te_examples} demonstrates this difference, showing two simulated systems from Model 3 \edr{(JWST-small)} with very similar Einstein radii, both with $\tE \sim 0.03''$, but very different masses, due to their different redshift configurations.

\subsubsection{Forecasts of detectable numbers of low-mass lenses}\label{sec:forecasts_detectable}
\begin{figure}
    \centering
    \includegraphics[width=\textwidth]{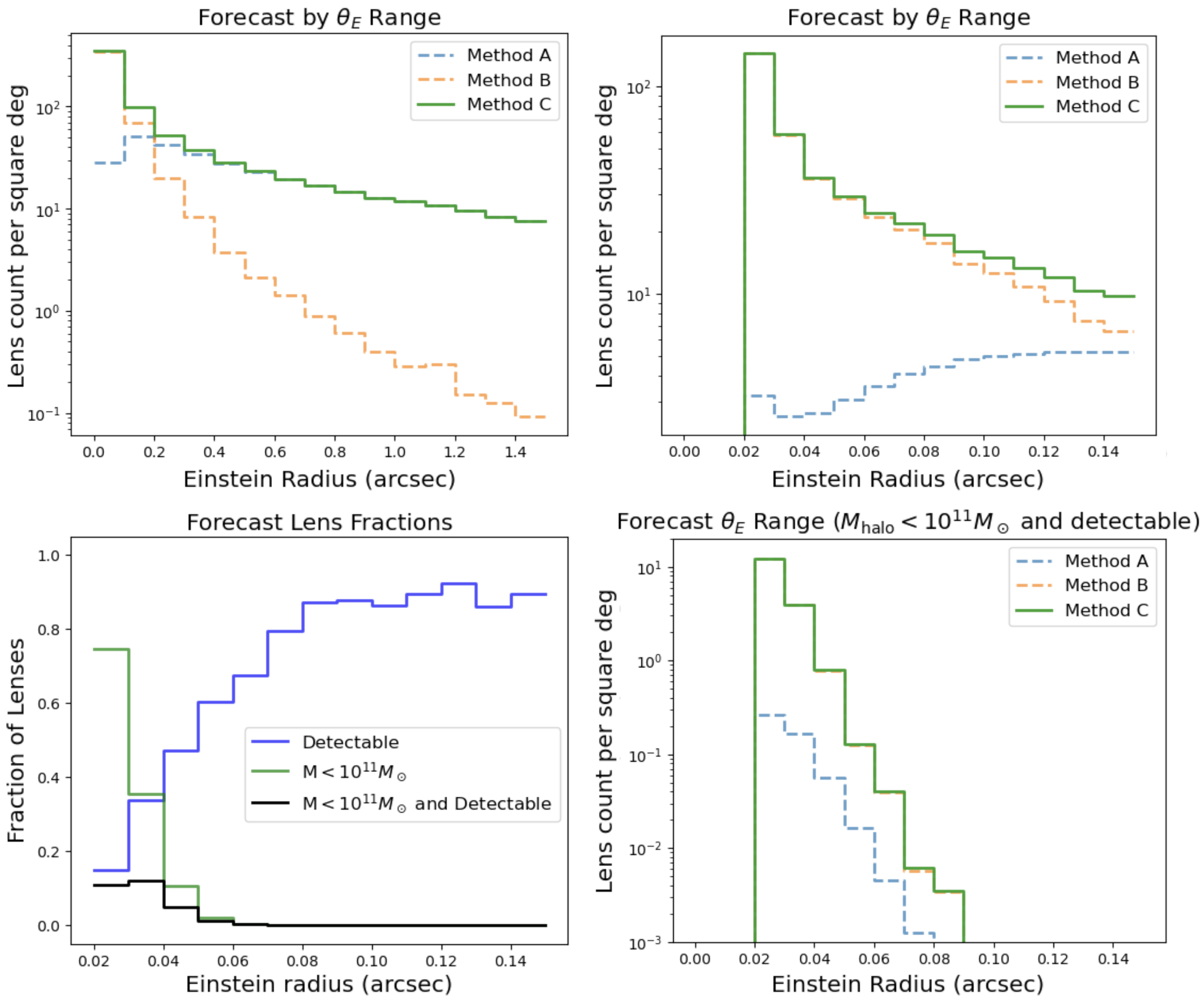}
    \caption{These four panels illustrate the numbers we forecast for low mass lenses \ed{($M_\mathrm{halo} <10^{11}M_\odot$)} that are detectable. 
    The top row shows histograms of the overall forecast of lens numbers (using the methods from Section \ref{sec:forecast}), as a function of Einstein radius. 
    The top left panel shows $0.02''<\theta_E<1.50''$ and top right panel shows the zoom-in on the range $0.02''<\theta_E<0.15''$. The bottom left and right panels shows the fractions and numbers, respectively, of lenses \ed{with $M_\mathrm{halo}<10^{11}M_\odot$} that are detectable by Model~3 \edr{(JWST-small)} at a 0.96 probability threshold \ed{(with zero false positives). Note that for the lower right panel, the forecast numbers from Method~A contribute minimally to this category, as expected.}} 
    % Finally, the bottom right panel shows the forecast lens numbers which are detectable by Model~3 at a 0.96 threshold and have a lens with $M<10^{11}M_\odot.$}
    \label{fig:forecast_detectable_lenses}
\end{figure}

\begin{deluxetable}{cccc}
\tablehead{\colhead{$\theta_E$ range} & \colhead{ Forecast (lenses/deg$^2$)} & \colhead{Detectable lenses/deg$^2$}\\ \colhead{} & \colhead{} & \colhead{with $M<10^{11}M_\odot$ (ResNet)}}
    \startdata
        $0.02''<\theta_E<0.05''$ & \numsmallthetae & \numlomassresnet \\
        % $0.05''<\theta_E<0.10''$ & 230 & 0.38 \\
        % $0.10''<\theta_E<0.15''$ & 150 & 0.0 \\
    \enddata
    \caption{Forecasts for the number of lenses with $0.02'' < \tE < 0.05''$ per square degree (\numsmallthetae, which is the same as the difference between the first two rows of Table \ref{tab:more_forecasts}), and how many of those with $M_\mathrm{halo}<10^{11}M_\odot$ are detectable (by the ResNet Model 3 \edr{(JWST-small)} at a 0.96 threshold), with zero false positives. Results are given to 2 significant figures.}
    \label{tab:forecast_detectable_lenses_no_unet}
\end{deluxetable}

\ed{By connecting the results of our forecasts in \S~\ref{sec:forecast} with the results of the model, we can predict the numbers of these low-mass lenses that we actually expect to be detectable. Figure \ref{fig:forecast_detectable_lenses} shows the steps in our calculations.}

\ed{We start with the overall forecasts of lens numbers, for the entire \tE range of 0.02$''$-$1.5''$ and the zoomed-in range of $0.02''$-$0.15''$, shown in the top left and top right of Figure~\ref{fig:forecast_detectable_lenses}, respectively. 
Note that Method A dominates the total counts of Method C at large Einstein radius (at $\theta_E\gtrsim0.25''$) but Method~B dominates at low Einstein radius (at $\theta_E\lesssim0.25''$),
as expected. 
Then we calculate the numbers and fractions of the lenses with $M_\mathrm{halo}<10^{11}M_\odot$.
The recall of Model~3 \edr{(JWST-small)} at a 0.96 probability threshold gives the fraction of lenses we can find with minimal false positive rates (see Section \ref{sec:purity}). 
\emph{This translates to 
%35 
\numlomassresnet
discoverable lenses per deg$^2$ with $M_\mathrm{halo}<10^{11}M_\odot$ above this threshold (Table \ref{tab:forecast_detectable_lenses_no_unet}).}
The bottom left and right panels in Figure~\ref{fig:forecast_detectable_lenses} show the resulting histogram in terms of numbers and fractions, respectively.
% For the smallest \tE bin of $0.02''\text{-}0.05''$, we forecast how many of those with $M_\mathrm{halo}<10^{11}M_\odot$ are detectable (by the ResNet Model 3 at a 0.96 threshold), with zero false positives 
% are also presented 
% in Table \ref{tab:forecast_detectable_lenses_no_unet}.
\emph{Therefore, with tens of detectable low-mass lenses per square degree forecast to be discoverable with \jwst, we expect to find them in real observations.}
% can be confident we 
%that even if the fraction of detectable lenses is much lower in real observations, a non-zero number of such systems could be found.
}

\subsubsection{Scaling and the neural network advantage}\label{sec:pixel_val_scaling}

\begin{figure}
    \centering
    \includegraphics[width=\textwidth]{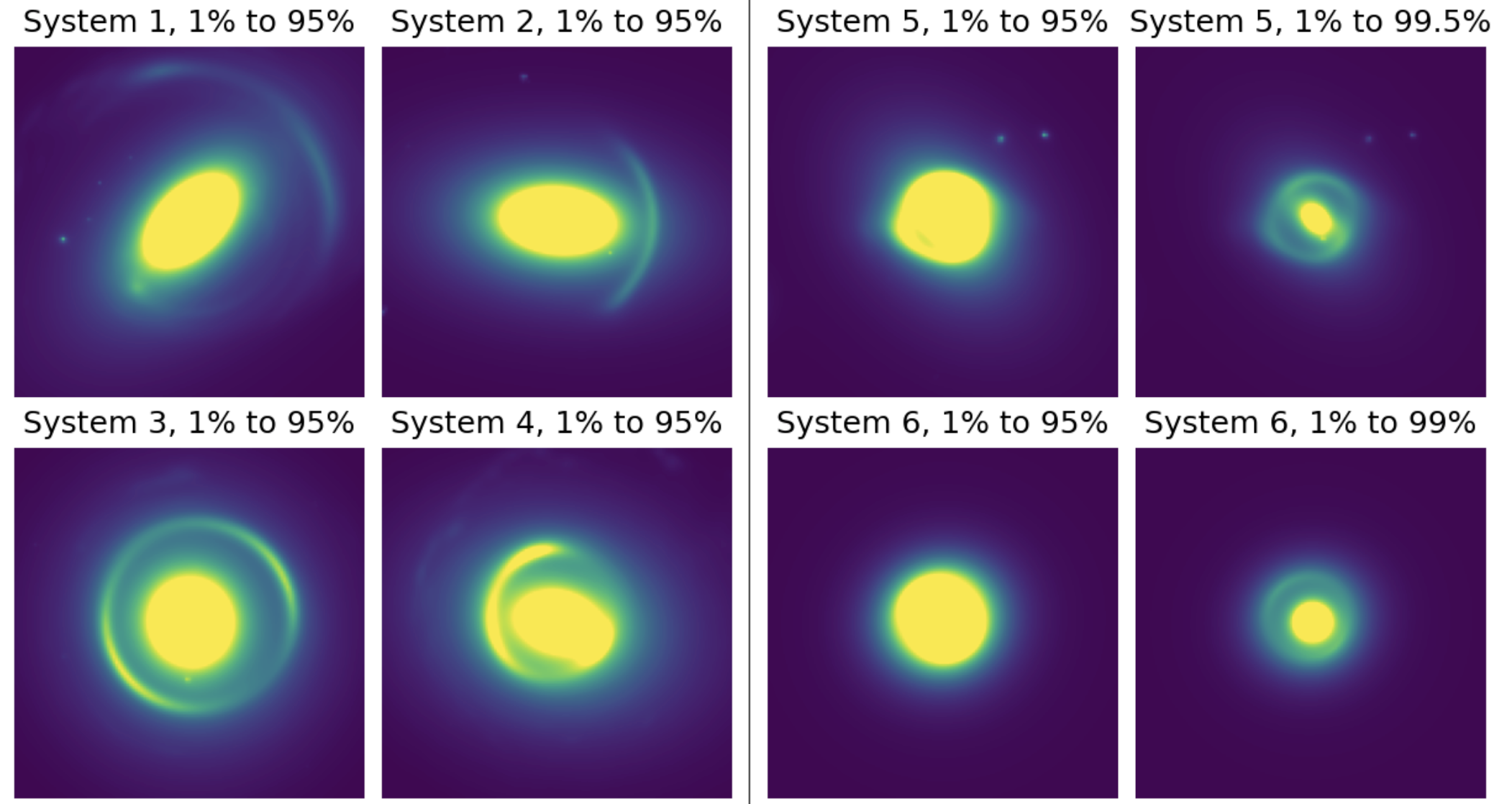}
    \caption{The left half of this figure shows four examples of simulated systems where a 1\% to 95\% pixel scaling allows for clearly identifying the lens.
    The right half of this image shows two examples of simulated systems where the lens cannot be clearly identified with a 1\% to 95\% pixel scaling, along with optimal scaling to clearly reveal the lensed arcs. The top right system shows a barely noticeable small gap at the lower left corner of the image, hinting at the lensing effect, while the bottom right system shows no clues.}
    % Three examples of simulated lenses in which altering the image pixel scaling strongly affects how easy it is to identify the lens.
    % The first row shows an example of a lens with a very faint arc, which requires brightening the image (scaling to a maximum of 95\% or 85\%).
    % The second row shows an example of a lens that is easiest to identify at fainter scalings, but the lowest scaling of 1\% to 85\% leaves the main arcs completely obscured.
    % The third row also shows an example of a lens easier to identify at fainter scalings, and this example cannot be identified at 95\%, but only at 99.9\%.}
    \label{fig:sim_scaling}
\end{figure}

One very important factor we find in this work is that the image pixel scaling can be significant when trying to identify lensing systems. 
In fact, image scaling is likely the dominant factor of why discoverable, large (i.e., conventional-sized) Einstein radius systems are not found by humans.
Figure~\ref{fig:sim_scaling} shows examples to demonstrate this: 
it is possible to find a scaling that will allow for clearly identifying most lenses, such as a 1\% to 99\% pixel scaling.
But there is not a single scaling setting that will work for every lensing system, such as the examples on the figure's right panel.

Even the best possible default scaling option would cause some arcs to be too faint to see, and other arcs to be so bright that they are no longer distinguishable from the lens.
Thus human identification would require trying many different options for image scaling. 
This process would of course be very time-consuming, 
making it impractical to identify all observable lenses from a large volume of images. 
The data volume will become much larger even in the near future (\euc, LSST), and therefore it is especially worth pointing out also that for human inspection, given time constraints, it is impossible to use cutout sizes as small as those shown in Figure~\ref{fig:sim_scaling}. 
\citet{Garvin_2022}, e.g., used cutouts images orders of magnitudes larger for their citizen science project. The much larger cutout size significantly exacerbates the pixel scaling issue, likely causing an even higher number of lens candidates to escape human detection.
In this work we show that our ResNet model can vastly improve not only speed, but also performance over human identification.
This is because the ResNet model is much less sensitive to pixel value scaling of an image.

In addition, the use of the ResNet model provides an advantage in observing images with small Einstein radius. 
In many cases, such as those seen in Figure \ref{fig:very_small_fn}, it may still be hard for a human to identify lenses at \emph{any} scaling.
% , if it is hard to find the \ed{location of the} small lens on the image. 
% This challenge of locating small lenses might even be the dominant factor impeding discovery, unlike for large Einstein radius systems. 
The ResNet model would be \ed{more than merely helpful and saving time in identifying systems that contain a small Einstein radius lens, but rather, practically, it likely is the only way to find such systems.} 
% , greatly saving the time needed to find these systems.

% It is possible that 
Broadly, lenses with small Einstein radius might be detectable in two ways. 
First, ``astrometrically'', where the lens causes a distortion of the shape of at least a part of a source.
% around the lens. 
Second,``photometrically'', where the lensing effect is more subtle and only causes a deviation in the light profile of the source galaxy. This latter type is likely especially difficult for the human eye to notice,
% might not show visible arcs to a human without subtracting the source light, and 
but can only be found with a neural network.

\subsubsection{New lens candidates}\label{sec:new-lenses}

\begin{figure}
    \centering
    \includegraphics[width=.485\textwidth]{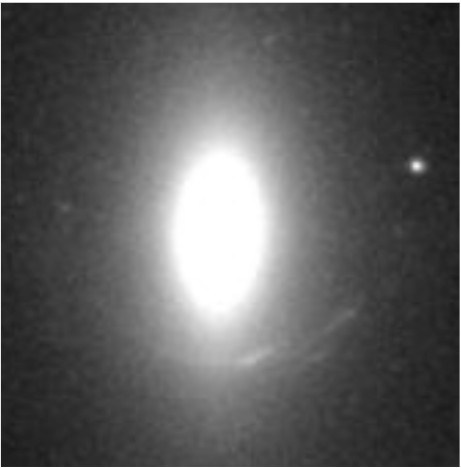}
    \includegraphics[width=.505\textwidth]{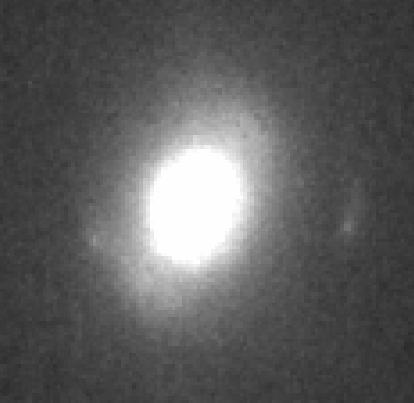}
    \caption{Two new lens candidates found in this work. The first is at $\alpha=$ 09:36:03.00, $\delta=$ 09:14:03.12 (\hst-144.0125+9.2342) and the second is at $\alpha=$ 12:02:00.41, $\delta=$ 47:42:16.92 (\hst-180.5017+47.7047).
    Note that our naming convention uses the decimal coordinates of the lens center.
    The second candidate, on the right, was first identified as a lens by the ResNet while humans had not initially recognized it as one.}
    \label{fig:new_candidates}
\end{figure}

The insensitivity of our ResNet model to image pixel scaling has significant implications to finding large $\theta_E$ lensing systems, too.
To illustrate this advantage, 
we show the identification of two \emph{new} strong lens candidates from \emph{\hst} archival images in Figure \ref{fig:new_candidates}.
\ed{Both of these lenses were missed by the comprehensive citizen science search using human inspectors by \cite{Garvin_2022}.}
\edr{Their work involved an initial identification of 2354 cutouts containing possible lenses by citizen scientists.\footnote{The cutouts can be found at \url{https://www.zooniverse.org/projects/sandorkruk/hubble-asteroid-hunter/talk/tags/gravitational_lens}, but to our knowledge, the coordinates are not provided.}
% (which cannot be easily searched).
Then, the visual inspection by the authors of \cite{Garvin_2022} selected 417 lens candidates, which did not include the two lens candidates found here.}

The first (\hst-144.0125+9.2342) was discovered by eye serendipitously. However, this system has a very bright lens compared to the arcs, and many choices of image pixel scaling of the system render the arcs invisible. 
We subsequently applied our trained model to this system and found a probability of  98.3\%.
For the second one (\hst-180.5017+47.7047), we picked a galaxy from an \hst image to serve as a non-lens for model testing, 
but our trained model consistently returned a very high probability of 99.8\% (we first mentioned this system in \S~\ref{sec:first_new_lens}). 
Eventually, we realized that when we changed the scaling on the image, there was a possible pair of a faint lensed arc and counterarc around a massive elliptical galaxy.
% , which the neural network was picking up on. 
The first example shows that, once again, the ResNet does not miss what the human eye identifies as a lens, even when the lens cannot be identified at all in some image pixel scalings.
As for the second example, to our knowledge, this is the first time when a neural net model identified a lens that  the human eye initially missed, performing at a ``super-human'' level.
\clearpage
\section{Detecting Small Einstein Radius Lenses with the U-Net}\label{sec:unet}
\edr{In the above section (\S~\ref{sec:sim-resnet}) we have shown that our ResNet model performs very well on finding all three categories of simulated lenses (``conventional-sized'' or``large'': $\tE > 0.5''$, ``intermediate'': $0.5''>\tE > 0.15''$, and ``small'': $0.15'' > \tE > 0.02''$).}

\ed{For systems with a small Einstein radius that are correctly classified by} the ResNet (Model~3 \edr{(JWST-small)}, as described in Section \ref{sec:methods_trained_class_model}), 
% identifies a cutout 
% as a strong lens, but  
we need to find a way to \edr{pinpoint} such a system.
In this section (\S\,\ref{sec:unet}), we \edr{therefore} focus on these \edr{``small''} systems with $\tE < 0.15''$.
As we have shown above, 
\ed{typically, such lenses}
% the lens in such a} system 
% can be 
are very difficult to detect by the human eye without knowing its location in the image cutout. We therefore resort to the U-Net model to detect the \emph{location} of such a candidate lens, for follow-up observation with the purpose of confirmation.

% In this section, 
We describe our methods (\S~\ref{sec:methods_unet}) and results (\S~\ref{sec:results_unet}) of training a U-Net on the same dataset as Model~3 \edr{(JWST-small)} for pinpointing the location of a lens with small Einstein radius. 
In our results, we also present evidence that our U-Net model is not merely assigning high probabilities to local maxima in the light distributions of cutout images, but rather appears to have ``learned'' what a small Einstein radius lens is.

\subsection{Methods: U-Net Model}\label{sec:methods_unet}

\subsubsection{Model Purpose and Selection} \label{sec:model_purpose}

Given an image that is a  candidate for a small Einstein-radius lens with $0.02''<\theta_E<0.15''$ (Model~3 \edr{(JWST-small)} for the ResNet, as described in Section \ref{sec:methods_trained_class_model}), we want to be able to
predict the location of the lens as an aid for human judgment and possible follow-up observations for the purpose of confirmation. 
% A U-Net model can not only help identify the spatial
% location of a lens candidate, but also perform the
% classification task on lenses by itself. 

We train a U-Net for this purpose,
% to detect the locations of foreground lensing halos in simulation data, 
as a necessary followup step to the ResNet positive classifications.
% and 2) increase our confidence in the lens classification task, in combination with the ResNet.
Our model assigns a detection probability to each pixel in an image. Since we are only interested in one lens candidate, we predict the pixel with the highest probability as the location of the lens.

\subsubsection{Image Preprocessing} \label{sec:image_preprocessing}

Training the U-Net model directly on given
simulation images after normalizing them to [0,~1] generally gives quite poor results because the distribution of pixel values in each image varies too much. For some images, the maximum pixel value is 10, while for others it can go up to about $2.5 \times 10^5$. Without added preprocessing, the U-Net model struggles to identify lenses. 
% when each image's pixel values scales differently. 
Since we want the model to learn to distinguish the relative brightness between pixels locally, inputting the image's pixel
values on a log scale or one similar to a log scale addresses this issue. 
\edr{We normalize each image by performing the following, represented as pseudocode:}

\edr{\begin{itemize}
    \item Get the 95th percentile pixel value of the image
    \item Divide each pixel in the image by the 95th percentile pixel value
    \item image = (image - min(image)) * 254 / (max(image) - min(image)) + 1
\end{itemize}}
\edr{The last line of the pseudocode amounts to linearly mapping each pixel in the image to a value in [1, 255].}

% \begin{lstlisting}[caption={Normalization for Visualization},label=vis-norm]
% function normalize(image):
%     image = image / get_percentile(image, 95)
%     image = (image - min(image)) * 254 / (max(image) - min(image) + 1)
%     return image
% \end{lstlisting}

% \begin{python}[caption={Normalizaton for Visualization},label=vis-norm]
% def normalize(image):
%     image = image / np.percentile(image, 95)
%     image = (image - image.min()) * 254 / (image.max() - image.min()) + 1
%     return image
% \end{python}

Note that this normalization process for the U-Net is \emph{different} from that for the ResNet as described in Section \ref{sec:noise_preprocessing}.
%The given normalization process gets the 95th percentile pixel value via the getspercentile function, divides each pixel in the image by the 95th percentile
%pixel, then linearly maps each pixel to a value in [1,~255]. 
Normalization typically maps
values to [0,~1]; the range [1,~255] is used for convenience to visualize images in
grayscale. 0 is not included in the range to avoid accidental division by 0 errors or
exploding log values when performing experiments.\footnote{Note that exploring different options of image normalization in preprocessing is often an important factor for the success of a ML algorithm in a wide variety of applications.}

With image normalization on each image performed before training, the model tends to overfit and
focus especially on bright pixels in the image, even though the brightest pixel in the image
does not always represent the ground truth. To address this, after normalization, we apply the
arcsinh function to each pixel to represent each pixel in log space. Upon applying arcsinh,
the image is normalized once again for image visualization purposes. \edr{Altogether, this preprocessing procedure is represented by the following pseudocode:
\begin{itemize}
    \item image = normalize(arcsinh(normalize(image))
\end{itemize}}
% The final data preprocessing procedure is given as follows:

% \newpage

% \begin{lstlisting}[caption={Normalizaton for U-Net Training},label=unet-norm]
% function preprocess(image):
%     image = normalize(arcsinh(normalize(image))
%     return image
% \end{lstlisting}

% \begin{python}[caption={Normalizaton for U-Net Training},label=unet-norm]
% def preprocess(image):
%     image = normalize(np.arcsinh(normalize(image)))
%     return image
% \end{python}

\edr{This data preprocessing technique was a crucial step in training a successful model for conventional lenses (the results for this project will be reported elsewhere). We continued this preprocessing technique for the lens detection in this project as well.}

\subsubsection{ResNet and U-Net Pipeline} \label{sec:detection_pipeline}

To build a pipeline that can process thousands to millions of images and select candidate lenses with
small Einstein radii, we utilize a combination of the ResNet and U-Net model. We follow a
series of steps, named the ResNet and U-Net (RUN) pipeline, to achieve this goal, we:

\begin{enumerate}
    \item Create a training and validation set using the same 80-20 split for both models. Train both models on the same training set.
    \item Decide on a desired level of purity, and set the probability threshold for the ResNet correspondingly.
    \item Use the chosen threshold, evaluate the ResNet on the validation set.
    \item Of the images ResNet predicts to be a lens, evaluate the U-Net. As mentioned in Section \ref{sec:model_purpose}, the predicted location of the lens for an image is the pixel with the highest probability. The detection outputs from the U-Net represent the final output of the pipeline.
\end{enumerate}

\FloatBarrier

\subsection{Results: U-Net Model}\label{sec:results_unet}

\subsubsection{Initial Training on Simulations Without Noise} \label{sec:initial_training}

% Even though we can use the ResNet to detect small Einstein radii lenses ($0.02''<\theta_E<0.15''$) in the simulated
% dataset, it is quite difficult, even with human inspection, to determine the true location of
% the lens. 
% As such, we decided to train a U-Net model that could learn the patterns of the shape of the tiny lenses as well as detect its location.

\begin{figure}
    \centering
         \includegraphics[scale=0.52]{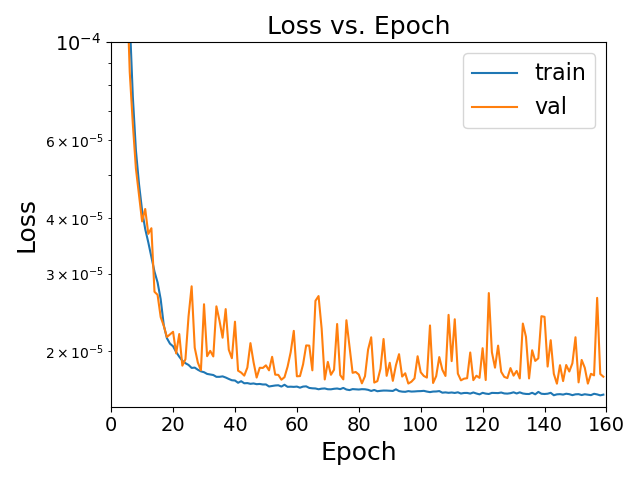}
         \includegraphics[scale=0.52]{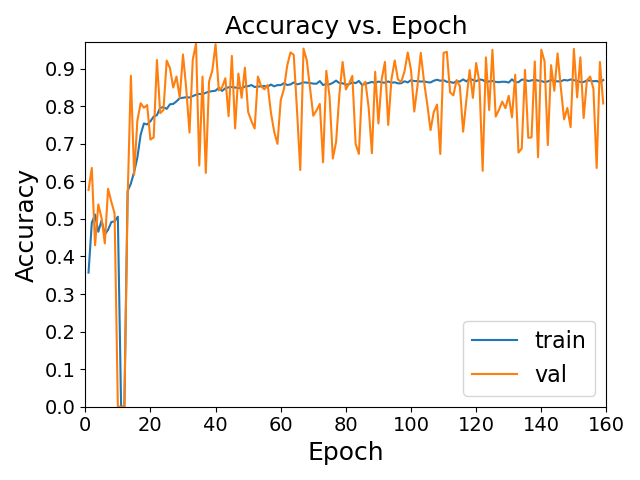}
     \caption{Loss and accuracy vs. training epochs. The loss curve drastically drops in the first 25 epochs. 
     It appears it may still be gradually decreasing afterward, 
     % However, after 25 epochs the model is still learning, 
     as the accuracy continues to slowly improve. 
     In our experiments, beyond approximately the 100th Epoch, 
     % the model seems to be unable to learn anything meaningful anymore, as 
     the validation accuracy tends to plateau, even when trained up to 400 Epochs. 
     Due to the nature of the detection problem, where the model predicts only one pixel in the entire image, the validation accuracy is quite volatile compared to the training accuracy.}
     \label{fig:training and loss curve}
\end{figure}

There is only one class of interest in this problem: small Einstein radius systems caused by low-mass lenses, so we use a standard binary cross-entropy
loss. Upon tuning hyperparameters, we settle on a learning rate of $1 \cdot 10^{-4}$, batch size of 16, and a weight
decay of $10^{-5}$\footnote{See, e.g., \url{https://pytorch.org/docs/stable/generated/torch.optim.Adam.html}} with the Adam optimizer. The model tends to converge at approximately 100
epochs. For validation, we use an 80:20 split (16000 images for training, 
4000
images for validation). Figure~\ref{fig:training and loss curve} displays the loss and accuracy vs.\ training epoch curves.

Directly training the U-Net on the simulation data shows promising results, with a detection
accuracy of up to approximately $95\%$. Training on both lenses and non-lenses achieves similar performance to training just on lenses in the dataset. As a result, we choose to train on lenses and non-lenses, as an additional probability prediction from the U-Net would potentially help filter out non-lenses that the ResNet does not recognize, aiding the ResNet in a lens search.
To standardize how we evaluate the model's performance, we consider a
lens to be correctly detected when the Euclidean distance between the ground truth location of the lens 
and the predicted location is at most $\sqrt{2}$ pixels, and incorrectly detected otherwise.
In other words, either the predicted pixel or the 8 neighboring pixels surrounding the prediction
must be the ground truth for the lens to be considered correctly detected.
We use this slightly more generous definition in order to identify as many small Einstein radius lenses as possible
% Though we use this somewhat relaxed definition of a correct detection, 
(though in practice, nearly all of the U-Net's correctly predicted lens locations exactly coincide with the ground truth). 
%However,since our objective is 
% , we want to include all cases where the U-Net correctly determines the 
% critical 
% location of the lens.
% the image for human inspection or follow-up observation.

When the model is trained on both lenses and non-lenses, the model's detection accuracy is no more than approximately 2-3\% worse than as training on only the 10000 lenses. 
However,
its \emph{classification} performance, while respectable, is still slightly worse than the ResNet's performance.

\subsubsection{Training on Simulations with Noise}

\begin{figure}
  \begin{center}
    $\includegraphics[scale=0.6]{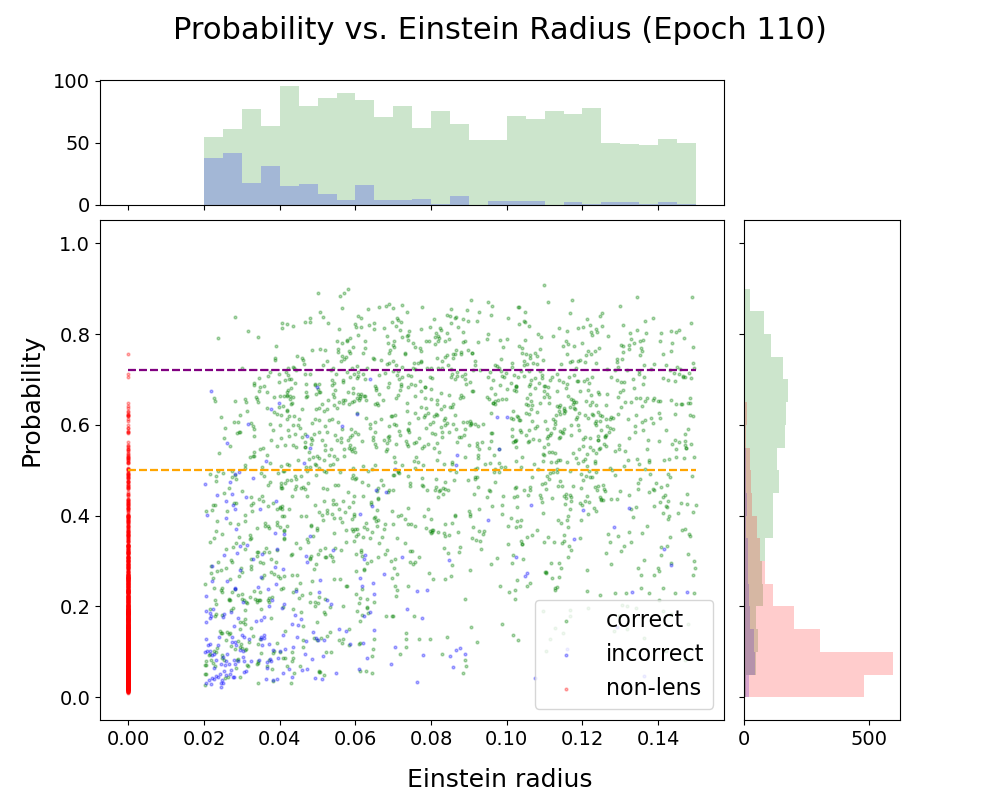}$
    \caption{Detection Probability vs. Einstein radius at U-Net training epoch 110, for the Einstein radius range of $0.02'' < \tE < 0.15''$ (as shown on the $x$-axis) and all halo mass values. 
    For an image with a lens, the pixel with the maximum probability represents the U-Net's prediction, with correct and incorrect detections (green and blue dots respectively) defined in \S\ref{sec:initial_training}. Non-lenses do not have an Einstein radius, but are displayed with an Einstein radius of 0 in this figure (red points).
    For approximately $25\%$ (1072 lenses) of the predictions with a probability over 0.5 in the validation set (orange dashed line), 
    the model achieves a 94.2\% pixel-level precision. 
    At a higher threshold, for 7.4\% (296 lenses) of the predictions with a probability over 0.72 (purple dashed line), the model achieves a 99.7\% pixel-level precision (only one false positive: the single red dot above the purpose dashed line). } 
    \label{fig: einstein radii}
  \end{center}
\end{figure}

After adding noise to the dataset using the same methods as training on the ResNet (see Section \ref{sec:noise_preprocessing}),
the model performance is still very strong. 
Figure \ref{fig: einstein radii} shows that
detection accuracy, while still decent, drops when the Einstein radius is very low ($0.02''<\theta_E<0.05''$). 
For all other ranges of
Einstein radii, however, the detection accuracy has very strong performance. Overall, as the
detection probability threshold is raised, the set of images the U-Net deems to be lenses becomes very pure. For
approximately $25\%$ (1072 lenses) of the predictions with a detection probability over 0.5 in the
validation set, the U-Net achieves a 94.2\% detection accuracy. Likewise, for about 7.4\%
(296 lenses) of the predictions with a detection probability over 0.72, the U-Net achieves a
99.7\% detection accuracy, with only one false positive.

\edr{Regarding high probability non-lenses and low probability $\tE > 0.08''$ (i.e. the top left and lower right corners of Figure \ref{fig: einstein radii}), we have not performed a systematic study on what properties of the image correlate with either non-lenses receiving high probability or $ \tE > 0.08''$ lenses receiving low probability. As mentioned before, interpretability is a well-known challenging aspect in machine learning.  In our experience, a bright environmental galaxy in the vicinity of the lens can “confuse” the U-Net and result in a low probability prediction. High probability non-lenses typically have local maximas, usually environmental galaxies, that are difficult to distinguish from a true positive.}

\begin{figure}
  \begin{center}
    $\includegraphics[scale=0.8]{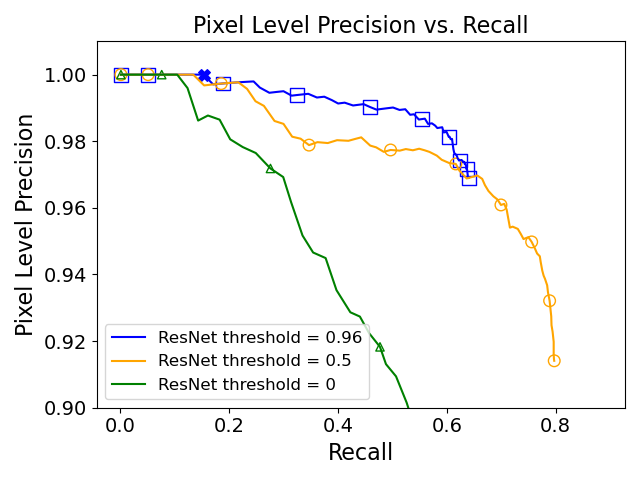}$
    \caption{Pixel Level Precision vs. Recall curve for various ResNet thresholds chosen at Step 2 of the ResNet and U-Net pipeline, for the Einstein radius range of $0.02'' < \tE < 0.15''$ and \emph{all halo mass values}. Colored shapes represent U-Net probability thresholds that are multiples of 0.1 (leftmost is 0.9), corresponding to the curve of the same color. Across all U-Net probability thresholds, the combination of the ResNet and U-Net (ResNet threshold = 0.5 or 0.96) is a strict improvement over the U-Net alone (ResNet threshold = 0) in pixel-level precision. From Section \ref{sec:purity}, a high classification threshold of 0.96 for the ResNet provides a precision of 1.0 (no false positives) on the testing set with no overlapping galaxies. 
    %the modified validation set with no overlapping galaxies. 
    Evaluating the U-Net at a detection threshold of 0.5 on this 
    % completely pure set (see text) 
    set yields a pixel-level precision of 99.0\% (sixth blue square from the right). To reach a pixel-level precision of 1.0 (no false positives) for this set, the U-Net requires a detection threshold of 0.72 (blue X-mark), which achieves a recall of 15.3\% (306 lenses). Note that the rightmost orange circle represents a U-Net probability threshold of 0.0. At this threshold, the pixel with the maximum probability is the U-Net's predicted location of the lens, regardless of what that probability is, and the pixel with the maximum probability can be arbitrarily small.}
    \label{fig:PRC}
  \end{center}
\end{figure}

Figure \ref{fig:PRC} displays the Precision-Recall curve (PRC) comparing the performance of the RUN
pipeline at various thresholds chosen at Step 2. We define pixel-level precision as the
proportion of lenses that are correctly classified \emph{and} detected out of the
number of lens predictions at or above the level of a specific probability threshold. 
When comparing the performance between the U-Net alone (ResNet thresold = 0) and the combination of the ResNet and the U-Net (ResNet threshold = 0.5 or 0.96), the ResNet strictly
improves the classification accuracy of the U-Net on the validation set, shown in Figure
\ref{fig:PRC}---\ed{the PRC sits higher for a higher ResNet threshold}. Across all U-Net probability thresholds, the combination of the ResNet and U-Net always
has an equal or higher pixel-level precision than the U-Net alone. As mentioned in Section \ref{sec:purity}, the ResNet has a precision of 1.0 (nominally, no false positives, although given the size of the  validation set, Poisson noise could still permit false positives, as discussed in \ref{sec:purity}) on the validation set at a high probability threshold of 0.96. 
Among this completely pure set of images, 
the U-Net achieves a very high detection precision of 99.0\% at a traditional threshold of 0.5 (sixth blue square from the right in Figure \ref{fig:PRC}). To reach a pixel-level precision of 1.0 (no false positives) on this set, the detection threshold needs to be set to 0.72 for the U-Net (blue X-mark in Figure \ref{fig:PRC}), which achieves a recall of 15.3\% (306 lenses).

\subsubsection{Forecasts of detectable numbers of low-halo-mass lenses} \label{sec:forecast_unet}
\begin{figure}
    \centering
    \includegraphics[width=.7\textwidth]{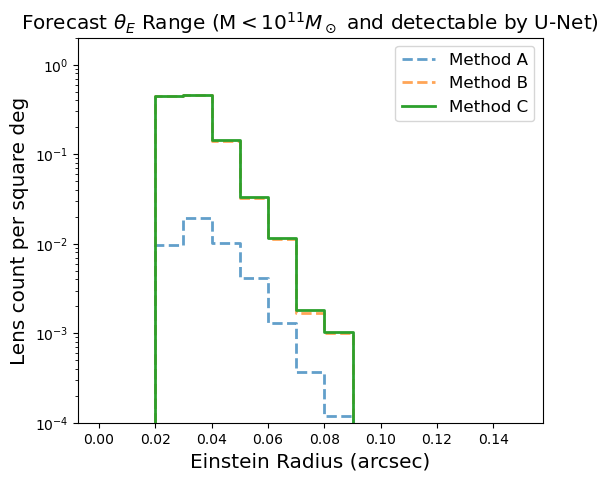}
    \caption{We show the forecast \ed{numbers for lenses with $M<10^{11}M_\odot$ that are correctly classified by the ResNet Model~3 \edr{(JWST-small)} at a 0.96 threshold (with zero false positives; see bottom right panel in Figure \ref{fig:forecast_detectable_lenses}) and are then correctly detected by the U-Net at a 0.72 threshold (again with zero false positives)}. Note that the forecast numbers from Method~A contribute minimally to this category, as expected.}
    \label{fig:unet_forecast}
\end{figure}

\begin{deluxetable}{cccc}
\tablehead{\colhead{$\theta_E$ range} & \colhead{ Forecast (lenses/deg$^2)$} & \colhead{Detectable lenses/deg$^2$} & \colhead{Detectable lenses (pixel level)/ deg$^2$} \\ \colhead{} & \colhead{} & \colhead{with $M<10^{11}M_\odot$ (ResNet)} & \colhead{with $M<10^{11}M_\odot$ (ResNet $+$ U-Net)}}
    \startdata
        $0.02''<\theta_E<0.05''$ & 490 & \numlomassresnet & \numlomassunet \\
        % $0.05''<\theta_E<0.10''$ & 230 & 0.38  & 0.099\\
        % $0.10''<\theta_E<0.15''$ & 150 & 0.0 & 0.0\\
    \enddata
    \caption{
    Forecasts for the number of lenses with $0.02'' < \tE < 0.05''$ per square degree, and how many of those with $M_\mathrm{halo}<10^{11}M_\odot$ are detectable (by the ResNet Model 3 \edr{(JWST-small)} at a 0.96 threshold and the U-Net model at a 0.72 threshold), \ed{with zero false positives in classification or lens location detection}. Results are given to 2 significant figures.}
    \label{tab:forecast_detectable_lenses}
\end{deluxetable}

\ed{With the results of the U-Net model, we can continue our results from Section \ref{sec:forecasts_detectable}. We want to know the number of low-halo-mass lenses that can be both classified by the ResNet model and detected by the U-Net model with high pixel-level precision.}

\ed{We take the forecasts shown in Figure \ref{fig:forecast_detectable_lenses}, 
% of the forecasts 
for Model 3 \edr{(JWST-small)} at a 0.96 threshold and multiply them by the recall of the U-Net in each Einstein radius bin at a 0.72 threshold, which results in no incorrect detections on the validation set (see Figure \ref{fig:PRC}). 
This results in detection forecasts for different Einstein radii shown in Figure~\ref{fig:unet_forecast}.}

% (also with zero false positives at the limit of the validation set).}

\begin{figure}
    \centering
    \includegraphics[scale=0.9]{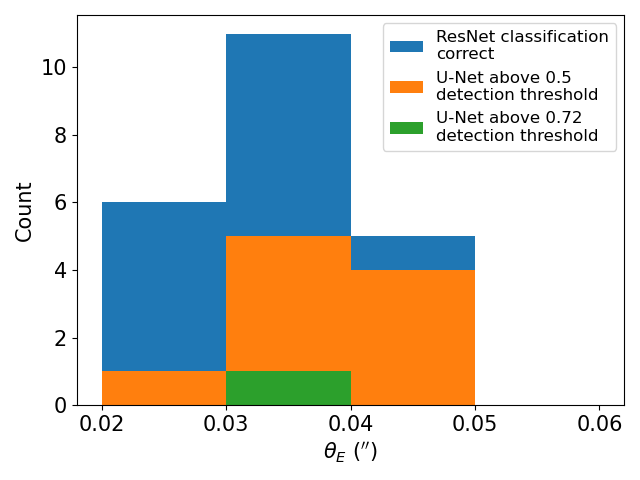}
    \caption{\ed{The blue histogram shows small Einstein radius lenses, with $M_\mathrm{halo} < 10^{11} M_\odot$ and $\theta_E<0.05''$, correctly classified by the ResNet (with threshold 0.96, see \S~\ref{sec:purity}). The orange histogram shows the number of lenses with a U-Net detection threshold above 0.5, and the green histogram shows the number of lenses with a detection threshold above 0.72. For both thresholds, every image has a correct lens detection. In fact, for $M_{halo} < 10^{11}M_\odot$ and $\theta_E<0.05''$, with either thresholds, the pixel-level precision remains at 100\%, albeit with low statistics. Note that in this figure, we simply show counts in  the vertical axis (and not in terms of per square degree).}}
    \label{fig:hist_detection_acc}
\end{figure}

\ed{
We therefore forecast about \numlomassunet low-halo-mass lenses per square degree detectable by the RUN pipeline\footnote{Note that we did not count the 0.099 systems from the \tE bin of $0.05''\text{-}0.10''$ in the last column of Table~\ref{tab:forecast_detectable_lenses}. 
As stated earlier (\S\,\ref{sec:small_thetaE_mass}), based on Figure~\ref{fig:forecast_masses}, we focus on the smallest \tE bin of $0.02''\text{-}0.05''$.}
with zero false positives in classification and location detection (Table~\ref{tab:forecast_detectable_lenses}, in which, for completeness, we show the ResNet results from Table~\ref{tab:forecast_detectable_lenses_no_unet} again).
% ---these are the numbers of lenses per deg$^2$ that are discoverable by the ResNet (with zero false positives at the limit of the validation set size) and their locations detectable by the U-Net, again with zero false positives).
% even with our most conservative, precise standard. 
This is down from the 
% 35 
\numlomassresnet
lenses/deg$^2$ detectable by the ResNet, or a factor 15 smaller, perhaps not surprisingly, as this is a very challenging task.
\emph{Nevertheless, this is an astonishingly large number of discoverable low-halo-mass lenses!}
As seen in Figure \ref{fig:PRC}, we can increase this number by using a lower U-Net threshold with a higher recall, and still reach high pixel-level precision. For example, with a U-Net threshold of 0.5, a pixel-level precision of 99.0\% is achieved, with a recall of $\sim46\%$. Therefore, at the cost of only $\sim 1\%$ reduction in precision (down from 100\%)
% 98.8\% pixel-level precision, 
%we calculate a prediction of
our forecast is about \numlomassunetninenine low-halo-mass halos per square degree detectable by the RUN pipeline, an increase by a factor of 7. In fact, for halo mass range $M_\mathrm{halo} < 10^{11}M_\odot$, with either threshold, the pixel-level precision is 100\% (Figure ~\ref{fig:hist_detection_acc}), albeit with low statistics.\footnote{
Given what is shown in Figure~\ref{fig:hist_detection_acc}, one may be tempted to claim 100\% precision at a U-Net threshold of 0.5 (or even lower), if it is possible to tell systems with lens halo mass $ > 10^{11} M_\odot$ from imaging. 
Figure~\ref{fig:small_te_examples} suggests this may be feasible, but even so it is just one example.
% , and one cannot be sure of being able to do so consistently. 
The results in Figure~\ref{fig:hist_detection_acc} are indeed encouraging. However, in our current validation  set, the numbers are quite small for the mass range of $M_\mathrm{halo} < 10^{11} M_\odot$, and therefore one needs to be careful to not overly generalize the results for this category. 
}}
% we leave a systematic study of this possibility to future work.

\edr{Exactly what is required for full confirmation of any of these observed lens candidates will take exploration, this being a new regime of lensing observation. 
Therefore, it is difficult to estimate the rates of confirmation at this point.
As with conventional lenses, high resolution imaging and spectroscopic redshift are necessary for definitive confirmation.
 % The starting point can be measurements of spectroscopic redshift,
% along with good-quality, high-resolution images of systems with 
We will first target candidates with the highest probabilities from the RUN pipeline and the most favorable configurations. For example, for systems where both the lens and source have a simple, smooth profile, the distortion due to lensing would be easier to recognize.
% it will be difficult to confirm lensing of very complicated sources compared to sources with  
Deep \hst and perhaps especially \jwst imaging\footnote{\edr{In some cases, very high-resolution observations from radio observatories such as ALMA may be possible. In addition, in the near future, \emph{Roman} can reach $\sim 0.05''$ in the bluest band, and the ELT, with a resolution of $0.005''$, using adaptive optics, is expected to have its first light in 2029.}} will likely be necessary.
Observations in multiple bands are particularly useful because the putative lensed source of such a candidate system may be much more prominent in one band, making it possible to clearly see the distortion due to lensing without the interference of the lens light, while the putative lens may be much more prominent in another band.
% \citep[e.g.,][ Figure~]{storfer2025a} 
% Of course, such imaging confirmation would be highly desirable, even if we would only be dealing with numbers in the single digits. 
Spectroscopic confirmation (e.g., using \jwst) is expensive, so we will first 
target candidates that are deemed very likely based on multi-band, high resolution deep imaging, with the brightest putative lenses.
That there is a real possibility that these systems can be unambiguously confirmed is an exciting prospect and sets them apart from the ``dark'' (sub)halos \citep[e.g.,][]{vegetti2012a, Sengul_2022}.} 

% In the abstract we stated that for a U-Net threshold of 0.5, the detection purity is 99\%. Now actually, for halo mass $< 10^{11}$, which is the range of interest for this paper, we can lower the threshold further for the U-Net (from 0.5 to 0.31) and still maintain 100\% purity. 

\subsection{Discussion: U-Net Model} \label{sec:pipeline_performance}

% Our validation
% set from an 80-20 split has only 2000 lenses, so the size of the dataset limits how many
% significant figures we can achieve in the model's pixel level precision. For example, using the same thresholds of 0.96 for the ResNet and 0.72 for the U-Net from Section \ref{sec:forecast_unet} yields a recall of roughly 15\%, or 300 lenses, so even with one
% incorrect classification or detection from either model would result in a pixel level precision of 99.7\%. However, following calculations from \ed{
% \S~\ref{sec:purity}}, at a probability threshold where the ResNet has a precision of 1, we expect a worst case ratio of 3:1 false positives to true positives in observation. Although we cannot expect a completely pure set of lens candidates from the ResNet and U-Net pipeline, a 3:1 false positive to true positive ratio is quite reasonable for human inspection or follow-up observation (if visual inspection is also inconclusive), for such valuable discoveries.
% especially with the aid of the U-Net’s prediction

% Since we are especially interested in low-mass lenses, we specifically examine the U-Net's results on small Einstein radius, low halo mass lenses in Figure \ref{fig:figure10}. 
%Although the U-Net's performance on small Einstein radii lenses is generally poorer, 
% \ed{Setting a reasonable probability threshold (specifically, 0.5 or 0.72) 
% allows the U-Net model to achieve a very high pixel level precision.}
% We believe that 
\ed{In this section (\S~\ref{sec:unet}), our goal is to detect the locations of low Einstein radius lenses ($0.02''$-$0.15''$) in a cutout image.
Here we present the RUN (ResNet and U-Net) pipeline.
Conditioned on a probability threshold of 0.96 from the ResNet, 
the performance of the U-Net is strong: above the threshold of 0.72, there are no false positives (i.e. no incorrect pixel detections), \emph{including lenses in the smallest \tE bin 
($0.02''$-$0.05''$) and in the lowest halo mass range available from CosmoDC ($<10^{11} M_\odot$).}}
\ed{It is possible that 
a U-Net model that is specially trained to detect lenses in the smallest Einstein radius bin and with lowest masses may lead to even better performance on these smallest lenses.
We leave this investigation to future work.}
% We specifically highlight the U-Net's strong performance on on small Einstein radius, low halo mass lenses.}

\begin{figure}
    \centering
    \includegraphics[scale=0.37]{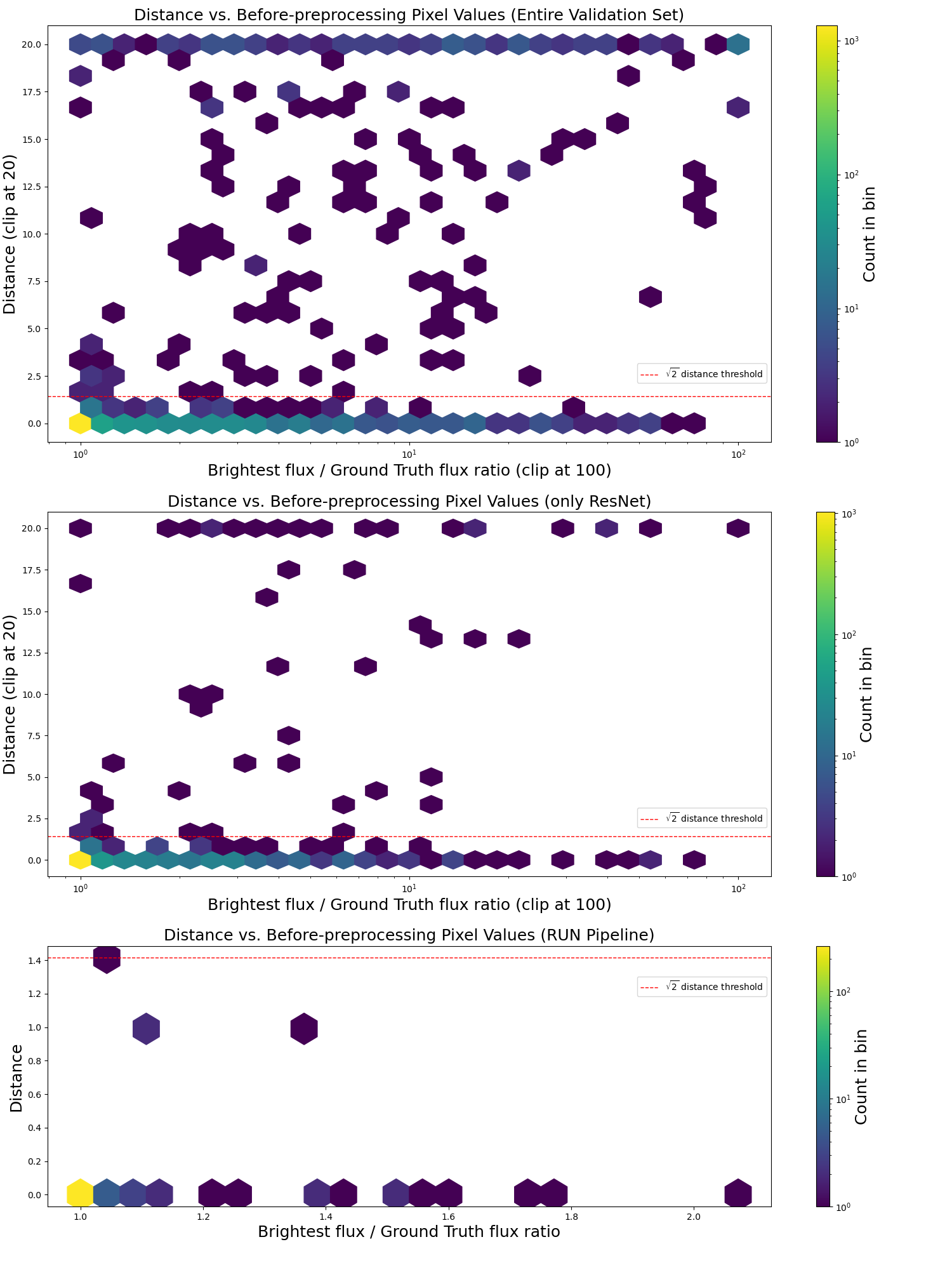}
    \caption{\edr{Hexbin plot of the distance between the ground truth and U-Net prediction versus the brightest pixel to ground truth flux ratio. The top panel (entire validation set) includes predictions on all images in the validation set, the middle panel (only ResNet) includes predictions on all images that have a ResNet classification threshold above 0.96, and the bottom panel (RUN pipeline) includes predictions on images that have a ResNet classification threshold above 0.96 and U-Net detection threshold above 0.72. At all stages of the RUN pipeline, the U-Net correctly detects (below the red dotted line of 1.4~pixels) numerous lenses that have a brightest pixel to ground truth flux ratio greater than 1. Even after applying the U-Net detection threshold of 0.72 (transitioning from middle to bottom panel), there are still several lenses with such a flux ratio being greater than 1, all of which the U-Net detects correctly. This demonstrates that the U-Net does not just predict the pixel with a local maximum brightness near the center of the image, but rather shows that it appears to ``understand'' the features of a lens.}}
    \label{fig:preprocess_ratios}
\end{figure}

\edr{In Figure \ref{fig:preprocess_ratios}, we show a hexbin plot of the distance between the ground truth and U-Net prediction versus the brightest pixel to ground truth ratio. At all stages of the RUN pipeline, the U-Net correctly detects numerous lenses that have a brightest pixel to ground truth flux ratio greater than 1. Even after applying the U-Net detection threshold of 0.72 (transitioning from middle to bottom plot), there are still several lenses with flux ratios greater than 1, all of which the U-Net detects correctly. This demonstrates that the U-Net does not just predict the pixel with a local maximum brightness near the center of the image, but rather shows that the U-Net seems to ``understand" the features of a lens.}

\begin{figure} 
    \centering
    \includegraphics[height=.8\textheight]{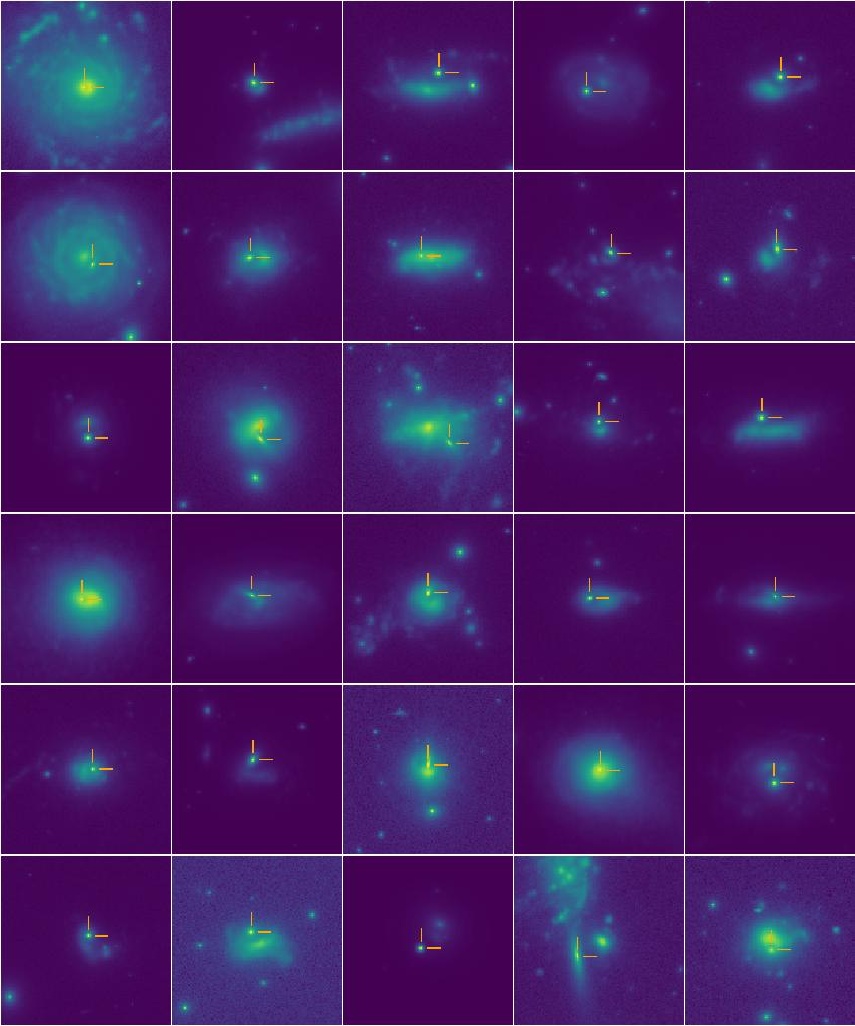} 
    \caption{Examples of lenses that are correctly classified and detected by the ResNet and U-Net models, respectively. All images have a ResNet classification probability above 0.96 and U-Net detection probability above 0.72, where the RUN pipeline achieves a pixel level precision of 1.0 (see Figure \ref{fig:PRC}). Orange cross hairs represent the location of the correctly detected lens. Out of these 30 examples, there are numerous complex lens systems in which the U-Net identifies the lens correctly, supporting the fact that the U-Net is not just predicting a local maximum near the source.}
    \label{fig:panel_tpc}
\end{figure}

\begin{figure}
    \centering
    \includegraphics[scale=0.45]{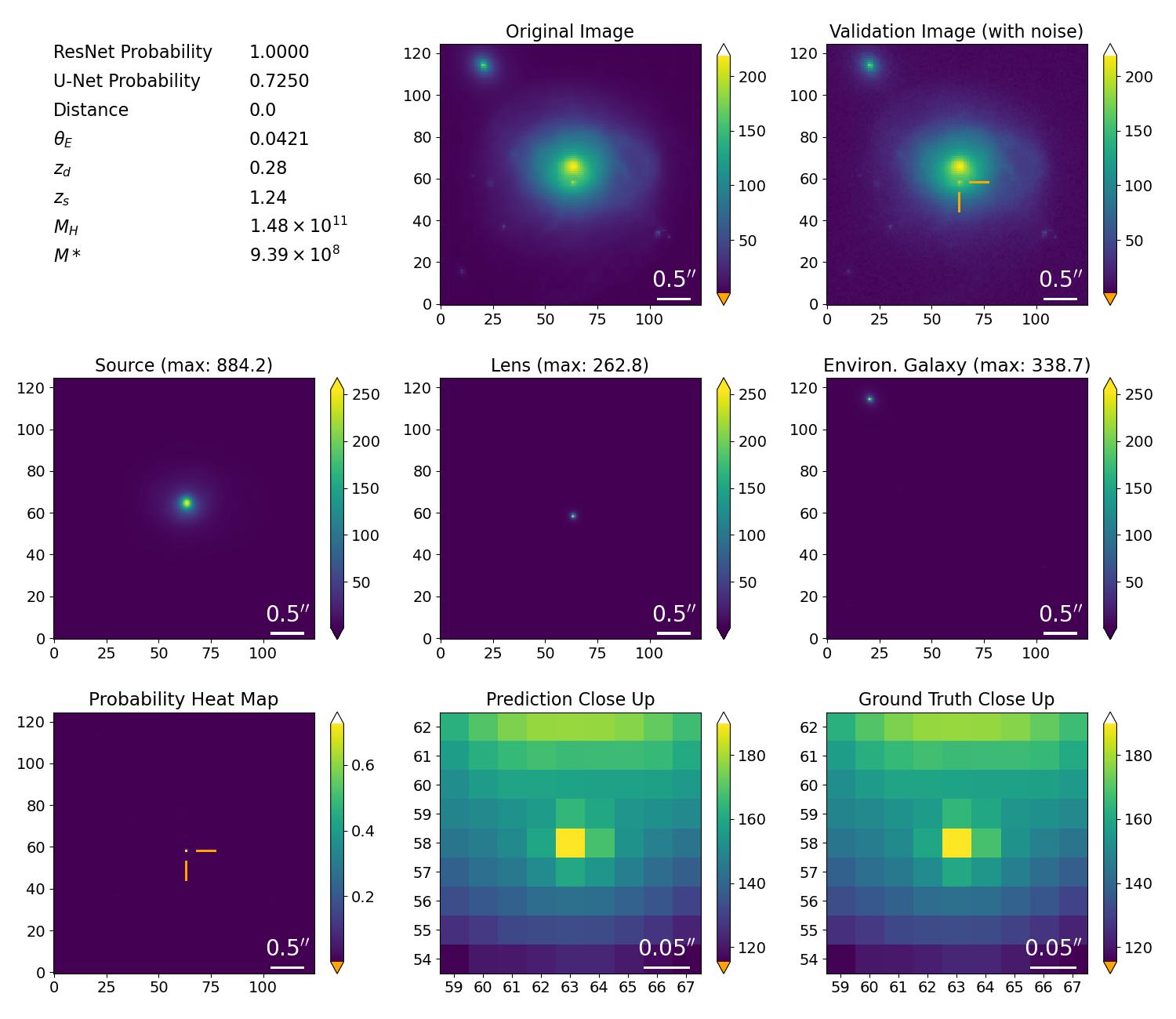}
    \caption{A relatively straightforward image for lens detection by the U-Net. 
    % There is a bright local maximum at the location of the lens. 
    \ed{The quantity ``Distance" at the top left of the figure indicates the separation between the predicted lens location and the true lens location in pixels.}
    The probability heat map (lower left) shows that the U-Net only focuses on one single pixel at (58, 63)---for the horizontal and vertical directions, respectively---giving a high detection probability of 0.725 for that pixel and much lower probabilities elsewhere. 
    \ed{The zoom-in panels (bottom row, middle and right) for the prediction and ground truth are identical because the prediction and the ground truth coincide.}}
    \label{fig:ex1_easy}
\end{figure}

\begin{figure}
    \centering
    \includegraphics[scale=0.45]{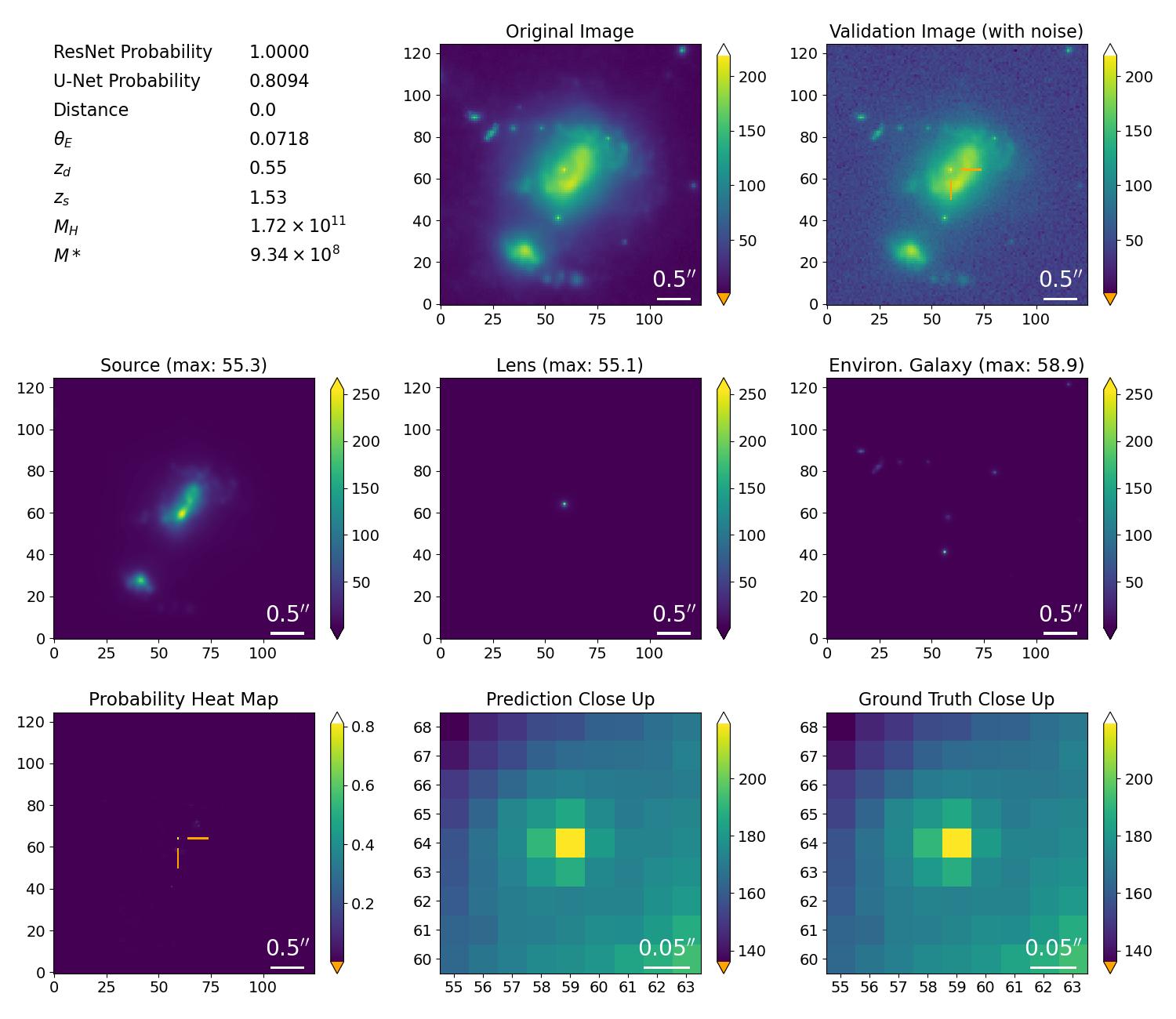}
    \caption{A challenging lens that the U-Net detects correctly. 
    % Although the U-Net's detection is correct, the model gives a low probability level. 
    There are many objects around the lens that make it difficult to detect the correct location of the lens.
    The U-Net nevertheless correctly predicts the lens location.
    % , as there are more candidates to choose from. 
    This challenging example provides promising evidence that the model is not simply predicting the brightest pixel: it clearly singles out the lens, compared with the many local maxima, as is clear from the heat map. }
    %The pixel with the second highest detection probability from the U-Net prediction, at (76, 42), has a detection probability of 0.18, significantly lower than the highest detection probability.} %``learned" to find the lens among many local probability maxima.
    \label{fig:ex2_hard}
\end{figure}

\begin{figure}
    \centering
    \includegraphics[scale=0.45]{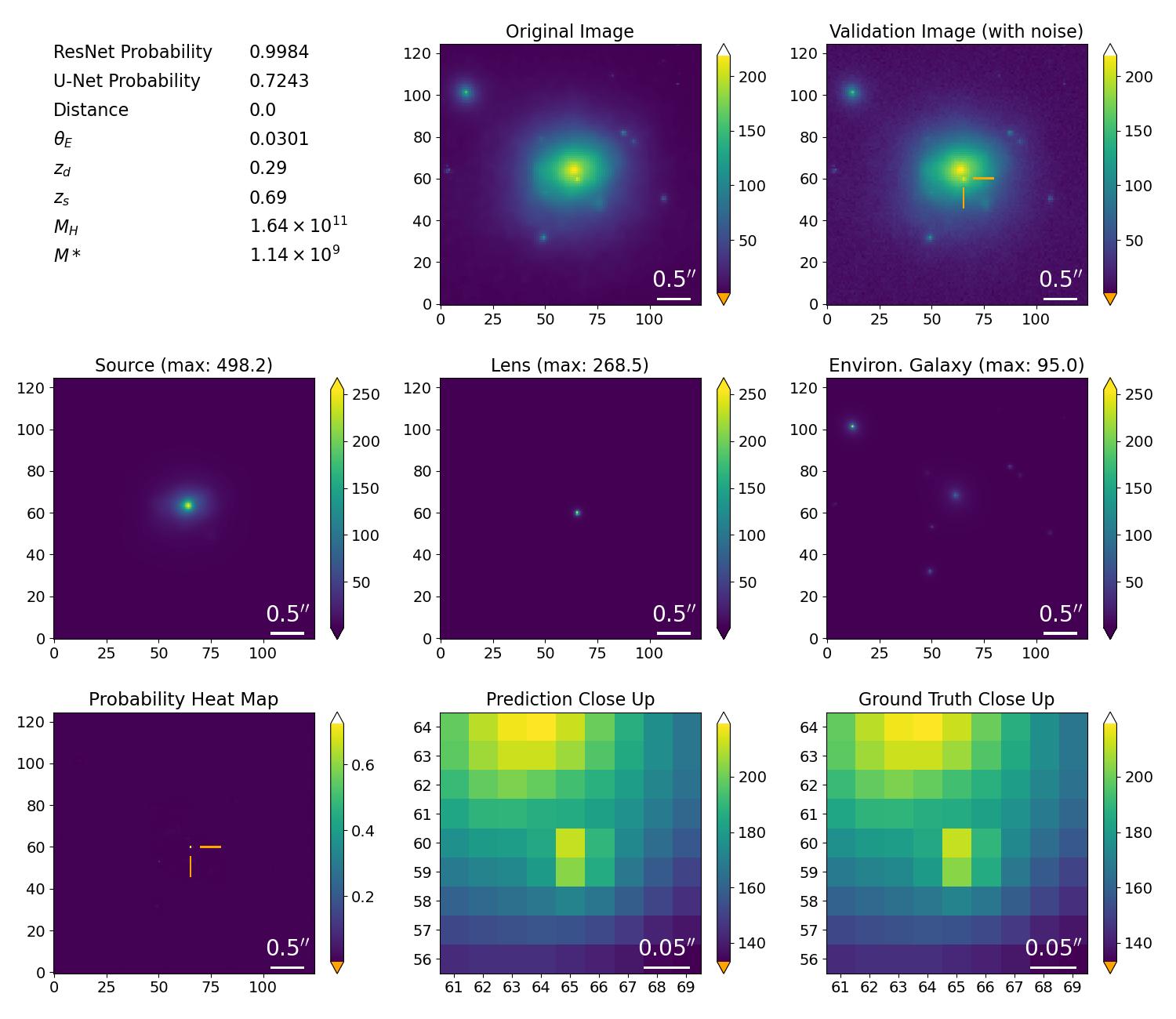}
    \caption{A lens with an Einstein radius of $\theta_E = 0.03''$. Even though the lens is not the brightest pixel in the image and has a very small Einstein radius, the U-Net is able to confidently detect the lens with a 0.7243 detection probability.}
    \label{fig:ex2_hard_}
\end{figure}

\begin{figure}
    \centering
    \includegraphics[scale=0.45]{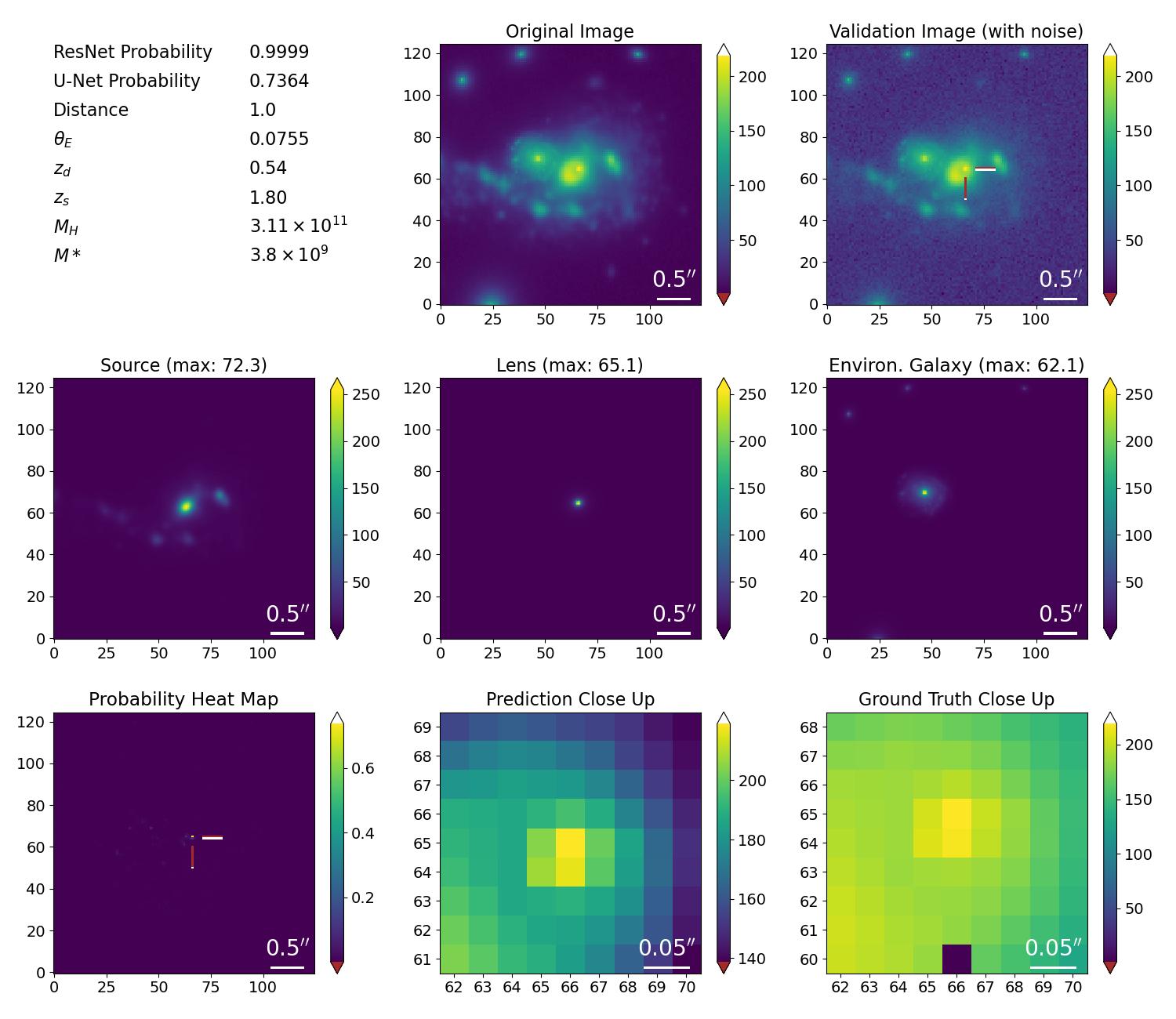}
    \caption{Even though the prediction (red cross hairs, barely visible) is one pixel away from the ground truth (white cross hairs), the detection is considered as correct (see Section \ref{sec:initial_training}), as the model would still identify the relevant location in the image for human inspection and follow-up observations.
    % in a lens search. 
    This situation occurs in 2.25\% (45) of the lenses in the validation set.}
    \label{fig:ex5_close_right}
\end{figure}

In Figure \ref{fig:panel_tpc}, we show many examples of correct detections on lenses in the validation set by the RUN pipeline. Orange cross hairs represent the location of the correctly detected lens.  
In particular, we show four examples in details (Figures \ref{fig:ex1_easy}-\ref{fig:ex5_close_right}).
Typically, in any given image there are at least several local pixel value maxima (including the global maximum). 
For lens images, our trained U-Net, when the threshold is set to be $> 0.72$, will pick the local maximum that corresponds to the correct location of the lens, by that we mean up to 1 pixel off for either or both coordinates (see \S\,\ref{sec:initial_training}).
For $\sim 2\%$ of the cases, the detection is one pixel off---when that is the case, the detection location is at the local maximum pixel. 
For the remaining 98\% of the cases, the detected location is at exactly the right pixel.
\ed{That is, for the U-Net, being a local maximum is a necessary condition for locating the lens to within 1 pixel, but not sufficient (in agreement with the expectation based on the ground truth in the simulations).
%\footnote{ We also perform an additional test to
%confirm that the U-Net is not just predicting the brightest local maximum near the center of the lens in each image, without ``understanding" what a small Einstein radius lens is. The results are presented in 
% In the Appendix Section 
%Appendix \ref{sec:lensing_effect_tests}.}}
For non-lens images, the predicted probabilities for all pixel are below (typically well below) this threshold.}

% We perform additional tests which provide evidence that the U-Net can identify features of a lens in Appendix~\ref{sec:lens_effect_tests}.
% When the detection threshold is $< 0.72$, some of the U-Net lens detection are incorrect. To provide a fuller picture of the U-Net's capabilities, we also show three examples of such
% % as well as more in-depth examples of correct and 
% incorrect detections in the Appendix Sections  and \ref{sec:individual_examples}, respectively.

Finally, for the completeness of this discussion, in Appendix Section \ref{sec:individual_examples}, we show a few examples of false positives \emph{below} our chosen threshold of 0.72, for illustration.

\clearpage
\section{Conclusions}

\edr{We have presented results on extending the current strong lens discovery space through the
combination of \jwst observations and machine learning (ML) techniques.
In this work, we define three categories of lenses: ``conventional-sized'' or``large'': $\tE > 0.5''$, ``intermediate'': $0.5''>\tE > 0.15''$, and ``small'': $0.15'' > \tE > 0.02''$.
We have forecast the number of lenses that could be observed by \jwst, including a new class of very small \tE lenses.
We have simulated very realistic strong lenses and used them to train residual neural network (ResNet) models to detect strong lenses.
Furthermore, we have trained a U-Net model to pinpoint the locations of the ``small'' lenses in images.
Combined in the RUN (ResNet and U-Net) pipeline, we can detect and pinpoint very small \tE lenses with very high precision.
Finally, we have performed many tests of the robustness of our results, which underscore profound implications from our results in strong lens discovery.
We summarize our results below:
\begin{itemize}
    \item Our \jwst forecast, combining the conventional strong lenses with very small $\theta_E$ lenses, estimates $29$ million lenses with $\theta_E>0.02''$ on the whole sky (690 per square degree), with $5.5$ million (130 per square degree) in the ``conventional-sized'' category. 
    These include $940,000$
    double source lenses (23 per square degree) due to the large source density with \jwst, nearly all of which have $\tE > 0.15''$ (see Table~\ref{tab:forecasts}).
    \item We have simulated the most realistic strong lenses to-date with  the lenses and sources drawn from the CosmoDC2 and the JAGUAR simulated catalog for \jwst, respectively, using the VELA simulations for realistic sources.
    \item We have trained residual neural network (ResNet) models for each of the three ranges of Einstein radii. We have also trained a U-Net model for the ``small'' \tE range.
    \item For large (conventional-sized) Einstein radii systems, by doing a small-scale deployment on real single-band \hst data, we demonstrate that, with space telescope resolution, our ResNet recommendations have near-100\% completeness and near-100\% purity in the sense that nearly every ResNet recommendation is a lens candidate.
    \begin{itemize}
    % \item With single band HST data, our ResNet recommendations are as pure as a citizen science identified lens images from a pool of neural network identified candidates” and “Our ResNet recommendations have also included two false negative cases missed in the citizen science identifications.
    \item With the ResNet model, we have discovered two new lenses that a comprehensive citizen science project \citep{Garvin_2022} using only human inspectors missed.
    \end{itemize}
    \item For small Einstein radii systems ($0.02'' < \tE < 0.15''$), the ResNet model can confidently identify nearly every lens that a human can discover. The ResNet model can even identify some of these systems that are impossible for the human eye to find.
    \begin{itemize}
    \item We can set probability thresholds for the small \tE model so that the ResNet false positive rates are low, even when applied to real data
    \item We report that, using the ResNet, \numlomassresnet lenses/deg$^2$ with the smallest Einstein radii ($0.02''$-$0.05''$) and in the lowest halo mass range ($<10^{11} M_{\odot}$) are discoverable with \jwst (see Figure \ref{fig:forecast_detectable_lenses}).
    \item Together with the U-Net in the RUN pipeline, for the purposes of follow-up observations, we can pinpoint the location in a cutout image of such a lens for \numlomassunet/deg$^2$ of the \numlomassresnet/deg$^2$, to within 1.4~pixels with 100\% precision (and \numlomassunetninenine/deg$^2$ with 99.0\% precision).
    \end{itemize}
    \item We have performed a series of tests (in the main text and in the Appendices), with a number of them using out-of-sample images, simulated (Appendix~\ref{sec:vela_source_test}) and real \hst (\S~\ref{sec:first_new_lens}, \S~\ref{sec:new-lenses}), which leads to the conclusion that our results are robust.
\end{itemize}}

Up to now, the main use of neural network models has been to make human inspection feasible, 
by significantly narrowing down the total volume of observations to a small selection to identify lenses. 
For example, \citet{Storfer_2024} showed that for the ground-based DESI Legacy Imaging Surveys data ($grz$ bands), about 1 in 25 ResNet ``recommendations" (image cutouts to which the ResNet model assigned a probability above a threshold) is a lens candidate. 
Here, we have demonstrated that on space-based data, the purity of the ResNet recommendations has so improved that every ResNet recommendation is a lens candidate.
Furthermore, we have discovered two new lens candidates missed by \cite{Garvin_2022}! 
(We believe they are mostly missed due to suboptimal pixel value scaling for human eyes, which the ResNet is not nearly as sensitive to.)
That is, our ResNet model has achieved ``superhuman" performance.
In fact, these preliminary results show 
% we have achieved near-100\% completeness and this will likely hold for applications to larger datasets.
that it is possible for the ResNet to achieve near-100\% purity and completeness, 
% with the ResNet using a reasonable threshold
and human inspection appears to be on the verge of becoming unnecessary.
In follow-up studies, we will further test if this is indeed the case with larger datasets.
% , if this result holds for larger datasets.
This has tremendous implication for lens discoveries using observations from current and future space telescope observations.
The \emph{Nancy Grace Roman Space Telescope} (\rst) will have 100 times the field of view of \hst with identical resolution in the Near-IR. 
In its lifetime, many strong lenses are expected to be discovered \citep[see, e.g., the white paper ``NANCY'' on an all-sky \rst survey with a resolution of 0.11$''$;][]{Han_2023}.
In addition, $\mathcal{O}(10^5)$ strong lenses are expected to be found from \euc observations, for which the VIS channel appears to deliver similar image quality as from \hst and \rst near-IR bands.
We have shown for the first time that a trained ResNet model will likely deliver discoveries truly in an automated fashion, 
with little or no human intervention.

The models for detecting small lenses with $\theta_E$ as low as 0.02$''$ in this work \emph{represents a new domain in strong lens discoveries.}
Many of these systems are very low halo mass lenses, with their lensing effects at current observational limits.
We have achieved a trained model that can successfully find them.
% quite accurately.
Firstly, as with large $\theta_E$ systems, 
nearly every lens that a human can discover, the ResNet can also identify (by assigning a high probability).
% As with large and intermediate \tE systems, 
Secondly, the vast majority of these systems are in fact nearly impossible for the human eye to identify both 
because the area around the lens often is only visible with careful image pixel value scaling and 
because these are very small lensing systems with
barely perceptible lensing effects (see \S\,\ref{sec:model3_val_performanace}).
But the ResNet model can readily identify many of them.
% even identify many lenses that are not possible for the human eye to discover at any pixel value scaling.
We can set acceptable probability thresholds so that the ResNet false positive rates are low \ed{(see \S~\ref{sec:purity})}.
% without exceedingly careful image scaling or modeling.
% Though we are not able to test on real observations, 
% Our validation performance on simulations demonstrates that these lenses represent a new class of strong lensing that is detectable. 
Even with possible reduction in performance level when applied to real data, our ResNet model should be able to find these small $\theta_E$ lenses, as long as an adequate exposure time is used (for $\theta_E\sim0.02''$, we only demonstrate good performance at 10,000~sec).

\ed{
Although our forecast is specific to \jwst, we emphasize that in general, with any current and future space telescopes that can achieve a resolution of $0.03''$ or better, combined with modern ML techniques, could open up a new class of lensing systems with much smaller Einstein radii than what is typically considered a strong lens today. 
For finding lenses using the ResNet, we have mainly focused on \jwst and \hst in this work because we can perform archival searches on their data and because for finding lenses with the smallest \tE and lowest halo mass, \jwst  can achieve the greatest depth and the highest resolution. 
For finding lenses with $\tE \gtrsim 0.15''$, in the near future, \rst has the potential to find many of these systems. \euc could also contribute to these future searches, with a $0.13''$ PSF FWHM for the VIS instrument \citep{Euclid_Mellier_2024}.}\footnote{\ed{In particular, \euc's VIS instrument has a requirement of a PSF FWHM smaller than $0.18''$, but has been measured at $0.13''$ on science exposures \citep[page 31]{Euclid_Mellier_2024}.}}

% Another possibility for future research is expanding on the simulations in this work by adding supernovae, to develop a model to find lensed supernovae. This idea is inspired by \cite{Rubin_2021}, which uses the VELA simulations as host galaxies for simulated supernovae, and a natural extension would be to use the VELA simulations as host galaxies for strongly lensed supernovae.

In recent years, there have been a number of improvements to the U-Net architecture for segmentation and detection tasks.
Given the small training sample, limited by the VELA simulation sample, any gain from a newer (and typically larger) architecture may be marginal.
However, hydrodynamic simulations will soon provide larger samples of realistic simulations.
Given the encouraging results shown here, it would be interesting to explore if newer architectures can improve the performance on larger training samples.

In this work, we have not fully explored the possibility of finding ``dark'' low-mass halos,
i.e., halos with $M_\mathrm{halo} \lesssim 10^9 M_{\odot}$ that either hosts a very small amount of baryonic matter or not at all.
This is due to the limitation of the CosmoDC2 catalog.  We discussed this in more detail in Section \ref{sec:low_mass_halos}.
Our work nevertheless shows great potential of using ResNet to classify followed with using U-Net to pinpoint these low-mass halos, of which the prevailing CDM model predicts a great abundance,
via their lensing distortion effect on background sources.\footnote{Note that three significant differences between our approach and that of \citet{Ostdiek_2022_1, Ostdiek_2022_2} are 1) our detection method does not require the presence a large Einstein radius lens ($\theta_E \gtrsim 0.5''$); 
2) the sources we use in our simulations are much more realistic; and 3) they do not use the ResNet in their pipeline.}
This is especially true for \textit{JWST} observations,
which contain a much higher density of high redshift (hence, for the purpose of lensing, ``background") galaxies.
In addition, future telescopes will push to even higher resolutions.
For example, the Thirty-Meter Telescope (TMT) and the Extremely Large Telescope (ELT) will potentially reach angular resolutions down to 0.005-0.015$''$ using Adaptive Optics \citep[e.g.,][]{Davies_2010, Wright_2016, Thatte_2021}. 
\ed{Using radio telescope arrays, even higher resolutions (micro-arcsecond scale) are achievable.}
\emph{It is all but certain that
dark low-mass halos with Einstein radii approaching these angular scales are only discoverable with the best machine learning techniques.}
% One of the many possible benefits of this future technology could be the ability to resolve 
Our work shows a path of finding these low-mass dark halos through their strong lensing effect using neural networks
%, as demonstrated in this research,
and potentially solving one of the greatest mysteries in cosmology.

% \ed{At a minimum, this is a proof of concept. As simulations improve -- i.e., becomes more realistic -- we are confident this pipeline will continue to be successful. Not to mention both image classification and segmentation (i.e. detection) algorithms will also continue to improve.}

\newpage
\bibliography{bibliography}

\begin{thebibliography}{}
\expandafter\ifx\csname natexlab\endcsname\relax\def\natexlab#1{#1}\fi
\providecommand{\url}[1]{\href{#1}{#1}}
\providecommand{\dodoi}[1]{doi:~\href{http://doi.org/#1}{\nolinkurl{#1}}}
\providecommand{\doeprint}[1]{\href{http://ascl.net/#1}{\nolinkurl{http://ascl.net/#1}}}
\providecommand{\doarXiv}[1]{\href{https://arxiv.org/abs/#1}{\nolinkurl{https://arxiv.org/abs/#1}}}

\bibitem[{Abazajian {et~al.}(2003)Abazajian, Adelman-McCarthy, Agüeros, Allam, Anderson, Annis, Bahcall, Baldry, Bastian, Berlind, Bernardi, Blanton, Blythe, John J.~Bochanski, Boroski, Brewington, Briggs, Brinkmann, Brunner, Budavári, Carey, Carr, Castander, Chiu, Collinge, Connolly, Covey, Csabai, Dalcanton, Dodelson, Doi, Dong, Eisenstein, Evans, Fan, Feldman, Finkbeiner, Friedman, Frieman, Fukugita, Gal, Gillespie, Glazebrook, Gonzalez, Gray, Grebel, Grodnicki, Gunn, Gurbani, Hall, Hao, Harbeck, Harris, Harris, Harvanek, Hawley, Heckman, Helmboldt, Hendry, Hennessy, Hindsley, Hogg, Holmgren, Holtzman, Homer, Hui, ichi Ichikawa, Ichikawa, Inkmann, Željko Ivezić, Jester, Johnston, Jordan, Jordan, Jorgensen, Jurić, Kauffmann, Kent, Kleinman, Knapp, Kniazev, Kron, Krzesiński, Kunszt, Kuropatkin, Lamb, Lampeitl, Laubscher, Lee, Leger, Li, Lidz, Lin, Loh, Long, Loveday, Lupton, Malik, Margon, McGehee, McKay, Meiksin, Miknaitis, Moorthy, Munn, Murphy, Nakajima, Narayanan, Nash, Eric H.~Neilsen, Newberg,
  Newman, Nichol, Nicinski, Nieto-Santisteban, Nitta, Odenkirchen, Okamura, Ostriker, Owen, Padmanabhan, Peoples, Pier, Pindor, Pope, Quinn, Rafikov, Raymond, Richards, Richmond, Rix, Rockosi, Schaye, Schlegel, Schneider, Schroeder, Scranton, Sekiguchi, Seljak, Sergey, Sesar, Sheldon, Shimasaku, Siegmund, Silvestri, Sinisgalli, Sirko, Smith, Smolčić, Snedden, Stebbins, Steinhardt, Stinson, Stoughton, Strateva, Strauss, SubbaRao, Szalay, Szapudi, Szkody, Tasca, Tegmark, Thakar, Tremonti, Tucker, Uomoto, Berk, Vandenberg, Vogeley, Voges, Vogt, Walkowicz, Weinberg, West, White, Wilhite, Willman, Xu, Yanny, Yarger, Yasuda, Yip, Yocum, York, Zakamska, Zehavi, Zheng, Zibetti, \& Zucker}]{Abazajian_2003}
Abazajian, K., Adelman-McCarthy, J.~K., Agüeros, M.~A., {et~al.} 2003, The Astronomical Journal, 126, 2081, \dodoi{10.1086/378165}

\bibitem[{{Asaki} {et~al.}(2023){Asaki}, {Maud}, {Francke}, {Nagai}, {Petry}, {Fomalont}, {Humphreys}, {Richards}, {Wong}, {Dent}, {Hirota}, {Fernandez}, {Takahashi}, \& {Hales}}]{Asaki_2023}
{Asaki}, Y., {Maud}, L.~T., {Francke}, H., {et~al.} 2023, \apj, 958, 86, \dodoi{10.3847/1538-4357/acf619}

\bibitem[{{Auger} {et~al.}(2010){Auger}, {Treu}, {Bolton}, {Gavazzi}, {Koopmans}, {Marshall}, {Moustakas}, \& {Burles}}]{auger2010a}
{Auger}, M.~W., {Treu}, T., {Bolton}, A.~S., {et~al.} 2010, \apj, 724, 511, \dodoi{10.1088/0004-637X/724/1/511}

\bibitem[{{Barnacka} {et~al.}(2016){Barnacka}, {Geller}, {Dell'Antonio}, \& {Zitrin}}]{Barnacka_2016}
{Barnacka}, A., {Geller}, M.~J., {Dell'Antonio}, I.~P., \& {Zitrin}, A. 2016, \apj, 821, 58, \dodoi{10.3847/0004-637X/821/1/58}

\bibitem[{{Benson} \& {Devereux}(2010)}]{Benson_2010}
{Benson}, A.~J., \& {Devereux}, N. 2010, \mnras, 402, 2321, \dodoi{10.1111/j.1365-2966.2009.16089.x}

\bibitem[{{Birrer} \& {Amara}(2018)}]{Birrer_2018}
{Birrer}, S., \& {Amara}, A. 2018, Physics of the Dark Universe, 22, 189, \dodoi{10.1016/j.dark.2018.11.002}

\bibitem[{{Birrer} {et~al.}(2021){Birrer}, {Shajib}, {Gilman}, {Galan}, {Aalbers}, {Millon}, {Morgan}, {Pagano}, {Park}, {Teodori}, {Tessore}, {Ueland}, {Van de Vyvere}, {Wagner-Carena}, {Wempe}, {Yang}, {Ding}, {Schmidt}, {Sluse}, {Zhang}, \& {Amara}}]{Birrer_2021}
{Birrer}, S., {Shajib}, A., {Gilman}, D., {et~al.} 2021, The Journal of Open Source Software, 6, 3283, \dodoi{10.21105/joss.03283}

\bibitem[{{Boddy} {et~al.}(2022){Boddy}, {Lisanti}, {McDermott}, {Rodd}, {Weniger}, {Ali-Ha{\"\i}moud}, {Buschmann}, {Cholis}, {Croon}, {Erickcek}, {Gluscevic}, {Leane}, {Mishra-Sharma}, {Mu{\~n}oz}, {Nadler}, {Natarajan}, {Price-Whelan}, {Vegetti}, \& {Witte}}]{boddy2022a}
{Boddy}, K.~K., {Lisanti}, M., {McDermott}, S.~D., {et~al.} 2022, Journal of High Energy Astrophysics, 35, 112, \dodoi{10.1016/j.jheap.2022.06.005}

\bibitem[{{Bolton} {et~al.}(2008){Bolton}, {Burles}, {Koopmans}, {Treu}, {Gavazzi}, {Moustakas}, {Wayth}, \& {Schlegel}}]{Bolton_2008}
{Bolton}, A.~S., {Burles}, S., {Koopmans}, L. V.~E., {et~al.} 2008, \apj, 682, 964, \dodoi{10.1086/589327}

\bibitem[{{Bolton} {et~al.}(2006){Bolton}, {Burles}, {Koopmans}, {Treu}, \& {Moustakas}}]{bolton2006a}
{Bolton}, A.~S., {Burles}, S., {Koopmans}, L.~V.~E., {Treu}, T., \& {Moustakas}, L.~A. 2006, \apj, 638, 703, \dodoi{10.1086/498884}

\bibitem[{{Brownstein} {et~al.}(2012){Brownstein}, {Bolton}, {Schlegel}, {Eisenstein}, {Kochanek}, {Connolly}, {Maraston}, {Pandey}, {Seitz}, {Wake}, {Wood-Vasey}, {Brinkmann}, {Schneider}, \& {Weaver}}]{Brownstein_2012}
{Brownstein}, J.~R., {Bolton}, A.~S., {Schlegel}, D.~J., {et~al.} 2012, \apj, 744, 41, \dodoi{10.1088/0004-637X/744/1/41}

\bibitem[{{Ca{\~n}ameras} {et~al.}(2021){Ca{\~n}ameras}, {Schuldt}, {Shu}, {Suyu}, {Taubenberger}, {Meinhardt}, {Leal-Taix{\'e}}, {Chao}, {Inoue}, {Jaelani}, \& {More}}]{canameras2021a}
{Ca{\~n}ameras}, R., {Schuldt}, S., {Shu}, Y., {et~al.} 2021, \aap, 653, L6, \dodoi{10.1051/0004-6361/202141758}

\bibitem[{{Caldeira} {et~al.}(2019){Caldeira}, {Wu}, {Nord}, {Avestruz}, {Trivedi}, \& {Story}}]{Caldeira_2019}
{Caldeira}, J., {Wu}, W.~L.~K., {Nord}, B., {et~al.} 2019, Astronomy and Computing, 28, 100307, \dodoi{10.1016/j.ascom.2019.100307}

\bibitem[{{Caminha} {et~al.}(2022){Caminha}, {Suyu}, {Grillo}, \& {Rosati}}]{caminha2022a}
{Caminha}, G.~B., {Suyu}, S.~H., {Grillo}, C., \& {Rosati}, P. 2022, \aap, 657, A83, \dodoi{10.1051/0004-6361/202141994}

\bibitem[{{{\c{C}}a{\v{g}}an {\c{S}}eng{\"u}l} {et~al.}(2020){{\c{C}}a{\v{g}}an {\c{S}}eng{\"u}l}, {Tsang}, {Diaz Rivero}, {Dvorkin}, {Zhu}, \& {Seljak}}]{cagansengul2020a}
{{\c{C}}a{\v{g}}an {\c{S}}eng{\"u}l}, A., {Tsang}, A., {Diaz Rivero}, A., {et~al.} 2020, \prd, 102, 063502, \dodoi{10.1103/PhysRevD.102.063502}

\bibitem[{{Cheng} {et~al.}(2020){Cheng}, {Li}, {Conselice}, {Arag{\'o}n-Salamanca}, {Dye}, \& {Metcalf}}]{cheng2020a}
{Cheng}, T.-Y., {Li}, N., {Conselice}, C.~J., {et~al.} 2020, \mnras, 494, 3750, \dodoi{10.1093/mnras/staa1015}

\bibitem[{{Collett} {et~al.}(2019){Collett}, {Montanari}, \& {R{\"a}s{\"a}nen}}]{collett2019a}
{Collett}, T., {Montanari}, F., \& {R{\"a}s{\"a}nen}, S. 2019, \prl, 123, 231101, \dodoi{10.1103/PhysRevLett.123.231101}

\bibitem[{{Collett}(2015)}]{Collett_2015}
{Collett}, T.~E. 2015, \apj, 811, 20, \dodoi{10.1088/0004-637X/811/1/20}

\bibitem[{{Collett} \& {Auger}(2014)}]{collett2014a}
{Collett}, T.~E., \& {Auger}, M.~W. 2014, \mnras, 443, 969, \dodoi{10.1093/mnras/stu1190}

\bibitem[{{Collett} {et~al.}(2012){Collett}, {Auger}, {Belokurov}, {Marshall}, \& {Hall}}]{Collett_2012}
{Collett}, T.~E., {Auger}, M.~W., {Belokurov}, V., {Marshall}, P.~J., \& {Hall}, A.~C. 2012, \mnras, 424, 2864, \dodoi{10.1111/j.1365-2966.2012.21424.x}

\bibitem[{{Collett} \& {Bacon}(2017)}]{collett2017a}
{Collett}, T.~E., \& {Bacon}, D. 2017, \prl, 118, 091101, \dodoi{10.1103/PhysRevLett.118.091101}

\bibitem[{{Collett} {et~al.}(2018){Collett}, {Oldham}, {Smith}, {Auger}, {Westfall}, {Bacon}, {Nichol}, {Masters}, {Koyama}, \& {van den Bosch}}]{collett2018a}
{Collett}, T.~E., {Oldham}, L.~J., {Smith}, R.~J., {et~al.} 2018, Science, 360, 1342, \dodoi{10.1126/science.aao2469}

\bibitem[{{Cornachione} {et~al.}(2018){Cornachione}, {Bolton}, {Shu}, {Zheng}, {Montero-Dorta}, {Brownstein}, {Oguri}, {Kochanek}, {Mao}, {P{\`e}rez-Fournon}, {Marques-Chaves}, \& {M{\`e}nard}}]{cornachione2018a}
{Cornachione}, M.~A., {Bolton}, A.~S., {Shu}, Y., {et~al.} 2018, \apj, 853, 148, \dodoi{10.3847/1538-4357/aaa412}

\bibitem[{{Costantin} {et~al.}(2021){Costantin}, {P{\'e}rez-Gonz{\'a}lez}, {M{\'e}ndez-Abreu}, {Huertas-Company}, {Dimauro}, {Alcalde-Pampliega}, {Buitrago}, {Ceverino}, {Daddi}, {Dom{\'\i}nguez-S{\'a}nchez}, {Espino-Briones}, {Hern{\'a}n-Caballero}, {Koekemoer}, \& {Rodighiero}}]{Constantin_2021}
{Costantin}, L., {P{\'e}rez-Gonz{\'a}lez}, P.~G., {M{\'e}ndez-Abreu}, J., {et~al.} 2021, \apj, 913, 125, \dodoi{10.3847/1538-4357/abef72}

\bibitem[{{{\c{S}}eng{\"u}l} {et~al.}(2022){{\c{S}}eng{\"u}l}, {Dvorkin}, {Ostdiek}, \& {Tsang}}]{Sengul_2022}
{{\c{S}}eng{\"u}l}, A.~{\c{C}}., {Dvorkin}, C., {Ostdiek}, B., \& {Tsang}, A. 2022, \mnras, 515, 4391, \dodoi{10.1093/mnras/stac1967}

\bibitem[{{Davies} {et~al.}(2010){Davies}, {Ageorges}, {Barl}, {Bedin}, {Bender}, {Bernardi}, {Chapron}, {Clenet}, {Deep}, {Deul}, {Drost}, {Eisenhauer}, {Falomo}, {Fiorentino}, {F{\"o}rster Schreiber}, {Gendron}, {Genzel}, {Gratadour}, {Greggio}, {Grupp}, {Held}, {Herbst}, {Hess}, {Hubert}, {Jahnke}, {Kuijken}, {Lutz}, {Magrin}, {Muschielok}, {Navarro}, {Noyola}, {Paumard}, {Piotto}, {Ragazzoni}, {Renzini}, {Rousset}, {Rix}, {Saglia}, {Tacconi}, {Thiel}, {Tolstoy}, {Trippe}, {Tromp}, {Valentijn}, {Verdoes Kleijn}, \& {Wegner}}]{Davies_2010}
{Davies}, R., {Ageorges}, N., {Barl}, L., {et~al.} 2010, in Society of Photo-Optical Instrumentation Engineers (SPIE) Conference Series, Vol. 7735, Ground-based and Airborne Instrumentation for Astronomy III, ed. I.~S. {McLean}, S.~K. {Ramsay}, \& H.~{Takami}, 77352A

\bibitem[{{De Lucia} {et~al.}(2012){De Lucia}, {Fontanot}, \& {Wilman}}]{De_Lucia_2012}
{De Lucia}, G., {Fontanot}, F., \& {Wilman}, D. 2012, \mnras, 419, 1324, \dodoi{10.1111/j.1365-2966.2011.19789.x}

\bibitem[{{Driver} {et~al.}(2022){Driver}, {Robotham}, {Obreschkow}, {Peacock}, {Baldry}, {Bellstedt}, {Bland-Hawthorn}, {Brough}, {Cluver}, {Holwerda}, {Hopkins}, {Lagos}, {Liske}, {Loveday}, {Phillipps}, \& {Taylor}}]{driver2022a}
{Driver}, S.~P., {Robotham}, A. S.~G., {Obreschkow}, D., {et~al.} 2022, \mnras, 515, 2138, \dodoi{10.1093/mnras/stac581}

\bibitem[{{Euclid Collaboration} {et~al.}(2024){Euclid Collaboration}, {Mellier}, {Abdurro'uf}, {Acevedo Barroso}, {Ach{\'u}carro}, {Adamek}, {Adam}, {Addison}, {Aghanim}, {Aguena}, {Ajani}, {Akrami}, {Al-Bahlawan}, {Alavi}, {Albuquerque}, {Alestas}, {Alguero}, {Allaoui}, {Allen}, {Allevato}, {Alonso-Tetilla}, {Altieri}, {Alvarez-Candal}, {Amara}, {Amendola}, {Amiaux}, {Andika}, {Andreon}, {Andrews}, {Angora}, {Angulo}, {Annibali}, {Anselmi}, {Anselmi}, {Arcari}, {Archidiacono}, {Aric{\`o}}, {Arnaud}, {Arnouts}, {Asgari}, {Asorey}, {Atayde}, {Atek}, {Atrio-Barandela}, {Aubert}, {Aubourg}, {Auphan}, {Auricchio}, {Aussel}, {Aussel}, {Avelino}, {Avgoustidis}, {Avila}, {Awan}, {Azzollini}, {Baccigalupi}, {Bachelet}, {Bacon}, {Baes}, {Bagley}, {Bahr-Kalus}, {Balaguera-Antolinez}, {Balbinot}, {Balcells}, {Baldi}, {Baldry}, {Balestra}, {Ballardini}, {Ballester}, {Balogh}, {Ba{\~n}ados}, {Barbier}, {Bardelli}, {Barreiro}, {Barriere}, {Barros}, {Barthelemy}, {Bartolo}, {Basset}, {Battaglia}, {Battisti}, {Baugh},
  {Baumont}, {Bazzanini}, {Beaulieu}, {Beckmann}, {Belikov}, {Bel}, {Bellagamba}, {Bella}, {Bellini}, {Benabed}, {Bender}, {Benevento}, {Bennett}, {Benson}, {Bergamini}, {Bermejo-Climent}, {Bernardeau}, {Bertacca}, {Berthe}, {Berthier}, {Bethermin}, {Beutler}, {Bevillon}, {Bhargava}, {Bhatawdekar}, {Bisigello}, {Biviano}, {Blake}, {Blanchard}, {Blazek}, {Blot}, {Bosco}, {Bodendorf}, {Boenke}, {B{\"o}hringer}, {Bolzonella}, {Bonchi}, {Bonici}, {Bonino}, {Bonino}, {Bonvin}, {Bon}, {Booth}, {Borgani}, {Borlaff}, {Borsato}, {Bosco}, {Bose}, {Botticella}, {Boucaud}, {Bouche}, {Boucher}, {Boutigny}, {Bouvard}, {Bouy}, {Bowler}, {Bozza}, {Bozzo}, {Branchini}, {Brau-Nogue}, {Brekke}, {Bremer}, {Brescia}, {Breton}, {Brinchmann}, {Brinckmann}, {Brockley-Blatt}, {Brodwin}, {Brouard}, {Brown}, {Bruton}, {Bucko}, {Buddelmeijer}, {Buenadicha}, {Buitrago}, {Burger}, {Burigana}, {Busillo}, {Busonero}, {Cabanac}, {Cabayol-Garcia}, {Cagliari}, {Caillat}, {Caillat}, {Calabrese}, {Calabro}, {Calderone}, {Calura}, {Camacho
  Quevedo}, {Camera}, {Campos}, {Canas-Herrera}, {Candini}, {Cantiello}, {Capobianco}, {Cappellaro}, {Cappelluti}, {Cappi}, {Caputi}, {Cara}, {Carbone}, {Cardone}, {Carella}, {Carlberg}, {Carle}, {Carminati}, {Caro}, {Carrasco}, {Carretero}, {Carrilho}, {Carron Duque}, {Carry}, {Carvalho}, {Carvalho}, {Casas}, {Casas}, {Casenove}, {Casey}, {Cassata}, {Castander}, {Castelao}, {Castellano}, {Castiblanco}, {Castignani}, {Castro}, {Cavet}, {Cavuoti}, {Chabaud}, {Chambers}, {Charles}, {Charlot}, {Chartab}, {Chary}, {Chaumeil}, {Cho}, {Chon}, {Ciancetta}, {Ciliegi}, {Cimatti}, {Cimino}, {Cioni}, {Claydon}, {Cleland}, {Cl{\'e}ment}, {Clements}, {Clerc}, {Clesse}, {Codis}, {Cogato}, {Colbert}, {Cole}, {Coles}, {Collett}, {Collins}, {Colodro-Conde}, {Colombo}, {Combes}, {Conforti}, {Congedo}, {Conseil}, {Conselice}, {Contarini}, {Contini}, {Conversi}, {Cooray}, {Copin}, {Corasaniti}, {Corcho-Caballero}, {Corcione}, {Cordes}, {Corpace}, {Correnti}, {Costanzi}, {Costille}, {Courbin}, {Courcoult Mifsud}, {Courtois},
  {Cousinou}, {Covone}, {Cowell}, {Cragg}, {Cresci}, {Cristiani}, {Crocce}, {Cropper}, {E Crouzet}, {Csizi}, {Cuby}, {Cucchetti}, {Cucciati}, {Cuillandre}, {Cunha}, {Cuozzo}, {Daddi}, {D'Addona}, {Dafonte}, {Dagoneau}, {Dalessandro}, {Dalton}, {D'Amico}, {Dannerbauer}, {Danto}, {Das}, {Da Silva}, {da Silva}, {Daste}, {Davies}, {Davini}, {de Boer}, {Decarli}, {De Caro}, {Degaudenzi}, {Degni}, {de Jong}, {de la Bella}, {de la Torre}, {Delhaise}, {Delley}, {Delucchi}, {De Lucia}, {Denniston}, {De Paolis}, {De Petris}, {Derosa}, {Desai}, {Desjacques}, {Despali}, {Desprez}, {De Vicente-Albendea}, {Deville}, {Dias}, {D{\'\i}az-S{\'a}nchez}, {Diaz}, {Di Domizio}, {Diego}, {Di Ferdinando}, {Di Giorgio}, {Dimauro}, {Dinis}, {Dolag}, {Dolding}, {Dole}, {Dom{\'\i}nguez S{\'a}nchez}, {Dor{\'e}}, {Dournac}, {Douspis}, {Dreihahn}, {Droge}, {Dryer}, {Dubath}, {Duc}, {Ducret}, {Duffy}, {Dufresne}, {Duncan}, {Dupac}, {Duret}, {Durrer}, {Durret}, {Dusini}, {Ealet}, {Eggemeier}, {Eisenhardt}, {Elbaz}, {Elkhashab}, {Ellien},
  {Endicott}, {Enia}, {Erben}, {Escartin Vigo}, {Escoffier}, {Escudero Sanz}, {Essert}, {Ettori}, {Ezziati}, {Fabbian}, {Fabricius}, {Fang}, {Farina}, {Farina}, {Farinelli}, {Farrens}, {Faustini}, {Feltre}, {Ferguson}, {Ferrando}, {Ferrari}, {Ferr{\'e}-Mateu}, {Ferreira}, {Ferreras}, {Ferrero}, {Ferriol}, {Ferruit}, {Filleul}, {Finelli}, {Finkelstein}, {Finoguenov}, {Fiorini}, {Flentge}, {Focardi}, {Fonseca}, {Fontana}, {Fontanot}, {Fornari}, {Fosalba}, {Fossati}, {Fotopoulou}, {Fouchez}, {Fourmanoit}, {Frailis}, {Fraix-Burnet}, {Franceschi}, {Franco}, {Franzetti}, {Freihoefer}, {Frittoli}, {Frugier}, {Frusciante}, {Fumagalli}, {Fumagalli}, {Fumana}, {Fu}, {Gabarra}, {Galeotta}, {Galluccio}, {Ganga}, {Gao}, {Garc{\'\i}a-Bellido}, {Garcia}, {Gardner}, {Garilli}, {Gaspar-Venancio}, {Gasparetto}, {Gautard}, {Gavazzi}, {Gaztanaga}, {Genolet}, {Genova Santos}, {Gentile}, {George}, {Ghaffari}, {Giacomini}, {Gianotti}, {Gibb}, {Gillard}, {Gillis}, {Ginolfi}, {Giocoli}, {Girardi}, {Giri}, {Goh}, {G{\'o}mez-Alvarez},
  {Gonzalez}, {Gonzalez}, {Gonzalez}, {Gouyou Beauchamps}, {Gozaliasl}, {Gracia-Carpio}, {Grandis}, {Granett}, {Granvik}, {Grazian}, {Gregorio}, {Grenet}, {Grillo}, {Grupp}, {Gruppioni}, {Gruppuso}, {Guerbuez}, {Guerrini}, {Guidi}, {Guillard}, {Gutierrez}, {Guttridge}, {Guzzo}, {Gwyn}, {Haapala}, {Haase}, {Haddow}, {Hailey}, {Hall}, {Hall}, {Hamaus}, {Haridasu}, {Harnois-D{\'e}raps}, {Harper}, {Hartley}, {Hasinger}, {Hassani}, {Hatch}, {Haugan}, {H{\"a}u{\ss}ler}, {Heavens}, {Heisenberg}, {Helmi}, {Helou}, {Hemmati}, {Henares}, {Herent}, {Hern{\'a}ndez-Monteagudo}, {Heuberger}, {Hewett}, {Heydenreich}, {Hildebrandt}, {Hirschmann}, {Hjorth}, {Hoar}, {Hoekstra}, {Holland}, {Holliman}, {Holmes}, {Hook}, {Horeau}, {Hormuth}, {Hornstrup}, {Hosseini}, {Hu}, {Hudelot}, {Hudson}, {Huertas-Company}, {Huff}, {Hughes}, {Humphrey}, {Hunt}, {Huynh}, {Ibata}, {Ichikawa}, {Iglesias-Groth}, {Ilbert}, {Ili{\'c}}, {Ingoglia}, {Iodice}, {Israel}, {Israelsson}, {Izzo}, {Jablonka}, {Jackson}, {Jacobson}, {Jafariyazani}, {Jahnke},
  {Jansen}, {Jarvis}, {Jasche}, {Jauzac}, {Jeffrey}, {Jhabvala}, {Jimenez-Teja}, {Jimenez Mu{\~n}oz}, {Joachimi}, {Johansson}, {Joudaki}, {Jullo}, {Kajava}, {Kang}, {Kannawadi}, {Kansal}, {Karagiannis}, {K{\"a}rcher}, {Kashlinsky}, {Kazandjian}, {Keck}, {Keih{\"a}nen}, {Kerins}, {Kermiche}, {Khalil}, {Kiessling}, {Kiiveri}, {Kilbinger}, {Kim}, {King}, {Kirkpatrick}, {Kitching}, {Kluge}, {Knabenhans}, {Knapen}, {Knebe}, {Kneib}, {Kohley}, {Koopmans}, {Koskinen}, {Koulouridis}, {Kou}, {Kov{\'a}cs}, {Kova\{{\v{c}}\}i{\'c}}, {Kowalczyk}, {Koyama}, {Kraljic}, {Krause}, {Kruk}, {Kubik}, {Kuchner}, {Kuijken}, {K{\"u}mmel}, {Kunz}, {Kurki-Suonio}, {Lacasa}, {Lacey}, {La Franca}, {Lagarde}, {Lahav}, {Laigle}, {La Marca}, {La Marle}, {Lamine}, {Lam}, {Lan{\c{c}}on}, {Landt}, {Langer}, {Lapi}, {Larcheveque}, {Larsen}, {Lattanzi}, {Laudisio}, {Laugier}, {Laureijs}, {Lavaux}, {Lawrenson}, {Lazanu}, {Lazeyras}, {Le Boulc'h}, {Le Brun}, {Le Brun}, {Leclercq}, {Lee}, {Le Graet}, {Legrand}, {Leirvik}, {Le Jeune}, {Lembo}, {Le
  Mignant}, {Lepinzan}, {Lepori}, {Lesci}, {Lesgourgues}, {Leuzzi}, {Levi}, {Liaudat}, {Libet}, {Liebing}, {Ligori}, {Lilje}, {Lin}, {Linde}, {Linder}, {Lindholm}, {Linke}, {Li}, {Liu}, {Lloro}, {Lobo}, {Lodieu}, {Lombardi}, {Lombriser}, {Lonare}, {Longo}, {L{\'o}pez-Caniego}, {Lopez Lopez}, {Alvarez}, {Loureiro}, {Loveday}, {Lusso}, {Macias-Perez}, {Maciaszek}, {Magliocchetti}, {Magnard}, {Magnier}, {Magro}, {Mahler}, {Mainetti}, {Maino}, {Maiorano}, {Maiorano}, {Malavasi}, {Mamon}, {Mancini}, {Mandelbaum}, {Manera}, {Manj{\'o}n-Garc{\'\i}a}, {Mannucci}, {Mansutti}, {Manteiga Outeiro}, {Maoli}, {Maraston}, {Marcin}, {Marcos-Arenal}, {Margalef-Bentabol}, {Marggraf}, {Marinucci}, {Marinucci}, {Markovic}, {Marleau}, {Marpaud}, {Martignac}, {Mart{\'\i}n-Fleitas}, {Martin-Moruno}, {Martin}, {Martinelli}, {Martinet}, {Martin}, {Martins}, {Marulli}, {Massari}, {Massey}, {Masters}, {Matarrese}, {Matsuoka}, {Matthew}, {Maughan}, {Mauri}, {Maurin}, {Maurogordato}, {McCarthy}, {McConnachie}, {McCracken}, {McDonald},
  {McEwen}, {McPartland}, {Medinaceli}, {Mehta}, {Mei}, {Melchior}, {Melin}, {M{\'e}nard}, {Mendes}, {Mendez-Abreu}, {Meneghetti}, {Mercurio}, {Merlin}, {Metcalf}, {Meylan}, {Migliaccio}, {Mignoli}, {Miller}, {Miluzio}, {Milvang-Jensen}, {Mimoso}, {Miquel}, {Miyatake}, {Mobasher}, {Mohr}, {Monaco}, {Mongui{\'o}}, {Montoro}, {Mora}, {Moradinezhad Dizgah}, {Moresco}, {Moretti}, {Morgante}, {Morisset}, {Moriya}, {Morris}, {Mortlock}, {Moscardini}, {Mota}, {Moustakas}, {Moutard}, {M{\"u}ller}, {Munari}, {Murphree}, {Murray}, {Murray}, {Musi}, {Nadathur}, {Nagam}, {Nagao}, {Naidoo}, {Nakajima}, {Nally}, {Natoli}, {Navarro-Alsina}, {Navarro Girones}, {Neissner}, {Nersesian}, {Nesseris}, {Nguyen-Kim}, {Nicastro}, {Nichol}, {Nielbock}, {Niemi}, {Nieto}, {Nilsson}, {Noller}, {Norberg}, {Nourizonoz}, {Ntelis}, {Nucita}, {Nugent}, {Nunes}, {Nutma}, {Ocampo}, {Odier}, {Oesch}, {Oguri}, {Magalhaes Oliveira}, {Onoue}, {Oosterbroek}, {Oppizzi}, {Ordenovic}, {Osato}, {Pacaud}, {Pace}, {Padilla}, {Paech}, {Pagano}, {Page},
  {Palazzi}, {Paltani}, {Pamuk}, {Pandolfi}, {Paoletti}, {Paolillo}, {Papaderos}, {Pardede}, {Parimbelli}, {Parmar}, {Partmann}, {Pasian}, {Passalacqua}, {Paterson}, {Patrizii}, {Pattison}, {Paulino-Afonso}, {Paviot}, {Peacock}, {Pearce}, {Pedersen}, {Peel}, {Peletier}, {Pellejero Ibanez}, {Pello}, {Penny}, {Percival}, {Perez-Garrido}, {Perotto}, {Pettorino}, {Pezzotta}, {Pezzuto}, {Philippon}, {Piersanti}, {Pietroni}, {Piga}, {Pilo}, {Pires}, {Pisani}, {Pizzella}, {Pizzuti}, {Plana}, {Polenta}, {Pollack}, {Poncet}, {P{\"o}ntinen}, {Pool}, {Popa}, {Popa}, {Popp}, {Porciani}, {Porth}, {Potter}, {Poulain}, {Pourtsidou}, {Pozzetti}, {Prandoni}, {Pratt}, {Prezelus}, {Prieto}, {Pugno}, {Quai}, {Quilley}, {Racca}, {Raccanelli}, {R{\'a}cz}, {Radinovi{\'c}}, {Radovich}, {Ragagnin}, {Ragnit}, {Raison}, {Ramos-Chernenko}, {Ranc}, {Raylet}, {Rebolo}, {Refregier}, {Reimberg}, {Reiprich}, {Renk}, {Renzi}, {Retre}, {Revaz}, {Reyl{\'e}}, {Reynolds}, {Rhodes}, {Ricci}, {Ricci}, {Riccio}, {Ricken}, {Rissanen}, {Risso}, {Rix},
  {Robin}, {Rocca-Volmerange}, {Rocci}, {Rodenhuis}, {Rodighiero}, {Rodriguez Monroy}, {Rollins}, {Romanello}, {Roman}, {Romelli}, {Romero-Gomez}, {Roncarelli}, {Rosati}, {Rosset}, {Rossetti}, {Roster}, {Rottgering}, {Rozas-Fern{\'a}ndez}, {Ruane}, {Rubino-Martin}, {Rudolph}, {Ruppin}, {Rusholme}, {Sacquegna}, {S{\'a}ez-Casares}, {Saga}, {Saglia}, {Sahl{\'e}n}, {Saifollahi}, {Sakr}, {Salvalaggio}, {Salvaterra}, {Salvati}, {Salvato}, {Salvignol}, {S{\'a}nchez}, {Sanchez}, {Sanders}, {Sapone}, {Saponara}, {Sarpa}, {Sarron}, {Sartori}, {Sassolas}, {Sauniere}, {Sauvage}, {Sawicki}, {Scaramella}, {Scarlata}, {Scharr{\'e}}, {Schaye}, {Schewtschenko}, {Schindler}, {Schinnerer}, {Schirmer}, {Schmidt}, {Schmidt}, {Schmidt}, {Schneider}, {Schneider}, {Schneider}, {Sch{\"o}neberg}, {Schrabback}, {Schultheis}, {Schulz}, {Schwartz}, {Sciotti}, {Scodeggio}, {Scognamiglio}, {Scott}, {Scottez}, {Secroun}, {Sefusatti}, {Seidel}, {Seiffert}, {Sellentin}, {Selwood}, {Semboloni}, {Sereno}, {Serjeant}, {Serrano}, {Shankar},
  {Sharples}, {Short}, {Shulevski}, {Shuntov}, {Sias}, {Sikkema}, {Silvestri}, {Simon}, {Sirignano}, {Sirri}, {Skottfelt}, {Slezak}, {Sluse}, {Smith}, {Smith}, {Smith}, {Smit}, {Soldano}, {Solheim}, {Sorce}, {Sorrenti}, {Soubrie}, {Spinoglio}, {Spurio Mancini}, {Stadel}, {Stagnaro}, {Stanco}, {Stanford}, {Starck}, {Stassi}, {Steinwagner}, {Stern}, {Stone}, {Strada}, {Strafella}, {Stramaccioni}, {Surace}, {Sureau}, {Suyu}, {Swindells}, {Szafraniec}, {Szapudi}, {Taamoli}, {Talia}, {Tallada-Cresp{\'\i}}, {Tanidis}, {Tao}, {Tarr{\'\i}o}, {Tavagnacco}, {Taylor}, {Taylor}, {Taylor}, {Teixeira}, {Tenti}, {Teodoro Idiago}, {Teplitz}, {Tereno}, {Tessore}, {Testa}, {Testera}, {Tewes}, {Teyssier}, {Theret}, {Thizy}, {Thomas}, {Toba}, {Toft}, {Toledo-Moreo}, {Tolstoy}, {Tommasi}, {Torbaniuk}, {Torradeflot}, {Tortora}, {Tosi}, {Tosti}, {Trifoglio}, {Troja}, {Trombetti}, {Tronconi}, {Tsedrik}, {Tsyganov}, {Tucci}, {Tutusaus}, {Uhlemann}, {Ulivi}, {Urbano}, {Vacher}, {Vaillon}, {Valdes}, {Valentijn}, {Valenziano},
  {Valieri}, {Valiviita}, {Van den Broeck}, {Vassallo}, {Vavrek}, {Venemans}, {Venhola}, {Ventura}, {Verdoes Kleijn}, {Vergani}, {Verma}, {Vernizzi}, {Veropalumbo}, {Verza}, {Vescovi}, {Vibert}, {Viel}, {Vielzeuf}, {Viglione}, {Viitanen}, {Villaescusa-Navarro}, {Vinciguerra}, {Visticot}, {Voggel}, {von Wietersheim-Kramsta}, {Vriend}, {Wachter}, {Walmsley}, {Walth}, {Walton}, {Walton}, {Wander}, {Wang}, {Wang}, {Weaver}, {Weller}, {Whalen}, {Wiesmann}, {Wilde}, {Williams}, {Winther}, {Wittje}, {Wong}, {Wright}, {Yankelevich}, {Yeung}, {Youles}, {Yung}, {Zacchei}, {Zalesky}, {Zamorani}, {Zamorano Vitorelli}, {Zanoni Marc}, {Zennaro}, {Zerbi}, {Zinchenko}, {Zoubian}, {Zucca}, \& {Zumalacarregui}}]{Euclid_Mellier_2024}
{Euclid Collaboration}, {Mellier}, Y., {Abdurro'uf}, {et~al.} 2024, arXiv e-prints, arXiv:2405.13491, \dodoi{10.48550/arXiv.2405.13491}

\bibitem[{{Fan} {et~al.}(2017){Fan}, {Liao}, {Biesiada}, {Pi{\'o}rkowska-Kurpas}, \& {Zhu}}]{fan2017a}
{Fan}, X.-L., {Liao}, K., {Biesiada}, M., {Pi{\'o}rkowska-Kurpas}, A., \& {Zhu}, Z.-H. 2017, \prl, 118, 091102, \dodoi{10.1103/PhysRevLett.118.091102}

\bibitem[{{Freedman} {et~al.}(2020){Freedman}, {Madore}, {Hoyt}, {Jang}, {Beaton}, {Lee}, {Monson}, {Neeley}, \& {Rich}}]{freedman2020a}
{Freedman}, W.~L., {Madore}, B.~F., {Hoyt}, T., {et~al.} 2020, \apj, 891, 57, \dodoi{10.3847/1538-4357/ab7339}

\bibitem[{{Garvin} {et~al.}(2022){Garvin}, {Kruk}, {Cornen}, {Bhatawdekar}, {Ca{\~n}ameras}, \& {Mer{\'\i}n}}]{Garvin_2022}
{Garvin}, E.~O., {Kruk}, S., {Cornen}, C., {et~al.} 2022, \aap, 667, A141, \dodoi{10.1051/0004-6361/202243745}

\bibitem[{{Goobar} {et~al.}(2023){Goobar}, {Johansson}, {Schulze}, {Arendse}, {Carracedo}, {Dhawan}, {M{\"o}rtsell}, {Fremling}, {Yan}, {Perley}, {Sollerman}, {Joseph}, {Hinds}, {Meynardie}, {Andreoni}, {Bellm}, {Bloom}, {Collett}, {Drake}, {Graham}, {Kasliwal}, {Kulkarni}, {Lemon}, {Miller}, {Neill}, {Nordin}, {Pierel}, {Richard}, {Riddle}, {Rigault}, {Rusholme}, {Sharma}, {Stein}, {Stewart}, {Townsend}, {Vinko}, {Wheeler}, \& {Wold}}]{Goobar_2023}
{Goobar}, A., {Johansson}, J., {Schulze}, S., {et~al.} 2023, Nature Astronomy, 7, 1098, \dodoi{10.1038/s41550-023-01981-3}

\bibitem[{{GRAVITY Collaboration} {et~al.}(2017){GRAVITY Collaboration}, {Abuter}, {Accardo}, {Amorim}, {Anugu}, {{\'A}vila}, {Azouaoui}, {Benisty}, {Berger}, {Blind}, {Bonnet}, {Bourget}, {Brandner}, {Brast}, {Buron}, {Burtscher}, {Cassaing}, {Chapron}, {Choquet}, {Cl{\'e}net}, {Collin}, {Coud{\'e} Du Foresto}, {de Wit}, {de Zeeuw}, {Deen}, {Delplancke-Str{\"o}bele}, {Dembet}, {Derie}, {Dexter}, {Duvert}, {Ebert}, {Eckart}, {Eisenhauer}, {Esselborn}, {F{\'e}dou}, {Finger}, {Garcia}, {Garcia Dabo}, {Garcia Lopez}, {Gendron}, {Genzel}, {Gillessen}, {Gonte}, {Gordo}, {Grould}, {Gr{\"o}zinger}, {Guieu}, {Haguenauer}, {Hans}, {Haubois}, {Haug}, {Haussmann}, {Henning}, {Hippler}, {Horrobin}, {Huber}, {Hubert}, {Hubin}, {Hummel}, {Jakob}, {Janssen}, {Jochum}, {Jocou}, {Kaufer}, {Kellner}, {Kendrew}, {Kern}, {Kervella}, {Kiekebusch}, {Klein}, {Kok}, {Kolb}, {Kulas}, {Lacour}, {Lapeyr{\`e}re}, {Lazareff}, {Le Bouquin}, {L{\`e}na}, {Lenzen}, {L{\'e}v{\^e}que}, {Lippa}, {Magnard}, {Mehrgan}, {Mellein}, {M{\'e}rand},
  {Moreno-Ventas}, {Moulin}, {M{\"u}ller}, {M{\"u}ller}, {Neumann}, {Oberti}, {Ott}, {Pallanca}, {Panduro}, {Pasquini}, {Paumard}, {Percheron}, {Perraut}, {Perrin}, {Pfl{\"u}ger}, {Pfuhl}, {Phan Duc}, {Plewa}, {Popovic}, {Rabien}, {Ram{\'\i}rez}, {Ramos}, {Rau}, {Riquelme}, {Rohloff}, {Rousset}, {Sanchez-Bermudez}, {Scheithauer}, {Sch{\"o}ller}, {Schuhler}, {Spyromilio}, {Straubmeier}, {Sturm}, {Suarez}, {Tristram}, {Ventura}, {Vincent}, {Waisberg}, {Wank}, {Weber}, {Wieprecht}, {Wiest}, {Wiezorrek}, {Wittkowski}, {Woillez}, {Wolff}, {Yazici}, {Ziegler}, \& {Zins}}]{GRAVITY_2017}
{GRAVITY Collaboration}, {Abuter}, R., {Accardo}, M., {et~al.} 2017, \aap, 602, A94, \dodoi{10.1051/0004-6361/201730838}

\bibitem[{Gómez {et~al.}(2022)Gómez, Traianou, Krichbaum, Lobanov, Fuentes, Lico, Zhao, Bruni, Kovalev, Lähteenmäki, Voitsik, Lisakov, Angelakis, Bach, Casadio, Cho, Dey, Gopakumar, Gurvits, Jorstad, Kovalev, Lister, Marscher, Myserlis, Pushkarev, Ros, Savolainen, Tornikoski, Valtonen, \& Zensus}]{Gomez_2022}
Gómez, J.~L., Traianou, E., Krichbaum, T.~P., {et~al.} 2022, The Astrophysical Journal, 924, 122, \dodoi{10.3847/1538-4357/ac3bcc}

\bibitem[{{Han} {et~al.}(2023){Han}, {Dey}, {Price-Whelan}, {Najita}, {Schlafly}, {Saydjari}, {Wechsler}, {Bonaca}, {Schlegel}, {Conroy}, {Raichoor}, {Drlica-Wagner}, {Kollmeier}, {Koposov}, {Besla}, {Rix}, {Goodman}, {Finkbeiner}, {Anand}, {Ashby}, {Bahr-Kalus}, {Beaton}, {Behera}, {Bell}, {Bellm}, {BenZvi}, {Beraldo e Silva}, {Birrer}, {Blanton}, {Bock}, {Broekgaarden}, {Brout}, {Brown}, {Brown}, {Bulbul}, {Calderon}, {Carlin}, {Carrillo}, {Castander}, {Chakraborty}, {Chandra}, {Chiang}, {Choi}, {Clark}, {Clarkson}, {Cooper}, {Crill}, {Cunha}, {Cunningham}, {Dalcanton}, {Danieli}, {Daylan}, {de Jong}, {DeRose}, {Dey}, {Dickinson}, {Dominguez}, {Dong}, {Eifler}, {El-Badry}, {Erkal}, {Escala}, {Fazio}, {Ferguson}, {Ferraro}, {Filion}, {Forero-Romero}, {Fu}, {Galbany}, {Garavito-Camargo}, {Gawiser}, {Geha}, {Gnedin}, {Gomez}, {Greene}, {Guy}, {Hadzhiyska}, {Hawkins}, {Heinrich}, {Hernquist}, {Hirata}, {Hora}, {Horowitz}, {Horta}, {Huang}, {Huang}, {Huanyuan}, {Hunt}, {Ibata}, {Jannuzi}, {Johnston}, {Jones},
  {Juneau}, {Kado-Fong}, {Kalari}, {Kallivayalil}, {Karim}, {Keeley}, {Khoperskov}, {Kim}, {Kov{\'a}cs}, {Krause}, {Kremer}, {Kremin}, {Krolewski}, {Kulkarni}, {Kuna}, {L'Huillier}, {Lacy}, {Lan}, {Lang}, {Leahy}, {Li}, {Lim}, {L{\'o}pez-Morales}, {Macri}, {Marc}, {Mau}, {McCarthy}, {McDonald}, {McQuinn}, {Meisner}, {Melnick}, {Merloni}, {Millard}, {Millon}, {Minchev}, {Montero-Camacho}, {Morales-Gutierrez}, {Morrell}, {Moustakas}, {Moustakas}, {Murray}, {Mutlu-Pakdil}, {Myeong}, {Myers}, {Nadler}, {Navarete}, {Ness}, {Nidever}, {Nikutta}, {Nushkia}, {Olsen}, {Pace}, {Pacucci}, {Padmanabhan}, {Parkinson}, {Pearson}, {Peng}, {Petric}, {Petric}, {Ratcliffe}, {Razieh}, {Reiprich}, {Rezaie}, {Ricci}, {Rich}, {Richstein}, {Riley}, {Rockosi}, {Rossi}, {Salvato}, {Samushia}, {Sanchez}, {Sand}, {E Sanderson}, {{\v{S}}ar{\v{c}}evi{\'c}}, {Sarkar}, {Savino}, {Schweizer}, {Shafieloo}, {Shengqi}, {Shields}, {Shipp}, {Simon}, {Siudek}, {Siwei}, {Slepian}, {Smith}, {Sobeck}, {Sohn}, {Som}, {Speagle}, {Spergel}, {Szabo},
  {Tan}, {Theissen}, {Tollerud}, {Tolls}, {Tran}, {Tsiane}, {Vacca}, {Valluri}, {Verberi}, {Warfield}, {Weaverdyck}, {Weiner}, {Weisz}, {Wetzel}, {White}, {Williams}, {Wolk}, {Wu}, {Wyse}, {Yang}, {Zaritsky}, {Zelko}, {Zhimin}, \& {Zucker}}]{Han_2023}
{Han}, J.~J., {Dey}, A., {Price-Whelan}, A.~M., {et~al.} 2023, arXiv e-prints, arXiv:2306.11784, \dodoi{10.48550/arXiv.2306.11784}

\bibitem[{{Hoffman} {et~al.}(2021){Hoffman}, {Mack}, {et~al.}}]{Hoffman_2021}
{Hoffman}, S.~L., {Mack}, J., {et~al.} 2021, The {DrizzlePac} {Handbook}, 2nd edn. (Baltimore: STScI)

\bibitem[{{Hoyle} {et~al.}(2012){Hoyle}, {Masters}, {Nichol}, {Jimenez}, \& {Bamford}}]{Hoyle_2012}
{Hoyle}, B., {Masters}, K.~L., {Nichol}, R.~C., {Jimenez}, R., \& {Bamford}, S.~P. 2012, \mnras, 423, 3478, \dodoi{10.1111/j.1365-2966.2012.21146.x}

\bibitem[{{Huang} {et~al.}(2020){Huang}, {Storfer}, {Ravi}, {Pilon}, {Domingo}, {Schlegel}, {Bailey}, {Dey}, {Gupta}, {Herrera}, {Juneau}, {Landriau}, {Lang}, {Meisner}, {Moustakas}, {Myers}, {Schlafly}, {Valdes}, {Weaver}, {Yang}, \& {Y{\`e}che}}]{Huang_2020}
{Huang}, X., {Storfer}, C., {Ravi}, V., {et~al.} 2020, \apj, 894, 78, \dodoi{10.3847/1538-4357/ab7ffb}

\bibitem[{{Huang} {et~al.}(2021){Huang}, {Storfer}, {Gu}, {Ravi}, {Pilon}, {Sheu}, {Venguswamy}, {Banka}, {Dey}, {Landriau}, {Lang}, {Meisner}, {Moustakas}, {Myers}, {Sajith}, {Schlafly}, \& {Schlegel}}]{Huang_2021}
{Huang}, X., {Storfer}, C., {Gu}, A., {et~al.} 2021, \apj, 909, 27, \dodoi{10.3847/1538-4357/abd62b}

\bibitem[{{Izquierdo-Villalba} {et~al.}(2023){Izquierdo-Villalba}, {Colpi}, {Volonteri}, {Spinoso}, {Bonoli}, \& {Sesana}}]{Izquierdo-Villalba_2023}
{Izquierdo-Villalba}, D., {Colpi}, M., {Volonteri}, M., {et~al.} 2023, \aap, 677, A123, \dodoi{10.1051/0004-6361/202347008}

\bibitem[{{Jacobs} {et~al.}(2017){Jacobs}, {Glazebrook}, {Collett}, {More}, \& {McCarthy}}]{jacobs2017a}
{Jacobs}, C., {Glazebrook}, K., {Collett}, T., {More}, A., \& {McCarthy}, C. 2017, \mnras, 471, 167, \dodoi{10.1093/mnras/stx1492}

\bibitem[{{Jacobs} {et~al.}(2019){Jacobs}, {Collett}, {Glazebrook}, {McCarthy}, {Qin}, {Abbott}, {Abdalla}, {Annis}, {Avila}, {Bechtol}, {Bertin}, {Brooks}, {Buckley-Geer}, {Burke}, {Carnero Rosell}, {Carrasco Kind}, {Carretero}, {da Costa}, {Davis}, {De Vicente}, {Desai}, {Diehl}, {Doel}, {Eifler}, {Flaugher}, {Frieman}, {Garc{\'{\i}}a-Bellido}, {Gaztanaga}, {Gerdes}, {Goldstein}, {Gruen}, {Gruendl}, {Gschwend}, {Gutierrez}, {Hartley}, {Hollowood}, {Honscheid}, {Hoyle}, {James}, {Kuehn}, {Kuropatkin}, {Lahav}, {Li}, {Lima}, {Lin}, {Maia}, {Martini}, {Miller}, {Miquel}, {Nord}, {Plazas}, {Sanchez}, {Scarpine}, {Schubnell}, {Serrano}, {Sevilla-Noarbe}, {Smith}, {Soares-Santos}, {Sobreira}, {Suchyta}, {Swanson}, {Tarle}, {Vikram}, {Walker}, {Zhang}, \& {Zuntz}}]{jacobs2019a}
{Jacobs}, C., {Collett}, T., {Glazebrook}, K., {et~al.} 2019, \mnras, 484, 5330, \dodoi{10.1093/mnras/stz272}

\bibitem[{{Jenkins} {et~al.}(2001){Jenkins}, {Frenk}, {White}, {Colberg}, {Cole}, {Evrard}, {Couchman}, \& {Yoshida}}]{jenkins2001a}
{Jenkins}, A., {Frenk}, C.~S., {White}, S.~D.~M., {et~al.} 2001, \mnras, 321, 372, \dodoi{10.1046/j.1365-8711.2001.04029.x}

\bibitem[{{Kaplinghat} {et~al.}(2016){Kaplinghat}, {Tulin}, \& {Yu}}]{kaplinghat2016a}
{Kaplinghat}, M., {Tulin}, S., \& {Yu}, H.-B. 2016, \prl, 116, 041302, \dodoi{10.1103/PhysRevLett.116.041302}

\bibitem[{{Kelly} {et~al.}(2023){Kelly}, {Rodney}, {Treu}, {Oguri}, {Chen}, {Zitrin}, {Birrer}, {Bonvin}, {Dessart}, {Diego}, {Filippenko}, {Foley}, {Gilman}, {Hjorth}, {Jauzac}, {Mandel}, {Millon}, {Pierel}, {Sharon}, {Thorp}, {Williams}, {Broadhurst}, {Dressler}, {Graur}, {Jha}, {McCully}, {Postman}, {Schmidt}, {Tucker}, \& {von der Linden}}]{kelly2023a}
{Kelly}, P.~L., {Rodney}, S., {Treu}, T., {et~al.} 2023, Science, 380, abh1322, \dodoi{10.1126/science.abh1322}

\bibitem[{{Korytov} {et~al.}(2019){Korytov}, {Hearin}, {Kovacs}, {Larsen}, {Rangel}, {Hollowed}, {Benson}, {Heitmann}, {Mao}, {Bahmanyar}, {Chang}, {Campbell}, {DeRose}, {Finkel}, {Frontiere}, {Gawiser}, {Habib}, {Joachimi}, {Lanusse}, {Li}, {Mandelbaum}, {Morrison}, {Newman}, {Pope}, {Rykoff}, {Simet}, {To}, {Vikraman}, {Wechsler}, {White}, \& {(The LSST Dark Energy Science Collaboration}}]{Korytov_2019}
{Korytov}, D., {Hearin}, A., {Kovacs}, E., {et~al.} 2019, \apjs, 245, 26, \dodoi{10.3847/1538-4365/ab510c}

\bibitem[{{Lang} {et~al.}(2016){Lang}, {Hogg}, \& {Mykytyn}}]{Lang_2016}
{Lang}, D., {Hogg}, D.~W., \& {Mykytyn}, D. 2016, {The Tractor: Probabilistic astronomical source detection and measurement}, Astrophysics Source Code Library, record ascl:1604.008

\bibitem[{{Lanusse} {et~al.}(2018){Lanusse}, {Ma}, {Li}, {Collett}, {Li}, {Ravanbakhsh}, {Mandelbaum}, \& {P{\'o}czos}}]{lanusse2018a}
{Lanusse}, F., {Ma}, Q., {Li}, N., {et~al.} 2018, \mnras, 473, 3895, \dodoi{10.1093/mnras/stx1665}

\bibitem[{{Li} {et~al.}(2024{\natexlab{a}}){Li}, {Collett}, {Krawczyk}, \& {Enzi}}]{li2024a}
{Li}, T., {Collett}, T.~E., {Krawczyk}, C.~M., \& {Enzi}, W. 2024{\natexlab{a}}, \mnras, 527, 5311, \dodoi{10.1093/mnras/stad3514}

\bibitem[{{Li} {et~al.}(2024{\natexlab{b}}){Li}, {Collett}, {Marshall}, {Erickson}, {Enzi}, {Oldham}, \& {Ballard}}]{li2024b}
{Li}, T., {Collett}, T.~E., {Marshall}, P.~J., {et~al.} 2024{\natexlab{b}}, arXiv e-prints, arXiv:2410.16171, \dodoi{10.48550/arXiv.2410.16171}

\bibitem[{{Liao} {et~al.}(2017){Liao}, {Fan}, {Ding}, {Biesiada}, \& {Zhu}}]{liao2017a}
{Liao}, K., {Fan}, X.-L., {Ding}, X., {Biesiada}, M., \& {Zhu}, Z.-H. 2017, Nature Communications, 8, 1148, \dodoi{10.1038/s41467-017-01152-9}

\bibitem[{{Linder}(2011)}]{linder2011a}
{Linder}, E.~V. 2011, \prd, 84, 123529, \dodoi{10.1103/PhysRevD.84.123529}

\bibitem[{{Linder}(2016)}]{Linder_2016}
---. 2016, \prd, 94, 083510, \dodoi{10.1103/PhysRevD.94.083510}

\bibitem[{{Marshall} {et~al.}(2009){Marshall}, {Hogg}, {Moustakas}, {Fassnacht}, {Brada{\v{c}}}, {Schrabback}, \& {Blandford}}]{marshall2009a}
{Marshall}, P.~J., {Hogg}, D.~W., {Moustakas}, L.~A., {et~al.} 2009, \apj, 694, 924, \dodoi{10.1088/0004-637X/694/2/924}

\bibitem[{{M{\'e}ndez-Abreu} {et~al.}(2017){M{\'e}ndez-Abreu}, {Ruiz-Lara}, {S{\'a}nchez-Menguiano}, {de Lorenzo-C{\'a}ceres}, {Costantin}, {Catal{\'a}n-Torrecilla}, {Florido}, {Aguerri}, {Bland-Hawthorn}, {Corsini}, {Dettmar}, {Galbany}, {Garc{\'\i}a-Benito}, {Marino}, {M{\'a}rquez}, {Ortega-Minakata}, {Papaderos}, {S{\'a}nchez}, {S{\'a}nchez-Blazquez}, {Spekkens}, {van de Ven}, {Wild}, \& {Ziegler}}]{Mendez-Abreu_2017}
{M{\'e}ndez-Abreu}, J., {Ruiz-Lara}, T., {S{\'a}nchez-Menguiano}, L., {et~al.} 2017, \aap, 598, A32, \dodoi{10.1051/0004-6361/201629525}

\bibitem[{{Meneghetti} {et~al.}(2020){Meneghetti}, {Davoli}, {Bergamini}, {Rosati}, {Natarajan}, {Giocoli}, {Caminha}, {Metcalf}, {Rasia}, {Borgani}, {Calura}, {Grillo}, {Mercurio}, \& {Vanzella}}]{meneghetti2020a}
{Meneghetti}, M., {Davoli}, G., {Bergamini}, P., {et~al.} 2020, Science, 369, 1347, \dodoi{10.1126/science.aax5164}

\bibitem[{{Metcalf} {et~al.}(2019){Metcalf}, {Meneghetti}, {Avestruz}, {Bellagamba}, {Bom}, {Bertin}, {Cabanac}, {Courbin}, {Davies}, {Decenci{\`e}re}, {Flamary}, {Gavazzi}, {Geiger}, {Hartley}, {Huertas-Company}, {Jackson}, {Jacobs}, {Jullo}, {Kneib}, {Koopmans}, {Lanusse}, {Li}, {Ma}, {Makler}, {Li}, {Lightman}, {Petrillo}, {Serjeant}, {Sch{\"a}fer}, {Sonnenfeld}, {Tagore}, {Tortora}, {Tuccillo}, {Valent{\'\i}n}, {Velasco-Forero}, {Verdoes Kleijn}, \& {Vernardos}}]{Metcalf_2019}
{Metcalf}, R.~B., {Meneghetti}, M., {Avestruz}, C., {et~al.} 2019, \aap, 625, A119, \dodoi{10.1051/0004-6361/201832797}

\bibitem[{{Narayan} \& {Bartelmann}(1996)}]{Narayan_1996}
{Narayan}, R., \& {Bartelmann}, M. 1996, arXiv e-prints, astro, \dodoi{10.48550/arXiv.astro-ph/9606001}

\bibitem[{{Navarro} {et~al.}(1996){Navarro}, {Frenk}, \& {White}}]{navarro1996a}
{Navarro}, J.~F., {Frenk}, C.~S., \& {White}, S. D.~M. 1996, \apj, 462, 563, \dodoi{10.1086/177173}

\bibitem[{{Newton} {et~al.}(2011){Newton}, {Marshall}, {Treu}, {Auger}, {Gavazzi}, {Bolton}, {Koopmans}, \& {Moustakas}}]{Newton_2011}
{Newton}, E.~R., {Marshall}, P.~J., {Treu}, T., {et~al.} 2011, \apj, 734, 104, \dodoi{10.1088/0004-637X/734/2/104}

\bibitem[{{Olofsson} {et~al.}(2011){Olofsson}, {Benisty}, {Augereau}, {Pinte}, {M{\'e}nard}, {Tatulli}, {Berger}, {Malbet}, {Mer{\'\i}n}, {van Dishoeck}, {Lacour}, {Pontoppidan}, {Monin}, {Brown}, \& {Blake}}]{Olofsson_2011}
{Olofsson}, J., {Benisty}, M., {Augereau}, J.~C., {et~al.} 2011, \aap, 528, L6, \dodoi{10.1051/0004-6361/201016074}

\bibitem[{{Ostdiek} {et~al.}(2022a){Ostdiek}, {Diaz Rivero}, \& {Dvorkin}}]{Ostdiek_2022_1}
{Ostdiek}, B., {Diaz Rivero}, A., \& {Dvorkin}, C. 2022a, \aap, 657, L14, \dodoi{10.1051/0004-6361/202142030}

\bibitem[{Ostdiek {et~al.}(2022b)Ostdiek, Rivero, \& Dvorkin}]{Ostdiek_2022_2}
Ostdiek, B., Rivero, A.~D., \& Dvorkin, C. 2022b, The Astrophysical Journal, 927, 83, \dodoi{10.3847/1538-4357/ac2d8d}

\bibitem[{{Pierel} {et~al.}(2023){Pierel}, {Arendse}, {Ertl}, {Huang}, {Moustakas}, {Schuldt}, {Shajib}, {Shu}, {Birrer}, {Bronikowski}, {Hjorth}, {Suyu}, {Agarwal}, {Agnello}, {Bolton}, {Chakrabarti}, {Cold}, {Courbin}, {Della Costa}, {Dhawan}, {Engesser}, {Fox}, {Gall}, {Gomez}, {Goobar}, {Jha}, {Jimenez}, {Johansson}, {Larison}, {Li}, {Marques-Chaves}, {Mao}, {Mazzali}, {Perez-Fournon}, {Petrushevska}, {Poidevin}, {Rest}, {Sheu}, {Shirley}, {Silver}, {Storfer}, {Strolger}, {Treu}, {Wojtak}, \& {Zenati}}]{Pierel_2023}
{Pierel}, J.~D.~R., {Arendse}, N., {Ertl}, S., {et~al.} 2023, \apj, 948, 115, \dodoi{10.3847/1538-4357/acc7a6}

\bibitem[{{Planck Collaboration} {et~al.}(2020){Planck Collaboration}, {Aghanim}, {Akrami}, {Ashdown}, {Aumont}, {Baccigalupi}, \& {Ballardini}}]{planck2020a}
{Planck Collaboration}, {Aghanim}, N., {Akrami}, Y., {et~al.} 2020, \aap, 641, A6, \dodoi{10.1051/0004-6361/201833910}

\bibitem[{{Pritchet} {et~al.}(2024){Pritchet}, {Thanjavur}, {Bottrell}, \& {Gao}}]{Pritchet_2024}
{Pritchet}, C., {Thanjavur}, K., {Bottrell}, C., \& {Gao}, Y. 2024, \aj, 167, 131, \dodoi{10.3847/1538-3881/ad234b}

\bibitem[{{R{\"a}s{\"a}nen} {et~al.}(2015){R{\"a}s{\"a}nen}, {Bolejko}, \& {Finoguenov}}]{rasanen2015a}
{R{\"a}s{\"a}nen}, S., {Bolejko}, K., \& {Finoguenov}, A. 2015, \prl, 115, 101301, \dodoi{10.1103/PhysRevLett.115.101301}

\bibitem[{{Refsdal}(1964)}]{refsdal1964a}
{Refsdal}, S. 1964, \mnras, 128, 307, \dodoi{10.1093/mnras/128.4.307}

\bibitem[{{Riess} {et~al.}(2019){Riess}, {Casertano}, {Yuan}, {Macri}, \& {Scolnic}}]{riess2019a}
{Riess}, A.~G., {Casertano}, S., {Yuan}, W., {Macri}, L.~M., \& {Scolnic}, D. 2019, \apj, 876, 85, \dodoi{10.3847/1538-4357/ab1422}

\bibitem[{{Ronneberger} {et~al.}(2015){Ronneberger}, {Fischer}, \& {Brox}}]{Ronnegerger_2015}
{Ronneberger}, O., {Fischer}, P., \& {Brox}, T. 2015, arXiv e-prints, arXiv:1505.04597, \dodoi{10.48550/arXiv.1505.04597}

\bibitem[{{Rubin} {et~al.}(2021){Rubin}, {Cikota}, {Aldering}, {Fruchter}, {Perlmutter}, \& {Sako}}]{Rubin_2021}
{Rubin}, D., {Cikota}, A., {Aldering}, G., {et~al.} 2021, \pasp, 133, 064001, \dodoi{10.1088/1538-3873/abf406}

\bibitem[{{Saintonge} {et~al.}(2005){Saintonge}, {Schade}, {Ellingson}, {Yee}, \& {Carlberg}}]{Saintonge_2005}
{Saintonge}, A., {Schade}, D., {Ellingson}, E., {Yee}, H.~K.~C., \& {Carlberg}, R.~G. 2005, \apjs, 157, 228, \dodoi{10.1086/427939}

\bibitem[{{Sawala} {et~al.}(2024){Sawala}, {Frenk}, {Jasche}, {Johansson}, \& {Lavaux}}]{Sawala_2024}
{Sawala}, T., {Frenk}, C., {Jasche}, J., {Johansson}, P.~H., \& {Lavaux}, G. 2024, Nature Astronomy, 8, 247, \dodoi{10.1038/s41550-023-02130-6}

\bibitem[{{S\'{e}rsic}(1968)}]{Sersic_1968}
{S\'{e}rsic}, J.~L. 1968, {Atlas de Galaxias Australes} (Córdoba, Argentina: Observatorio Astronomico)

\bibitem[{{Sesana} {et~al.}(2014){Sesana}, {Barausse}, {Dotti}, \& {Rossi}}]{Sesana_2014}
{Sesana}, A., {Barausse}, E., {Dotti}, M., \& {Rossi}, E.~M. 2014, \apj, 794, 104, \dodoi{10.1088/0004-637X/794/2/104}

\bibitem[{{Shajib} {et~al.}(2021){Shajib}, {Treu}, {Birrer}, \& {Sonnenfeld}}]{shajib2021a}
{Shajib}, A.~J., {Treu}, T., {Birrer}, S., \& {Sonnenfeld}, A. 2021, \mnras, 503, 2380, \dodoi{10.1093/mnras/stab536}

\bibitem[{{Sharma} {et~al.}(2023){Sharma}, {Collett}, \& {Linder}}]{Sharma_2023}
{Sharma}, D., {Collett}, T.~E., \& {Linder}, E.~V. 2023, \jcap, 2023, 001, \dodoi{10.1088/1475-7516/2023/04/001}

\bibitem[{{Sharma} \& {Linder}(2022)}]{Sharma_2022}
{Sharma}, D., \& {Linder}, E.~V. 2022, \jcap, 2022, 033, \dodoi{10.1088/1475-7516/2022/07/033}

\bibitem[{{Sheu} {et~al.}(2024){Sheu}, {Cikota}, {Huang}, {Glazebrook}, {Storfer}, {Agarwal}, {Schlegel}, {Suzuki}, {Barone}, {Bian}, {Jeltema}, {Jones}, {Kacprzak}, {O'Donnell}, \& {Keerthi Vasan}}]{sheu2024a}
{Sheu}, W., {Cikota}, A., {Huang}, X., {et~al.} 2024, \apj, 973, 3, \dodoi{10.3847/1538-4357/ad65d3}

\bibitem[{Shu {et~al.}(2018)Shu, Bolton, Mao, Kang, Li, \& Soraisam}]{Shu_2018}
Shu, Y., Bolton, A.~S., Mao, S., {et~al.} 2018, The Astrophysical Journal, 864, 91, \dodoi{10.3847/1538-4357/aad5ea}

\bibitem[{{Shu} {et~al.}(2022){Shu}, {Ca{\~n}ameras}, {Schuldt}, {Suyu}, {Taubenberger}, {Inoue}, \& {Jaelani}}]{shu2022a}
{Shu}, Y., {Ca{\~n}ameras}, R., {Schuldt}, S., {et~al.} 2022, \aap, 662, A4, \dodoi{10.1051/0004-6361/202243203}

\bibitem[{{Shuntov} {et~al.}(2022){Shuntov}, {McCracken}, {Gavazzi}, {Laigle}, {Weaver}, {Davidzon}, {Ilbert}, {Kauffmann}, {Faisst}, {Dubois}, {Koekemoer}, {Moneti}, {Milvang-Jensen}, {Mobasher}, {Sanders}, \& {Toft}}]{Shuntov_2022}
{Shuntov}, M., {McCracken}, H.~J., {Gavazzi}, R., {et~al.} 2022, \aap, 664, A61, \dodoi{10.1051/0004-6361/202243136}

\bibitem[{{Simard} {et~al.}(2011){Simard}, {Mendel}, {Patton}, {Ellison}, \& {McConnachie}}]{Simard_2011}
{Simard}, L., {Mendel}, J.~T., {Patton}, D.~R., {Ellison}, S.~L., \& {McConnachie}, A.~W. 2011, \apjs, 196, 11, \dodoi{10.1088/0067-0049/196/1/11}

\bibitem[{{Simons} {et~al.}(2019){Simons}, {Kassin}, {Snyder}, {Primack}, {Ceverino}, {Dekel}, {Hayward}, {Mandelker}, {Mantha}, {Pacifici}, {de la Vega}, \& {Wang}}]{Simons_2019}
{Simons}, R.~C., {Kassin}, S.~A., {Snyder}, G.~F., {et~al.} 2019, \apj, 874, 59, \dodoi{10.3847/1538-4357/ab07c9}

\bibitem[{{Snyder}(2018)}]{Snyder_2018}
{Snyder}, G. 2018, Vela-{Sunrise} {Mock} {Observations}, \dodoi{doi:10.17909/t9-ge0b-jm58}

\bibitem[{Snyder {et~al.}(2015)Snyder, Lotz, Moody, Peth, Freeman, Ceverino, Primack, \& Dekel}]{Snyder_2015}
Snyder, G.~F., Lotz, J., Moody, C., {et~al.} 2015, Monthly Notices of the Royal Astronomical Society, 451, 4290, \dodoi{10.1093/mnras/stv1231}

\bibitem[{{Springel} {et~al.}(2008){Springel}, {Wang}, {Vogelsberger}, {Ludlow}, {Jenkins}, {Helmi}, {Navarro}, {Frenk}, \& {White}}]{springel2008a}
{Springel}, V., {Wang}, J., {Vogelsberger}, M., {et~al.} 2008, \mnras, 391, 1685, \dodoi{10.1111/j.1365-2966.2008.14066.x}

\bibitem[{{Storfer} {et~al.}(2024){Storfer}, {Huang}, {Gu}, {Sheu}, {Banka}, {Dey}, {Jain}, {Kwon}, {Lang}, {Lee}, {Meisner}, {Moustakas}, {Myers}, {Tabares-Tarquinio}, {Schlafly}, \& {Schlegel}}]{Storfer_2024}
{Storfer}, C., {Huang}, X., {Gu}, A., {et~al.} 2024, ApJS (accepted), arXiv:2206.02764, \dodoi{10.48550/arXiv.2206.02764}

\bibitem[{{Stoughton} {et~al.}(2002){Stoughton}, {Lupton}, {Bernardi}, {Blanton}, {Burles}, {Castander}, {Connolly}, {Eisenstein}, {Frieman}, {Hennessy}, {Hindsley}, {Ivezi{\'c}}, {Kent}, {Kunszt}, {Lee}, {Meiksin}, {Munn}, {Newberg}, {Nichol}, {Nicinski}, {Pier}, {Richards}, {Richmond}, {Schlegel}, {Smith}, {Strauss}, {SubbaRao}, {Szalay}, {Thakar}, {Tucker}, {Vanden Berk}, {Yanny}, {Adelman}, {Anderson}, {Anderson}, {Annis}, {Bahcall}, {Bakken}, {Bartelmann}, {Bastian}, {Bauer}, {Berman}, {B{\"o}hringer}, {Boroski}, {Bracker}, {Briegel}, {Briggs}, {Brinkmann}, {Brunner}, {Carey}, {Carr}, {Chen}, {Christian}, {Colestock}, {Crocker}, {Csabai}, {Czarapata}, {Dalcanton}, {Davidsen}, {Davis}, {Dehnen}, {Dodelson}, {Doi}, {Dombeck}, {Donahue}, {Ellman}, {Elms}, {Evans}, {Eyer}, {Fan}, {Federwitz}, {Friedman}, {Fukugita}, {Gal}, {Gillespie}, {Glazebrook}, {Gray}, {Grebel}, {Greenawalt}, {Greene}, {Gunn}, {de Haas}, {Haiman}, {Haldeman}, {Hall}, {Hamabe}, {Hansen}, {Harris}, {Harris}, {Harvanek}, {Hawley}, {Hayes},
  {Heckman}, {Helmi}, {Henden}, {Hogan}, {Hogg}, {Holmgren}, {Holtzman}, {Huang}, {Hull}, {Ichikawa}, {Ichikawa}, {Johnston}, {Kauffmann}, {Kim}, {Kimball}, {Kinney}, {Klaene}, {Kleinman}, {Klypin}, {Knapp}, {Korienek}, {Krolik}, {Kron}, {Krzesi{\'n}ski}, {Lamb}, {Leger}, {Limmongkol}, {Lindenmeyer}, {Long}, {Loomis}, {Loveday}, {MacKinnon}, {Mannery}, {Mantsch}, {Margon}, {McGehee}, {McKay}, {McLean}, {Menou}, {Merelli}, {Mo}, {Monet}, {Nakamura}, {Narayanan}, {Nash}, {Neilsen}, {Newman}, {Nitta}, {Odenkirchen}, {Okada}, {Okamura}, {Ostriker}, {Owen}, {Pauls}, {Peoples}, {Peterson}, {Petravick}, {Pope}, {Pordes}, {Postman}, {Prosapio}, {Quinn}, {Rechenmacher}, {Rivetta}, {Rix}, {Rockosi}, {Rosner}, {Ruthmansdorfer}, {Sandford}, {Schneider}, {Scranton}, {Sekiguchi}, {Sergey}, {Sheth}, {Shimasaku}, {Smee}, {Snedden}, {Stebbins}, {Stubbs}, {Szapudi}, {Szkody}, {Szokoly}, {Tabachnik}, {Tsvetanov}, {Uomoto}, {Vogeley}, {Voges}, {Waddell}, {Walterbos}, {Wang}, {Watanabe}, {Weinberg}, {White}, {White}, {Wilhite},
  {Wolfe}, {Yasuda}, {York}, {Zehavi}, \& {Zheng}}]{Stoughton_2002}
{Stoughton}, C., {Lupton}, R.~H., {Bernardi}, M., {et~al.} 2002, \aj, 123, 485, \dodoi{10.1086/324741}

\bibitem[{{Suyu} {et~al.}(2024){Suyu}, {Goobar}, {Collett}, {More}, \& {Vernardos}}]{suyu2024a}
{Suyu}, S.~H., {Goobar}, A., {Collett}, T., {More}, A., \& {Vernardos}, G. 2024, \ssr, 220, 13, \dodoi{10.1007/s11214-024-01044-7}

\bibitem[{{Taylor} {et~al.}(2022){Taylor}, {Bezanson}, {van der Wel}, {Pearl}, {Bell}, {D'Eugenio}, {Franx}, {Maseda}, {Muzzin}, {Sobral}, {Straatman}, {Whitaker}, \& {Wu}}]{Taylor_2022}
{Taylor}, L., {Bezanson}, R., {van der Wel}, A., {et~al.} 2022, \apj, 939, 90, \dodoi{10.3847/1538-4357/ac9796}

\bibitem[{{Thatte} {et~al.}(2021){Thatte}, {Tecza}, {Schnetler}, {Neichel}, {Melotte}, {Fusco}, {Ferraro-Wood}, {Clarke}, {Bryson}, {O'Brien}, {Mateo}, {Garcia Lorenzo}, {Evans}, {Bouch{\'e}}, {Arribas}, \& {HARMONI Consortium}}]{Thatte_2021}
{Thatte}, N., {Tecza}, M., {Schnetler}, H., {et~al.} 2021, The Messenger, 182, 7, \dodoi{10.18727/0722-6691/5215}

\bibitem[{{Treu} \& {Marshall}(2016)}]{treu2016a}
{Treu}, T., \& {Marshall}, P.~J. 2016, Astronomy and Astrophysics Review, 24, 11, \dodoi{10.1007/s00159-016-0096-8}

\bibitem[{{van Dokkum}(2001)}]{van_Dokkum_2001}
{van Dokkum}, P.~G. 2001, \pasp, 113, 1420, \dodoi{10.1086/323894}

\bibitem[{{Vanzella} {et~al.}(2020){Vanzella}, {Meneghetti}, {Caminha}, {Castellano}, {Calura}, {Rosati}, {Grillo}, {Dijkstra}, {Gronke}, {Sani}, {Mercurio}, {Tozzi}, {Nonino}, {Cristiani}, {Mignoli}, {Pentericci}, {Gilli}, {Treu}, {Caputi}, {Cupani}, {Fontana}, {Grazian}, \& {Balestra}}]{vanzella2020a}
{Vanzella}, E., {Meneghetti}, M., {Caminha}, G.~B., {et~al.} 2020, \mnras, 494, L81, \dodoi{10.1093/mnrasl/slaa041}

\bibitem[{{Vegetti} {et~al.}(2014){Vegetti}, {Koopmans}, {Auger}, {Treu}, \& {Bolton}}]{vegetti2014a}
{Vegetti}, S., {Koopmans}, L.~V.~E., {Auger}, M.~W., {Treu}, T., \& {Bolton}, A.~S. 2014, \mnras, 442, 2017, \dodoi{10.1093/mnras/stu943}

\bibitem[{{Vegetti} {et~al.}(2012){Vegetti}, {Lagattuta}, {McKean}, {Auger}, {Fassnacht}, \& {Koopmans}}]{vegetti2012a}
{Vegetti}, S., {Lagattuta}, D.~J., {McKean}, J.~P., {et~al.} 2012, \nat, 481, 341, \dodoi{10.1038/nature10669}

\bibitem[{Vegetti {et~al.}(2024)Vegetti, Birrer, Despali, Fassnacht, Gilman, Hezaveh, Perreault~Levasseur, McKean, Powell, O'Riordan, \& Vernardos}]{vegetti2024a}
Vegetti, S., Birrer, S., Despali, G., {et~al.} 2024, Space Science Reviews, 220, 58, \dodoi{10.1007/s11214-024-01087-w}

\bibitem[{{Wagner-Carena} {et~al.}(2023){Wagner-Carena}, {Aalbers}, {Birrer}, {Nadler}, {Darragh-Ford}, {Marshall}, \& {Wechsler}}]{Wagner-Carena_2022}
{Wagner-Carena}, S., {Aalbers}, J., {Birrer}, S., {et~al.} 2023, \apj, 942, 75, \dodoi{10.3847/1538-4357/aca525}

\bibitem[{{Warren} {et~al.}(2006){Warren}, {Abazajian}, {Holz}, \& {Teodoro}}]{warren2006a}
{Warren}, M.~S., {Abazajian}, K., {Holz}, D.~E., \& {Teodoro}, L. 2006, \apj, 646, 881, \dodoi{10.1086/504962}

\bibitem[{{Wei} \& {Wu}(2017)}]{wei2017a}
{Wei}, J.-J., \& {Wu}, X.-F. 2017, \mnras, 472, 2906, \dodoi{10.1093/mnras/stx2210}

\bibitem[{Williams {et~al.}(2018)Williams, Curtis-Lake, Hainline, Chevallard, Robertson, Charlot, Endsley, Stark, Willmer, Alberts, Amorin, Arribas, Baum, Bunker, Carniani, Crandall, Egami, Eisenstein, Ferruit, Husemann, Maseda, Maiolino, Rawle, Rieke, Smit, Tacchella, \& Willott}]{Williams_2018}
Williams, C.~C., Curtis-Lake, E., Hainline, K.~N., {et~al.} 2018, The Astrophysical Journal Supplement Series, 236, 33, \dodoi{10.3847/1538-4365/aabcbb}

\bibitem[{{Wilman} {et~al.}(2013){Wilman}, {Fontanot}, {De Lucia}, {Erwin}, \& {Monaco}}]{Wilman_2013}
{Wilman}, D.~J., {Fontanot}, F., {De Lucia}, G., {Erwin}, P., \& {Monaco}, P. 2013, \mnras, 433, 2986, \dodoi{10.1093/mnras/stt941}

\bibitem[{{Wright} {et~al.}(2016){Wright}, {Walth}, {Do}, {Marshall}, {Larkin}, {Moore}, {Adamkovics}, {Andersen}, {Armus}, {Barth}, {Cote}, {Cooke}, {Chisholm}, {Davidge}, {Dunn}, {Dumas}, {Ellerbroek}, {Ghez}, {Hao}, {Hayano}, {Liu}, {Lopez-Rodriguez}, {Lu}, {Mao}, {Marois}, {Pandey}, {Phillips}, {Schoeck}, {Subramaniam}, {Subramanian}, {Suzuki}, {Tan}, {Terai}, {Treu}, {Simard}, {Weiss}, {Wincentsen}, {Wong}, \& {Zhang}}]{Wright_2016}
{Wright}, S.~A., {Walth}, G., {Do}, T., {et~al.} 2016, in Society of Photo-Optical Instrumentation Engineers (SPIE) Conference Series, Vol. 9909, Adaptive Optics Systems V, ed. E.~{Marchetti}, L.~M. {Close}, \& J.-P. {V{\'e}ran}, 990905

\bibitem[{{Yue} {et~al.}(2022){Yue}, {Fan}, {Yang}, \& {Wang}}]{Yue_2022}
{Yue}, M., {Fan}, X., {Yang}, J., \& {Wang}, F. 2022, \apj, 925, 169, \dodoi{10.3847/1538-4357/ac409b}

\bibitem[{{Zahid} {et~al.}(2018){Zahid}, {Sohn}, \& {Geller}}]{Zahid_2018}
{Zahid}, H.~J., {Sohn}, J., \& {Geller}, M.~J. 2018, \apj, 859, 96, \dodoi{10.3847/1538-4357/aabe31}

\end{thebibliography}

\appendix
\section{Estimating fractions of lensing galaxies}\label{sec:lens_fractions}

\begin{deluxetable}{cccc}   \label{tab:lens_fractions}
    \caption{Strong Lenses as a Fraction of Elliptical Galaxies}
\tablehead{\colhead{Parameter} & \colhead{Added Cut on Foreground Galaxies} & \colhead{Added Cut on Lens Systems} & \colhead{Ratio}}
    \startdata
        Fiducial \emph{JWST} forecast & \nodata &  $\theta_E>0.5''$ & $1/72$ \\
        Velocity dispersion & $\sigma_v>100\textrm{km/s}$ & \nodata & $1/50$ \\
        Source magnitude & \nodata & $i_{s}<26$ ($i_{s}<27)\textsuperscript{\textdagger}$& $1/883\textrm{ }(1/348)\textsuperscript{\textdagger}$ \\
        Lens redshift & $z_l<0.4$ & \nodata & $1/53$ \\
        Source redshift & \nodata & $z_s<1.0$ & $1/248$ \\
        Einstein radius & \nodata & $\theta_E<1.5''$ & $1/281$ \\
        SFR & \nodata & SFR $>0.7M_\odot/\textrm{yr}$ & $1/1102$\\
    \enddata
    \tablecomments{\ed{This table estimates the fraction of elliptical galaxies that are strong lenses, as parameter cuts are cumulatively applied. We remind the reader that for our forecasts, $I>5\sigma_{BKG}$ for sources (\S~\ref{sec:methods_integration}) and $z_l<3.0$ 
    (\S~\ref{sec:methods_lens_sampling}). 
    The first row shows the fiducial \emph{JWST} forecast with $\theta_E>0.5''$ to focus on conventional galaxy-scale strong lensing. Then, we apply the cut $\sigma_v>100\textrm{km/s}$ to match C15. Then, $i_s<26$, $z_l<0.4$, and $z_s<1.0$ to match \cite{Newton_2011} (see text). $\theta_E<1.5''$ and $\textrm{SFR}>0.7M_\odot/\textrm{yr}$ are cuts specific to the SLACS lenses. The middle two columns represent what the parameter cuts are applied. The third column cuts are only applied to potential lenses, so they decrease the ratio, 
    while the second column parameter cuts are applied to the parent population of galaxies as well as potential lenses, so they could increase or decrease the ratio (here they happen to increase it compared with the previous step). \\\textsuperscript{\textdagger} This represents the source magnitude cut $i_s<27$ to match C15.}}
\end{deluxetable}

% \tablehead{\colhead{Parameter} & \colhead{Added Cut} & \colhead{Ratio}}
%     \startdata
%         Fiducial \emph{JWST} forecast & $\theta_E>0.5''$ & $1/72$ \\
%         Lens redshift & $z_l<0.4$ & $1/7^*$ \\
%         Velocity dispersion & $\sigma_v>100\textrm{km/s}$ & $1/5^*$ \\
%         Source magnitude & $i<27$ & $1/24$ \\
%         Source magnitude & $i<26$ & $1/53$ \\
%         Einstein radius & $\theta_E<1.5''$ & $1/65$ \\
%         Source redshift & $z_s<1.0$ & $1/281$ \\
%         SFR & SFR $>0.7M_\odot/\textrm{yr}$ & $1/1102$\\
%     \enddata

\ed{\cite{bolton2006a} noted that based on single-fiber strong lens searches using the spectroscopic observations of SDSS, the empirical frequency of an early-type galaxy being a lens is about $1/1000$. 
As a consistency check of our lens number forecasts, we apply reasonable selection cuts appropriate for the SLACS sample.
These cuts lead to a lens frequency that is in agreement with \cite{bolton2006a}.
% In order to better understand the selection effects that lead to this observation, we performed forecasts to try and roughly replicate this ratio. 
The results are summarized in Table~\ref{tab:lens_fractions}.}

\ed{To calculate the fraction of galaxies that are expected to be lenses, we must first decide the parent population. As with C15, we use a lower bound on velocity dispersion of $\sigma_v>100$ km/s. The observed SDSS sample of LRG's is flux-limited at $z\sim0.4$ \citep{Stoughton_2002}. Based on confirmed SLACS lenses, only one is at $z_l\sim0.5$, and the rest are at $z_l\lesssim0.4$ \citep{Newton_2011}, so we use a lens redshift bound of $z_l<0.4$ in Table~\ref{tab:lens_fractions}.}
%lens redshift bounds of $z_l<0.4$ and $z_l<0.5$ in Table~\ref{tab:lens_fractions}. This fraction of lenses becomes smaller for $z_l<0.5$ (compared with $z_l<0.4$. This is because for foreground galaxies with between redshifts of $0.4$ and $0.5$, with the cut of $z_s<1.0$, many relatively nearby sources no longer have $theta_E>0.5''$. Thus, the galaxies with $0.4<z_l<0.5$ have a lower fraction of lenses.

\ed{Based on the 3$''$ diameter aperture of the SDSS spectroscopic fibers, we use an Einstein radius range of $0.5''<\theta_E<1.5''$ as the requirement for lenses. We make other parameter cuts based on the observed SLACS lenses \citep{Newton_2011}. This includes an $i$-band source magnitude cut of $i<26$, because out of 46 definite lenses, only 4 are between 26$^\textrm{th}$ and 27$^\textrm{th}$ magnitude in $i$-band and each has a large magnification between 27 and 45, which are atypically high for galaxy-scale strong lenses. For source redshift, one is at $z_s=1.3$, but the rest are at $z_s\lesssim 1$, so we use a cut of $z_s<1.0$.}

\ed{Finally, a cut on the star-formation rate (SFR) is necessary for the lenses to be spectroscopically identified. From the minimum SFR of the SLACS lenses, we use a cut of ${\textrm{SFR}>0.7M_\odot/\textrm{yr}}$  \citep[][Figure~2]{Shu_2018}.\footnote{The minimum SFR increases slightly with redshift. If we apply a crude linear relationship of ${\textrm{SFR}>(2.6\times z_s-0.6)M_\odot/\textrm{yr}}$, which limits $\textrm{SFR}>0.7M_\odot/\textrm{yr}$ at $z_s=0.5$ and $\textrm{SFR}>2.0M_\odot/\textrm{yr}$ at $z_s=1.0$, this results in slightly lower ratios (e.g., for the last line of Table \ref{tab:lens_fractions}, we would obtain 1/1865).}}

\ed{We perform these calculations with the same methods as in Section $\ref{sec:forecast}$, using the C15 method of sampling velocity dispersion. As seen in Table~\ref{tab:lens_fractions}, 
% we are able to reach ratios close to the 
our results are consistent with the 1/1000 empirical estimate from \cite{bolton2006a}. However, we have only implemented these selection effects in a relatively simple manner, as this is intended to be an order of magnitude estimate. 
A more careful investigation would be needed to determine if simulations can fully replicate the observed fractions of early-type galaxies that are strong lenses.}

\section{Testing on single-exposure images}\label{sec:single_exp_test}

Here, we present our results on single-exposure \textit{HST} images. For this test, we use the observations from the \textit{HST} 
% observation ``Dark-matter halos and evolution of high-z early-type galaxies'' (ID: 
GO-10174 (PI: L. Koopmans), with a single exposure of just 420~sec, in F435W of WFC1.

Because these images are only single-exposure, we cannot use Drizzlepac to drizzle the images and remove cosmic rays. Instead, we use the LACosmic technique \citep{van_Dokkum_2001}\footnote{This is included in the Python library ccdproc,  \url{https://github.com/astropy/ccdproc}.} to identify and remove cosmic rays. The parameter ``sigclip'', which determines the cutoff for which pixels are counted as cosmic rays, was set to 2.5 through trial and error that determined it to be the value that most effectively removes the cosmic rays without significantly affecting non-cosmic ray pixels.

In order to account for the higher levels of noise, it was necessary to train a model using simulations with more noise than for the multi-exposure images (see Section \ref{sec:noise_preprocessing}). We call this model, Model~1b \edr{(HST-short)}. In Model 1b \edr{(HST-short)}, for single-exposure \textit{HST} images, $$\sigma_{BKG}\sim U(0.75,0.9)\textrm{ and }t_{exp}\sim 10^{U(1,4)}s$$

\begin{figure}
    \centering
    \includegraphics[width=.9\textwidth]{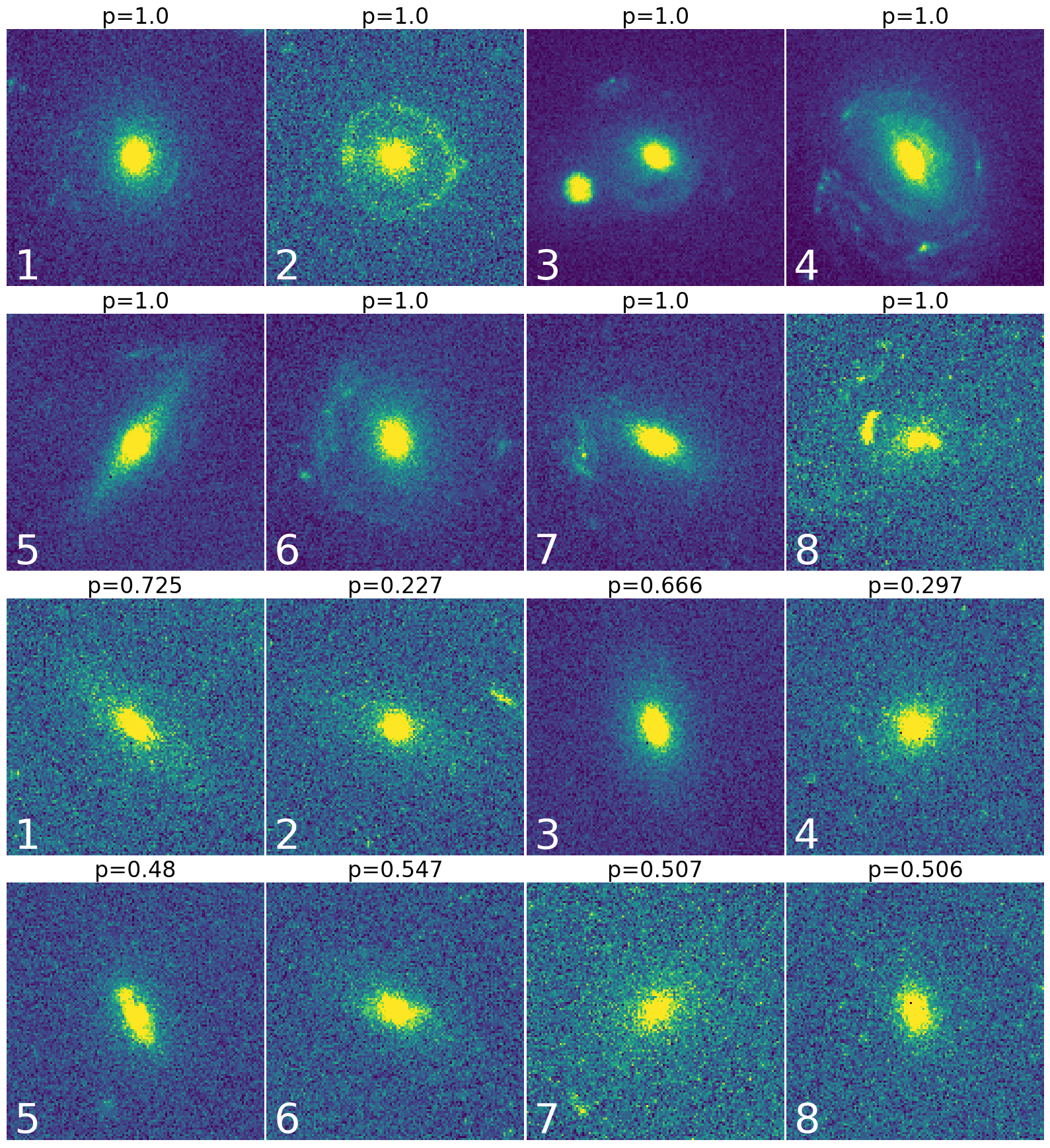}
    \caption{Trained Model 1b \edr{(HST-short)} tested on single-exposure lenses (top two rows) and non-lenses (bottom two rows). From \textit{HST}-GO 10174, with exposure time of 420s. The probability the model predicted is shown for each image.}
    \label{fig:single_exp}
\end{figure}

The result of Model 1b's \edr{(HST-short)} performance is shown in Figure \ref{fig:single_exp}. Although the model does not perform well with a probability threshold of $p=0.5$, it does still separate the range of probabilities for lenses and non-lenses, with the highest-probability non-lens at a value of 66.6\% and the lowest probability lens at a value of 99.96\%, suggesting a properly scaled threshold of anywhere from $p\sim0.9$ to $p\sim0.999$ would work well.
%which might be ideal.

One variable that can affect the performance of the model is the selection of non-lenses.
The non-lens images used in the training sample each have an elliptical galaxy at the center.
For \textit{HST} images with long exposure times ($\gtrsim$ 4800 sec), 
the model performs well even when ``out-of-sample'' non-lenses (i.e., those that are very different from the simulated ellipticals used in the training sample) are included.
For example, non-elliptical galaxies, very faint galaxies, and galaxies with other bright environmental galaxies nearby, some of which are shown in Figure \ref{fig:vegetti}. 
For single-exposure images, the model appears more sensitive to deviations from the simulated elliptical non-lenses. 
For this set of non-lenses, we only select elliptical galaxies and impose a brightness cut of 
% an upper magnitude bound of 
$<22$-mag (in the F435W filter), which is still fainter than the faintest lens, but we want to exclude non-lenses that are significantly fainter than the lens sample.
The morphology and magnitude cuts can be easily implemented for observed data.

\ed{Here we emphasize that a) all images in this test are out of sample, given that for these single-exposure images, cosmic rays had to be removed by LACosmic and b) these images all have a very short exposure exposure time of 420~sec on a rather old instrument. 
The vast majority of the \hst observations have higher quality images. Furthermore, we can expect all future observations on \hst, \jwst and \rst will have much higher quality images than these. We performed this test to see how far we can push our ResNet model, and we demonstrate its superb capability: even on this set, it performs well.}

\section{Truncated Lens Models}\label{sec:trunc_lens_models}

\subsection{Motivation and Methods}

One possible concern one could have for our smallest Einstein radius ResNet models and the U-Net model is whether they are properly distinguishing the lenses from non-lenses using the lensing effect. In particular, one non-physical issue with the singular isothermal models currently used in this work and studies from many other groups is that they cause non-vanishing lensing effects far from the lens. In the spherical SIS model, the magnitude of the deflection angle is a constant ($=\theta_E$) even far from the lens \citep[e.g.,][]{Narayan_1996}; this represents an unphysical artifact. 
Thus, for very small lenses ($\theta_E\lesssim 0.1''$), parts of the source many times $\theta_E$ away from the lens would be subject to this artifact.
% For large lenses ($\theta_E\sim1.0''$), this is not much of a concern because the lensed source will be only $\sim \theta_E$ away from the lens. However, for very small lenses ($\theta_E\lesssim 0.1''$), parts of the source could be many times $\theta_E$ away from the lens.

In order to test whether the ResNet or U-Net is being biased by the presence of these non-physical far-away lensing effects, once again, without retraining, we evaluate our ResNet and U-Net models on truncated lens simulations, which have lensing effect fully removed beyond some distance.

\begin{figure}
    \centering
    \includegraphics[width=.9\textwidth]{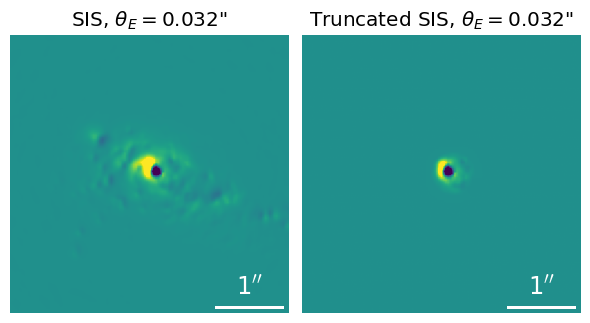}
    \caption{Example of the lensing effect of a lens with Einstein radius $\theta_E=0.032''$, shown using a standard SIS and a truncated SIS lens model.}
    \label{fig:trunc_sis_example}
\end{figure}

Lenstronomy does not provide a truncated SIE model, so for this test, we simply use the truncated SIS model, and truncate at a radius of $r_{\textrm{trunc}}=5\times\theta_E$. Figure \ref{fig:trunc_sis_example} shows an example of the lensing effect of a lens using the SIS and truncated SIS models. The lensing effect is calculated as the difference between the image with lensing/gravity turned on, and the image with lensing/gravity turned off. This example shows how for the un-truncated SIS, the lensing effect continues far beyond the small Einstein radius.

We generate a validation set that is the same as that for Model 3 \edr{(JWST-small)} (see Section \ref{sec:methods_trained_class_model} for the simulation details) except for the change from SIE to a truncated SIS lens model. Then we evaluate our trained Model 3 \edr{(JWST-small)} and U-Net on this dataset.

\begin{figure}
    \centering
    \includegraphics[width=.9\textwidth]{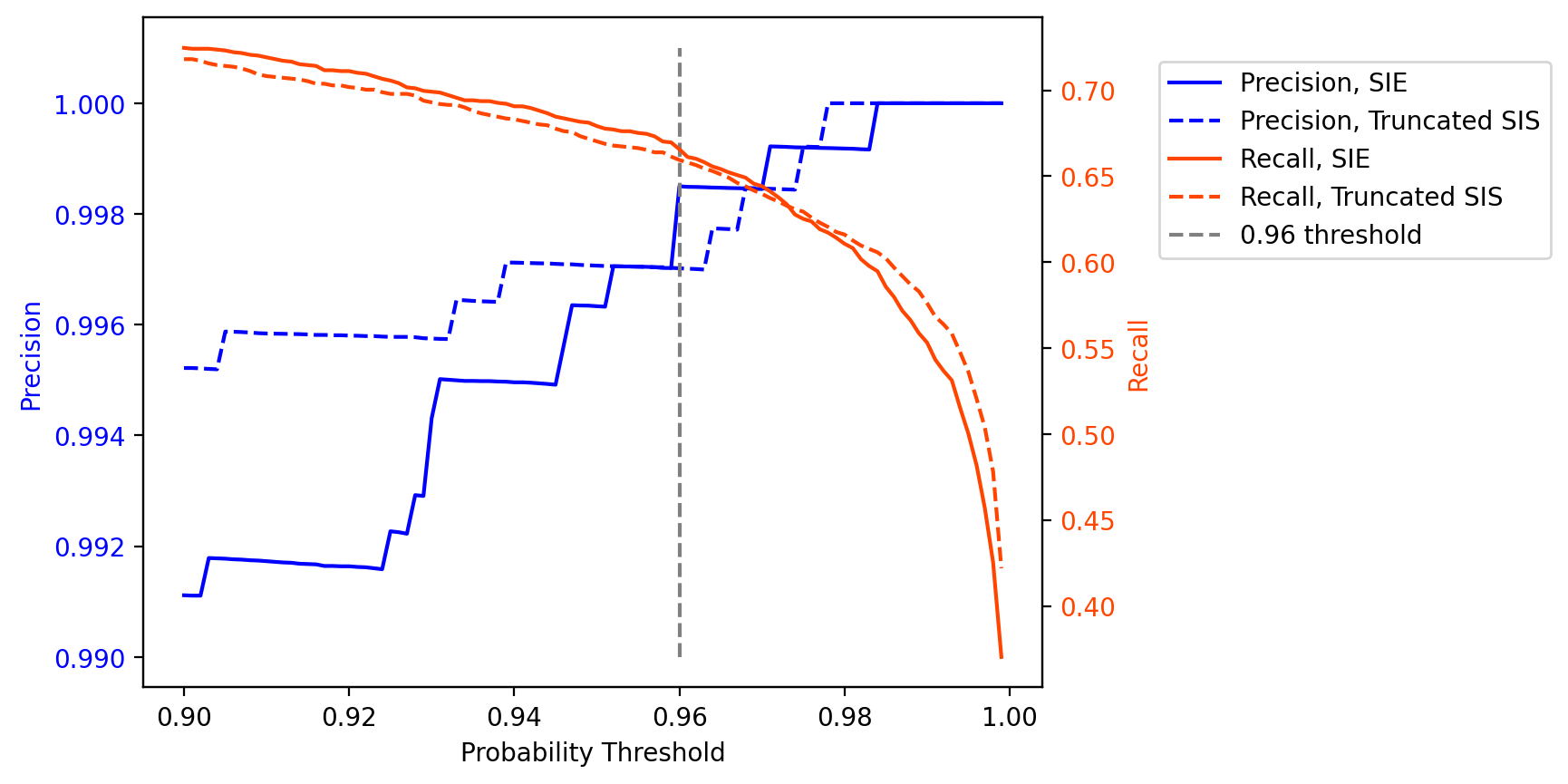}
    \caption{Precision and recall for Model 3 \edr{(JWST-small)} as a function of probability threshold. The solid lines show the performance on the original validation set with SIE lenses, and the dotted lines show the performance on the validation set with truncated SIS lenses.}
    \label{fig:trunc_sis_prc}
\end{figure}

\begin{figure}
    \centering
    \includegraphics[width=.7\textwidth]{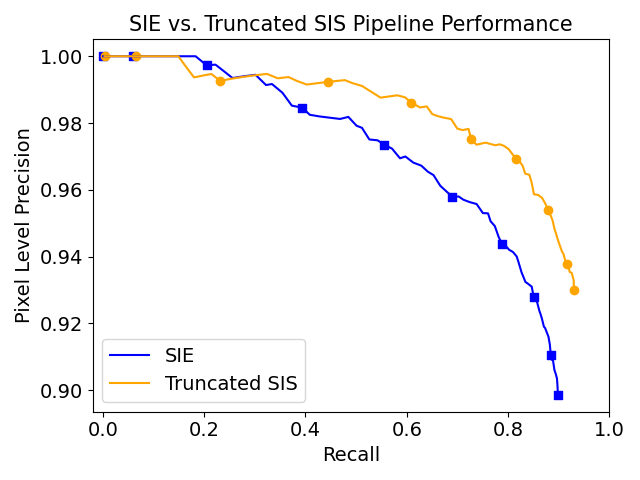}
    \caption{Precision-Recall curve for the ResNet and U-Net pipeline, evaluated on the SIS and truncated SIS simulations. Colored shapes represent U-Net probability thresholds that are multiples of 0.1 (leftmost is 0.9), corresponding to the curve of the same color. Across most probability thresholds, the pipeline performs slightly better on the truncated SIS simulation, suggesting that the U-Net model is not dependent on distortions that are far away from the lens when identifying the location of a lens.}
    % and detecting lens candidates.}
    \label{fig:trunc_SIS_comparison}
\end{figure}

\subsection{Results}
We show the resulting PRC curve from evaluating Model 3 \edr{(JWST-small)} and the ResNet and U-Net pipeline on the truncated SIS images in Figure \ref{fig:trunc_sis_prc} and Figure \ref{fig:trunc_SIS_comparison}, respectively. Overall, the ResNet results are similar at high probability thresholds to the original precision and recall curves for Model 3 \edr{(JWST-small)} shown here from Figure \ref{fig:new_precision_recall}. The recall with truncated SIS images is slightly lower at low probability thresholds, and slightly higher at high thresholds, and the precision is similar at a threshold of $0.96$ or above.

Because our results are not significantly degraded in high precision region that we are interested in when switching to a truncated SIS model, we feel satisfied that our ResNet and U-Net models are not relying on unphysical lensing effects for classification or detection. 
However, to ensure performance results based on simulations are robust, particularly for these very small Einstein radius systems, more physical models (such as truncated SIE) would be advisable
in future simulations.
% of these very small Einstein radius lenses would be made more accurate by considering truncated lens models, such as a truncated SIE model. 
More consideration and testing may be given to the lens model and parameters such as truncation radius. There is certainly more work to be done in understanding how to best model lenses with small Einstein radii in the most physical ways.

\section{Testing VELA Source Galaxies}\label{sec:vela_source_test}
\edr{
% One issue with our simulations is that the VELA simulations we use for source galaxies (and 20\% of environmental galaxies) only include 
In this work, as a remainder to the reader, we use the VELA simulations for source galaxies and 20\% of environmental galaxies. The VELA simulations include 34 distinct galaxies. Although the catalog provides images of these galaxies at many different redshifts and viewing angles, a possible concern might be that the model is only learning the appearance of these 34 galaxies, i.e. that we are ``overfitting'' to these particular galaxies. This is of greatest concern for the performance of Model~3 \edr{(JWST-small)}, where the details of the source galaxy matter the most.}
% are most prominent.}

\begin{figure*}
    \centering
    \includegraphics[width=0.8\linewidth]{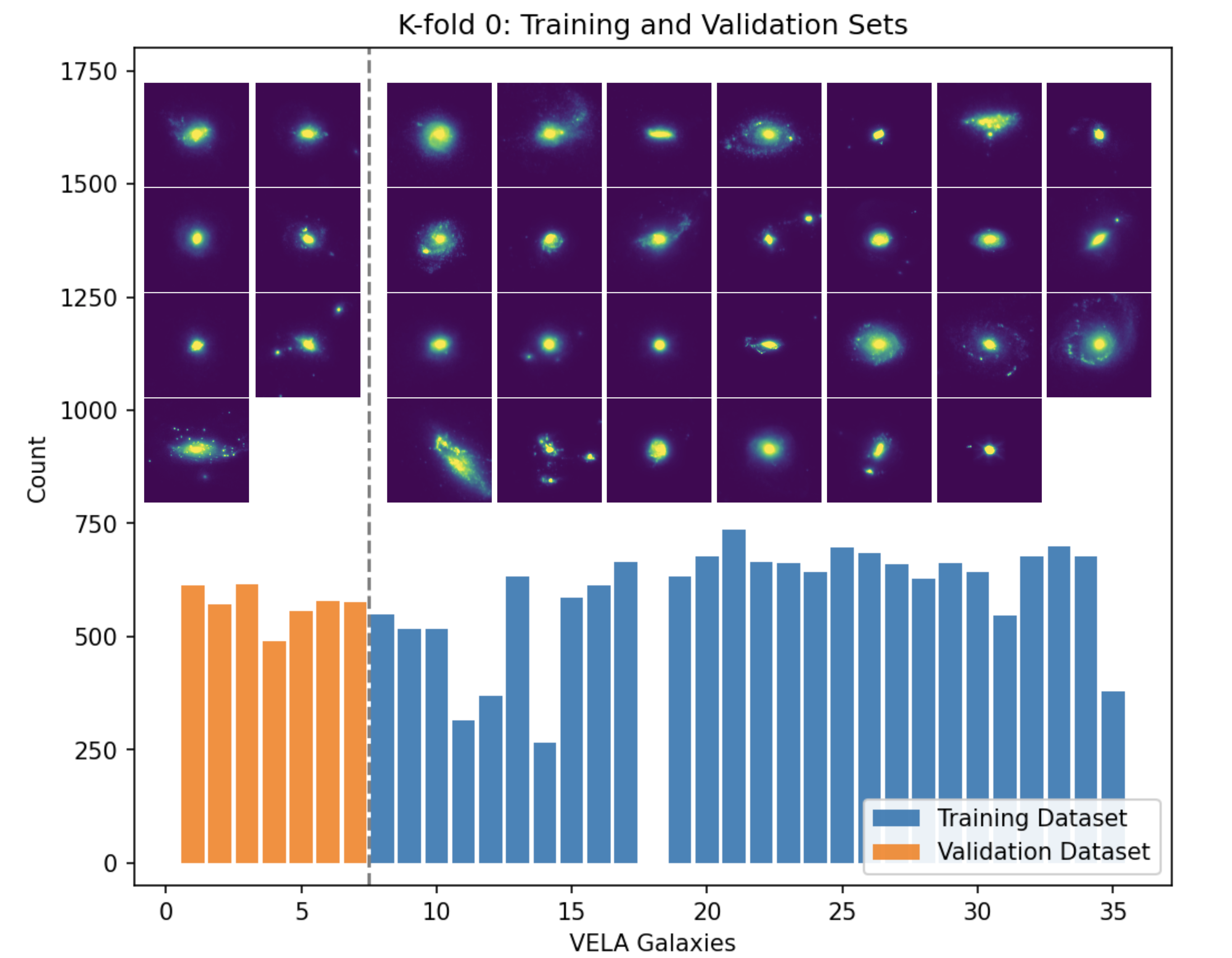}
    \caption{\edr{Histogram of the VELA galaxies used in the training (blue bars) and validation (orange bars) datasets for the $k$-fold 0th split. VELA images of the galaxies at $z=1$ with a randomly chosen angle are shown above the histogram.}}
    \label{fig:kfold_splits}
\end{figure*}

\edr{
% In order to test this, 
To address this question, we perform a $k$-fold split of the 34 VELA galaxy simulations into 5 groups of 6-7 galaxies each. For each split, we generate two datasets: the training dataset uses four sets (or 27-28 galaxies) for source and environmental galaxies, and the validation dataset uses the remaining set (6-7 galaxies). An example of the galaxies included in the training and validation sets for the first $k$-fold split is shown in Figure \ref{fig:kfold_splits}.\footnote{\edr{Note that there is no VELA galaxy 18. This is because there were originally 35 galaxy simulations, but galaxy 18 was unsuitable to generate mock images from (see: \url{https://archive.stsci.edu/prepds/vela/)}.}}}

\edr{For each split, we retrain Model~3 \edr{(JWST-small)} only using the training dataset without the group reserved for validation.
% any of the 6-7 set aside galaxies from appearing. 
We evaluate the performance of both the original baseline Model~3 \edr{(JWST-small)} and the retrained model on the group set aside for validation. 
If the baseline model (in \S~\ref{sec:model3_val_performanace}), which has seen some versions (at certain redshifts, with certain viewing angles) of the validation galaxies during training, perform a little better than the $k$-fold model, that is to be expected; but if its performance is significantly better, this could be an indication of overfitting in the baseline model.}
% , which did not see those galaxies during training.}

\begin{figure*}
    \centering
    \includegraphics[width=0.6\linewidth]{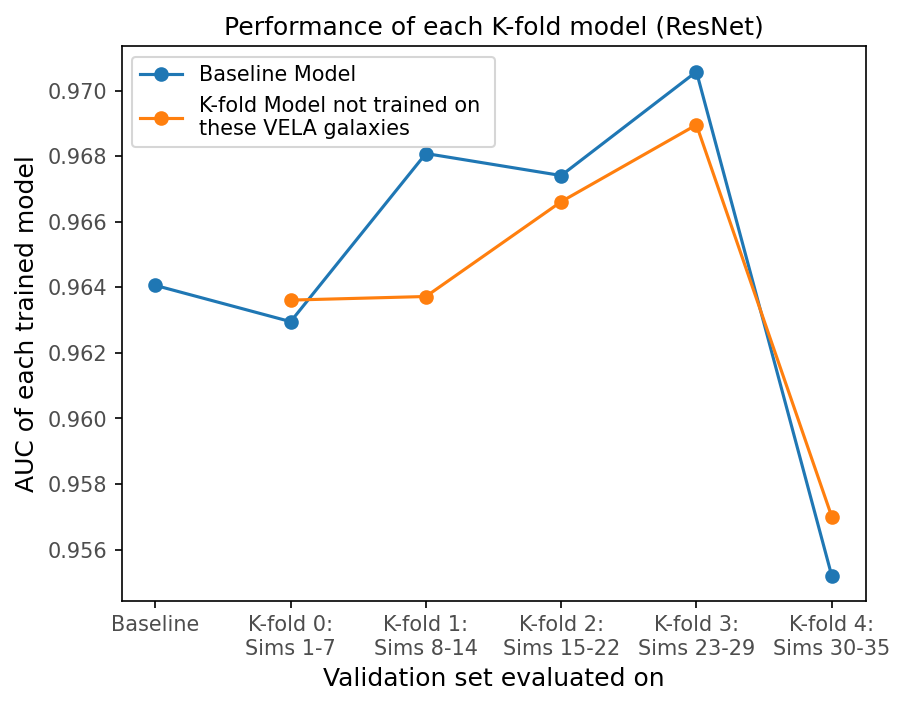}
    \caption{\edr{The AUC performance of each $k$-fold model evaluated on its corresponding validation set compared to the baseline model's performance (from \S~\ref{sec:model3_val_performanace}) for the ResNet model. The horizontal axis indicates the validation set used.}}
    \label{fig:kfold_results}
\end{figure*}

\begin{figure*}
    \centering
    \includegraphics[width=0.6\linewidth]{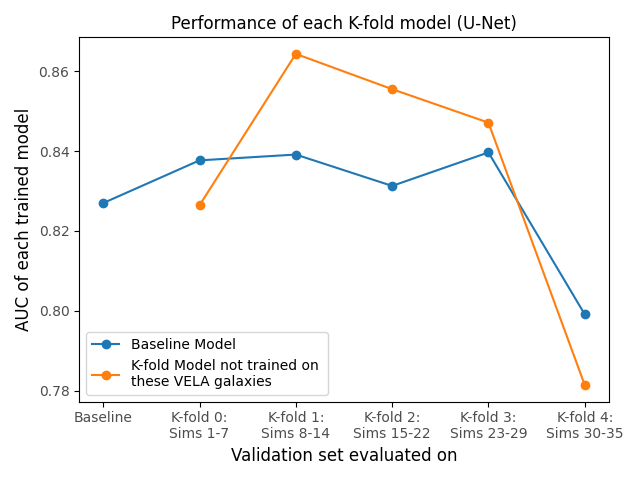}
    \caption{\edr{The AUC performance of each $k$-fold model evaluated on its corresponding validation set compared to the baseline model's performance (from \S~\ref{sec:model3_val_performanace}) for the U-Net model. The horizontal axis indicates the validation set used. Note that the AUC is calculated using the pixel-level precision recall curve, rather than the traditional PRC curve (see \S~\ref{sec:detection_pipeline}).} }
    \label{fig:kfold_results_unet}
\end{figure*}

\begin{figure}
    \centering
    \includegraphics[scale=0.45]{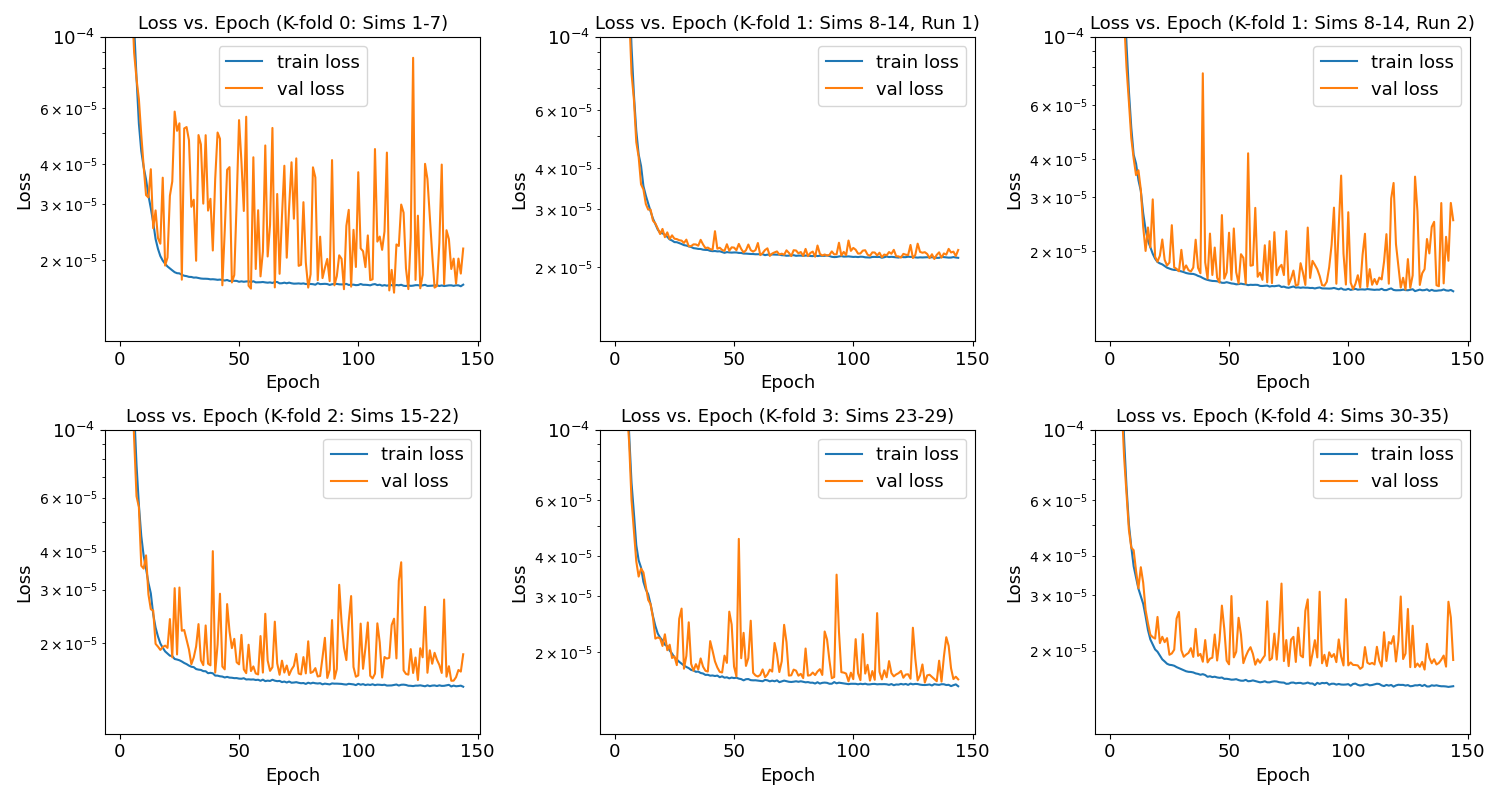}
    \caption{\edr{While training each $k$-fold model for Figure \ref{fig:kfold_results_unet}, we noticed that sometimes the validation loss has significantly less fluctuation than is typical and worse performance, even when trained using the same hyperparameters and dataset (compare the top middle and top right loss charts). This result is not specific to $k$-fold 1, as we also observed a similar instance for $k$-fold 3 (not shown in figure). Since learning rate is constant across all of these models, we suspect that the loss function calculation may be subject to high volatility when predicting for only one pixel in the image, but requires more exploration of the hyperparameter space to confirm. From our findings, we recommend performing a thorough search of the hyperparameter space, including learning rate, such that one can find a set of hyperparameters that delivers more consistent training results from one training session to the next. Due to the stochastic nature of neural network training, we also recommend training multiple models with the same hyperparameters and selecting the best one. While we did perform hyperparameter tuning for the U-Net model, a fuller exploration of the hyperparameter space is out of scope for this paper.}}
    %  When this phenomenon occurs, we believe that adjusting the learning rate could benefit model training. While we did perform hyperparameter tuning for the U-Net model, a fuller exploration of learning rate is out of scope for this paper. We recommend training multiple models and performing model selection to achieve best results.
    %  , resulting in a less robust exploration of the parameter space and therefore, reduced performance.
    \label{fig:unet_training_results}
\end{figure}

\edr{The results of this test are shown in Figure \ref{fig:kfold_results}. We observe that the retrained $k$-fold models do mostly perform a little worse than the baseline model, but the difference is minor. While the performance decreases for some $k$-fold splits, the baseline model performs similarly on these validation sets, indicating that some VELA source galaxies are either a little easier or harder for both trained models to classify. Therefore, we conclude that ``overfitting'' is not a significant issue with using these 34 VELA simulations. 
Of course, a greater variety and a larger number of detailed 
galaxy simulations remains highly desirable.
% still improve performance.
}

\section{Testing VELA and S\'{e}rsic Environmental galaxies}
\label{sec:sersic_env_test}

\begin{figure*}
    \centering
    \includegraphics[width=0.6\linewidth]{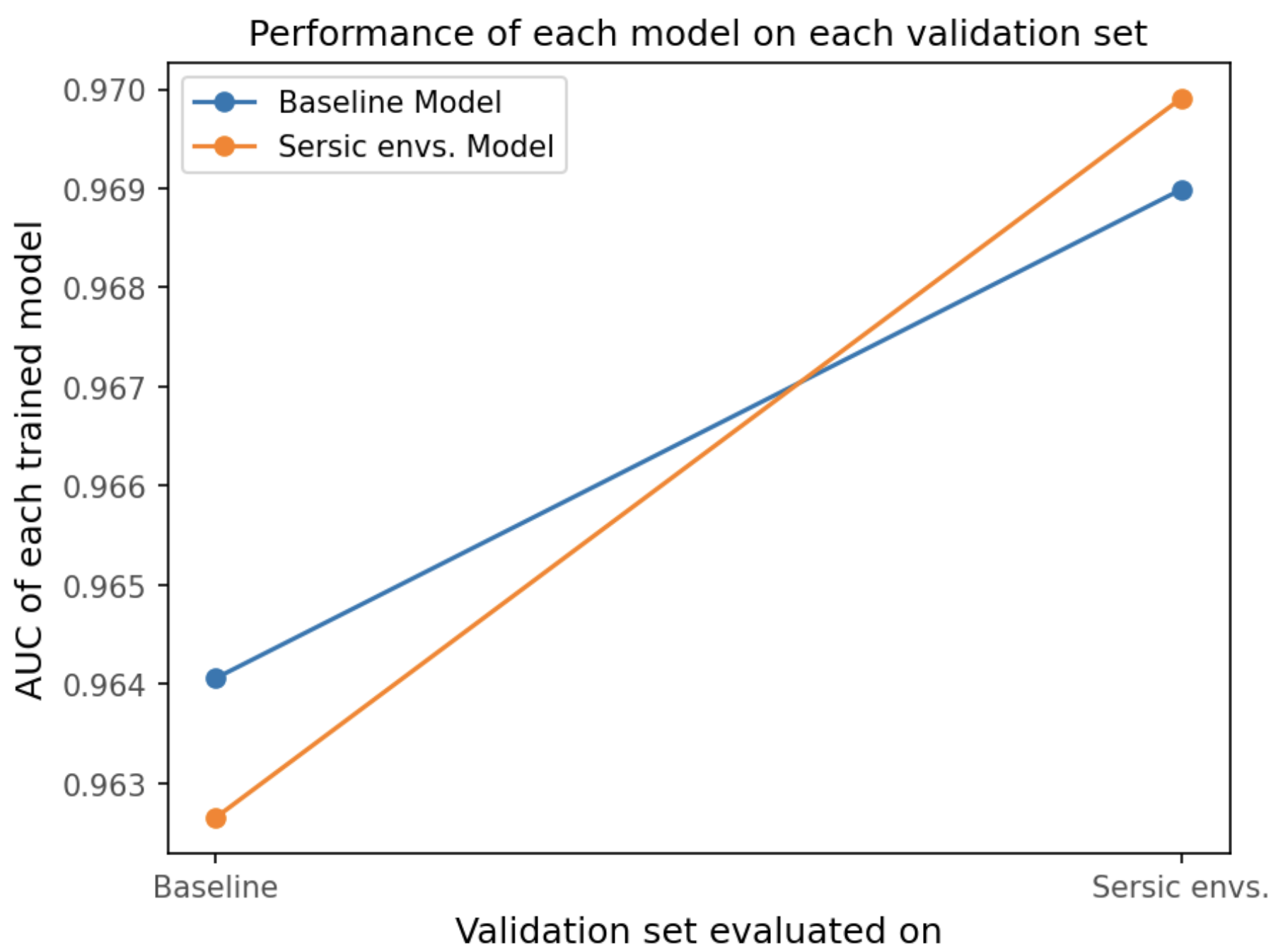}
    \caption{\edr{The AUC performance of the model trained with only S\'{e}rsic environmental galaxies compared to the baseline model (from \S~\ref{sec:model3_val_performanace}).}}
    \label{fig:sersic_env_test}
\end{figure*}

\edr{As discussed in Section \ref{sec:env_gals}, in our simulations, the environmental galaxies are 80\% S\'{e}rsic galaxies, and 20\% VELA galaxies. In order to test how much of a difference this choice makes on the performance of the trained model, we retrained a version of Model~3 \edr{(JWST-small)} with only S\'{e}rsic environmental galaxies.
The performance of the model compared to the baseline Model~3 \edr{(JWST-small)} is shown in Figure \ref{fig:sersic_env_test}. Overall, the models have similar performance. 
Unsurprisingly, the baseline model is slightly better when evaluated on the baseline validation set, and the retrained model is slightly better when evaluated on the validation set with only S\'{e}rsic environmental galaxies. However, for both validation sets, their AUC scores only vary by about $\sim0.01$.
Thus we determine that the performance of our Model~3 \edr{(JWST-small)} is insensitive to the particular choice of 80\% S\'{e}rsic galaxies and 20\% VELA galaxies.}
% ,so the difference is quite minor.}

\section{Lens Detection Examples} \label{sec:individual_examples}

% Below, we provide some examples of lens images from the U-Net predictions on the validation set from the
% ResNet and U-Net pipeline in Figures \ref{fig:ex1_easy}-\ref{fig:ex6_far}.

First we want to make it clear that in this section all three systems shown have a lower U-Net detection probability than our chosen threshold of 0.72. 
\emph{We remind the reader that there are no false positives above that threshold for the RUN pipeline (see Figure \ref{fig:PRC})}.

\begin{figure}
    \centering
    \includegraphics[scale=0.45]{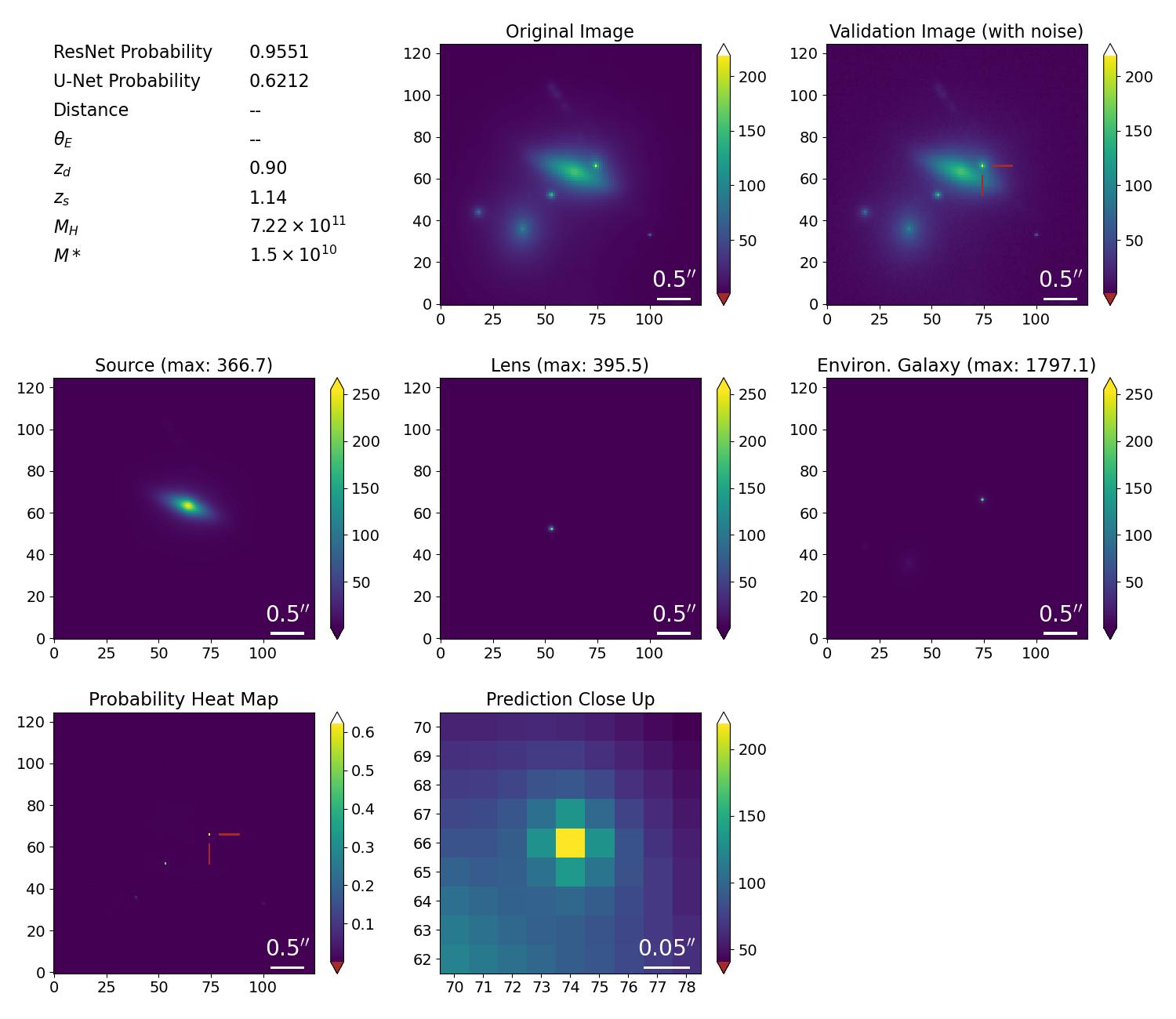}
    \caption{A non-lens that is classified as a lens by the ResNet and U-Net model (false positive). We note that for this system, the predicted probabilities are below the chosen thresholds of 0.96 and 0.72 for ResNet and U-Net, respectively. 
    There is a very bright environmental galaxy at the location of the prediction that caused 
    % The environmental galaxy is deceivingly similar to a lens, causing 
    the U-Net and even the ResNet to incorrectly predict a lens. The U-Net also outputs a detection probability of 0.494 for the environmental galaxy located at (53, 52).}
    \label{fig:ex3_fp}
\end{figure}

\begin{figure}
    \centering
    \includegraphics[scale=0.45]{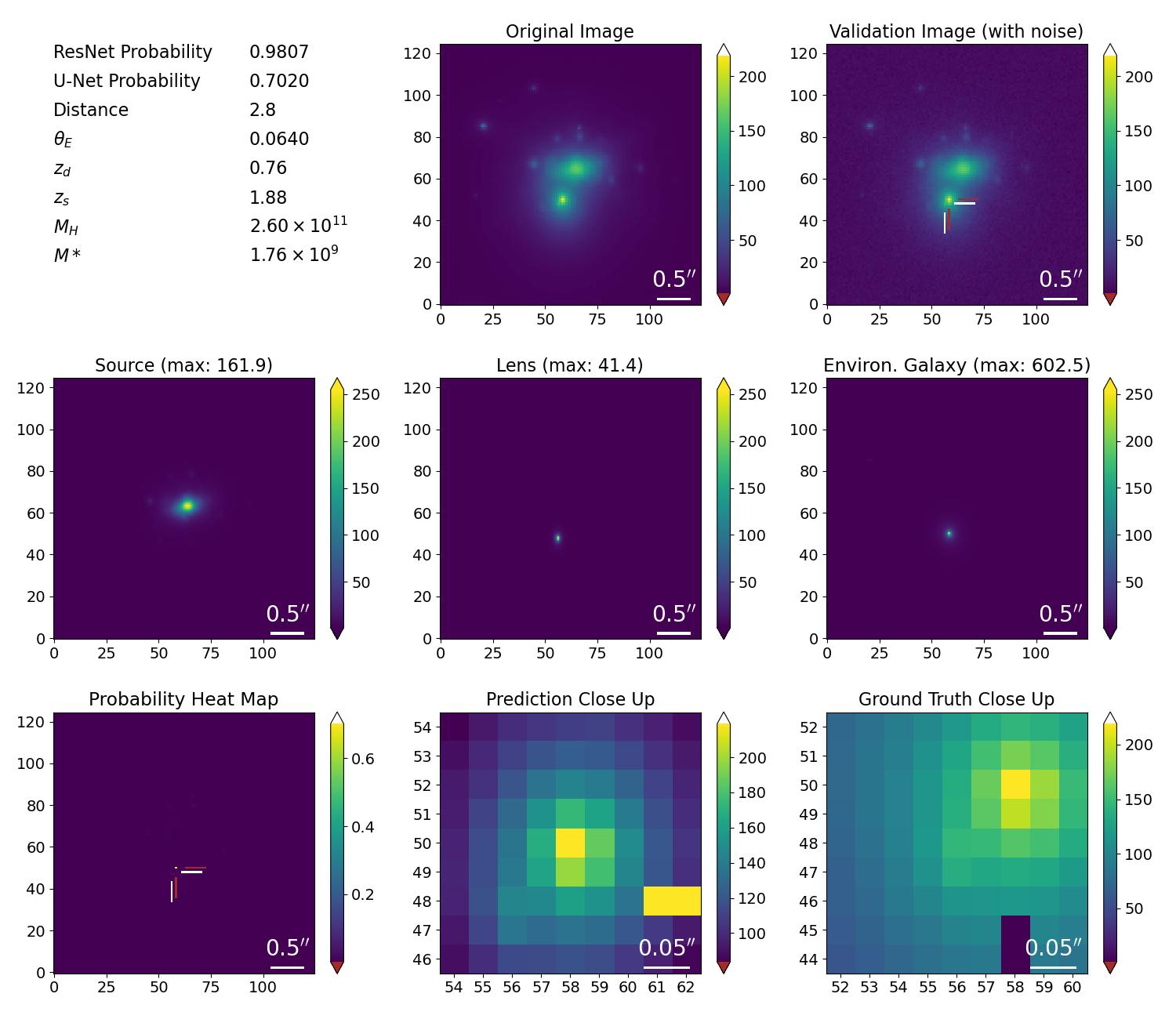}
    \caption{A lens that is correctly classified by the ResNet (with a probability higher than the chosen threshold of 0.96), but incorrectly detected by the U-Net (of course, with a probability lower than the chosen threshold of 0.72). 
    Since an environmental galaxy in the image is much brighter than the lens and very close to it, \ed{with a separation of only 2.8 pixels (see the ``Distance'' quantity at the top left)}, a local maximum does not occur at the location of the lens, but at the center of the environmental galaxy. Generally, a local maximum is a necessary but not sufficient condition for detecting a lens (see the discussion in \S\,\ref{sec:pipeline_performance}).
    % , which deceives the model. 
    % we verified that the model is not just predicting the brightest pixel in the image (see Figure \ref{fig:panel_tpc}), 
    % as well as the fact that 
    We have verified that the U-Net model is indeed confused 
    % deceived 
    by the environmental galaxy:
    % by predicting on the image 
    without the environmental galaxy, the U-Net predicts the correct coordinate with a probability of 0.576. 
    % Practically speaking, due to the fact that 
    % the overlap between 
    % the lens and the environmental galaxy virtually overlap, 
    % it would be extremely difficult to inspect this lens regardless of whether the U-Net is able to detect it correctly or not.
    However, one expects this kind of coincidence---the lens and the environmental galaxy virtually overlap---to be very rare.}
    \label{fig:ex4_close_wrong}
\end{figure}

\begin{figure}
    \centering
    \includegraphics[scale=0.45]{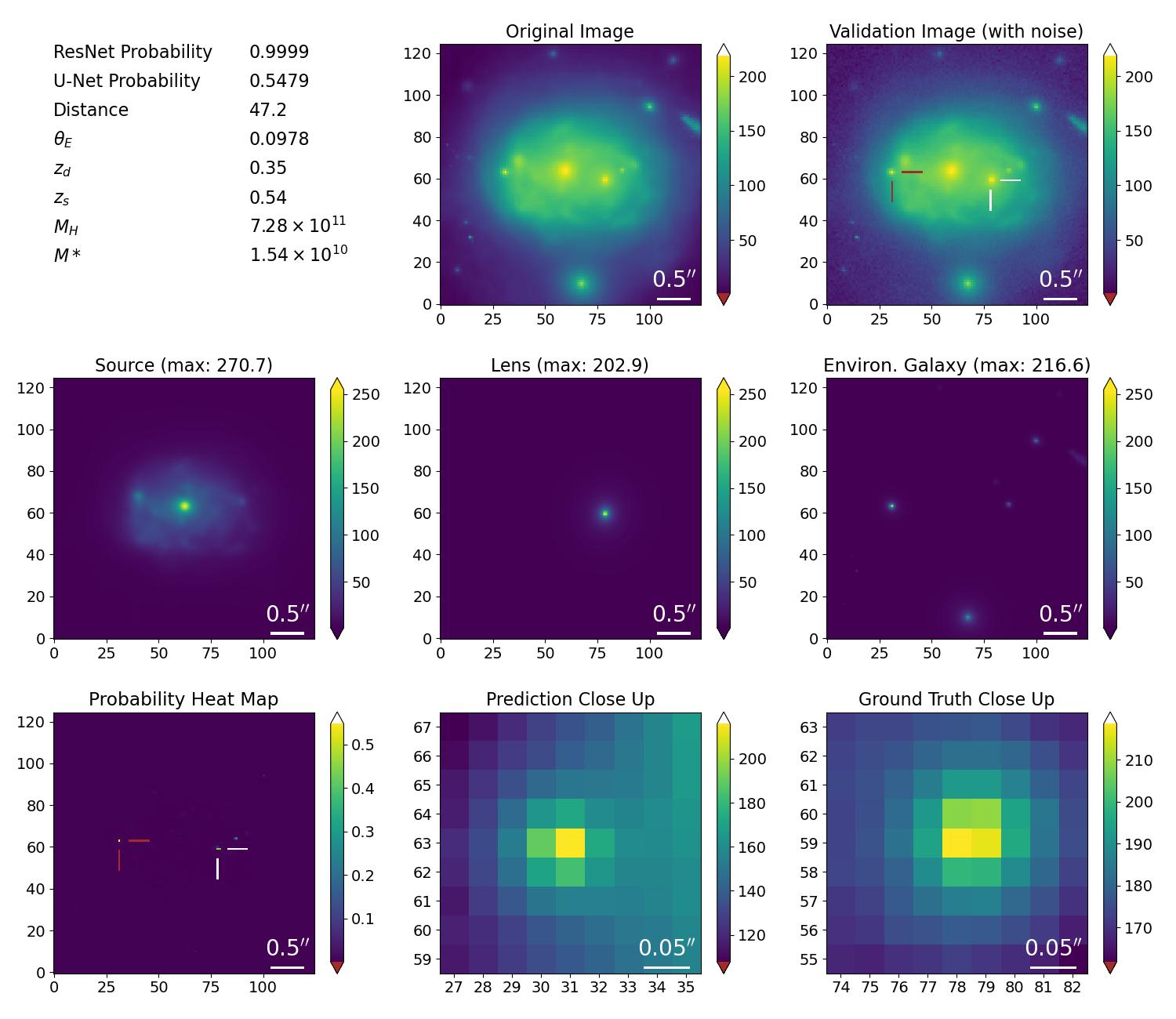}
    \caption{A lens that is correctly classified by the ResNet (with a probability higher than the chosen threshold of 0.96), but incorrectly detected by the U-Net (again, with a probability lower than the chosen threshold of 0.72).
    In this case, the model's prediction (red crosshairs) is far away from the ground truth (white crosshairs), with a separation (``Distance") of 47.2 pixels. Most of these cases occur when there is a bright environmental galaxy paired with a dim lens in a complex system.}
    % , making it quite difficult for the model to identify a lens correctly.}
    \label{fig:ex6_far}
\end{figure}

In Figures \ref{fig:ex3_fp}, \ref{fig:ex4_close_wrong}, and \ref{fig:ex6_far}, we show examples where bright environmental galaxies deceive the U-Net (and also the ResNet in the case of Figure \ref{fig:ex3_fp}), which are the largest source of detection false positives.
However, we expect this kind of coincidence to be rare.
Each image's scaling methodology is the
same as described in preprocessing (see Section \ref{sec:image_preprocessing}), meaning that pixel values cannot be directly compared across images. 
% Orange and white cross hairs represent
% predictions and ground truths, respectively, for lens cutout images.
Red cross hairs represent false positives for non-lens cutout images. 
For these figures, recall that ``Distance'' is defined as the Euclidean distance between the ground truth and model prediction; 
therefore ``Distance'', and Einstein radius ($\theta_E$), are irrelevant when the image is not a lens.

% 1, 5, 2, 3, 4, 6

% Figure \ref{fig:ex1_easy} shows the most common type of detection from the U-Net, a relatively straightforward detection for the model. Figure \ref{fig:ex5_close_right} shows a U-Net detection that is one pixel away from the ground truth, which is considered correct (see Section  \ref{sec:initial_training}). Figure \ref{fig:ex2_hard} shows a correct detection where the U-Net does not just predict a very bright local maximum in the image, providing promising evidence that the U-Net can identify features of a lens.\footnote{We also performed other tests to show that the U-Net can identify the lensing effect in Figure \ref{fig:high_confidence_lens}.}

\end{document}